\documentclass[fleqn]{2020SCGE}
\usepackage{amsfonts}
\usepackage{mathrsfs}
\usepackage{amsmath}
\usepackage{amssymb}
\usepackage{framed}
\usepackage{multirow, tabularx}
\usepackage{orcidlink}
\usepackage[normalem]{ulem}
\usepackage{extarrows}
\usepackage{slashed}
\usepackage{isodateo}
\usepackage{graphicx}
\usepackage{xcolor}

\graphicspath{{fig/}}

\usepackage{mdframed}

\newmdenv[skipabove=0pt,%
          skipbelow=5pt,%
          leftmargin=0pt,%
          rightmargin=0pt,%
          innertopmargin=-5pt,%
          innerbottommargin=7pt,%
          innerleftmargin=2pt,%
          innerrightmargin=2pt,%
          splittopskip=0pt,%
          splitbottomskip=0pt,%
          linewidth=0pt,%
          nobreak=true]%
          {keyeqn2}

\newmdenv[backgroundcolor=gray!15,%
          skipabove=0pt,%
          skipbelow=5pt,%
          leftmargin=0pt,%
          rightmargin=0pt,%
          innertopmargin=-5pt,%
          innerbottommargin=7pt,%
          innerleftmargin=2pt,%
          innerrightmargin=2pt,%
          splittopskip=0pt,%
          splitbottomskip=0pt,%
          linewidth=0pt,%
          nobreak=true]%
          {keyeqn}

\begin{document}

\ensubject{subject}

%Letter to the Editor??Article%??????
\ArticleType{Article}%??Article
%\SpecialTopic{SPECIAL TOPIC: Forum}%???????
\Year{2026}
\Month{January}
\Vol{69}
\No{1}
\DOI{10.1007/s11433-025-2740-8}
\ArtNo{210401}
%\ReceiveDate{Jan, 2025}
%\AcceptDate{**, 2025}
%\OnlineDate{**, 2025}

\title{Gravitational wave cosmology}

%\vspace{20mm}
\author[1]{Ligong Bian}{}%{lgbycl@cqu.edu.cn}
\author[2]{Rong-Gen Cai}{{cairg@itp.ac.cn}}
\author[3,4]{Yu-Qi Dong}{}%{dongyq2023@lzu.edu.cn}
\author[5]{Qing Gao\,\orcidlink{0000-0003-3797-4370}}{}%{{gaoqing1024@swu.edu.cn}}
\author[2]{Yungui Gong \orcidlink{0000-0001-5065-2259}}{gongyungui@nbu.edu.cn}
\author[6,7,8]{\\Zong-Kuan Guo}{guozk@itp.ac.cn}
\author[6,7,8]{Qing-Guo Huang}{huangqg@itp.ac.cn}
\author[9]{Zhaofeng Kang}{}%{zhaofengkang@gmail.com}
\author[6,7,8]{Li Li}{}%{liliphy@itp.ac.cn}
\author[10]{Jing Liu}{}%{liujing@ucas.ac.cn}
\author[11]{\\Lang Liu}{}%{liulang@bnu.edu.cn}
\author[3,4]{Yu-Xiao Liu}{}%{liuyx@lzu.edu.cn}
\author[2]{Xuchen Lu}{}
\author[11]{Zhi-Zhang Peng}{}%{pengzhizhang@bnu.edu.cn}
\author[7]{Jin Qiao}{}%{qiaojin@ucas.ac.cn}
\author[12]{Puxun Wu}{}%{pxwu@hunnu.edu.cn}
\author[10,7]{\\Yue-Liang Wu}{ylwu@ucas.edu.cn}
\author[6,7,8]{Jiang-Hao Yu}{}%{jhyu@itp.ac.cn}
\author[13]{Chen Yuan}{}%{chenyuan@tecnico.ulisboa.pt}
\author[2]{Chao Zhang}{}
\author[14,15]{Tao Zhu \orcidlink{0000-0003-2286-9009}}{zhut05@zjut.edu.cn}

\AuthorCitation{Ligong Bian, Rong-Gen Cai, Yu-Qi Dong, et. al., Gravitational wave cosmology}

\address[1]{Department of Physics and Chongqing Key Laboratory for Strongly Coupled Physics, Chongqing University, Chongqing 401331, China}
\address[2]{Institute of Fundamental Physics and Quantum Technology, Department of Physics, School of Physical Science and Technology, \\ Ningbo University, Ningbo, Zhejiang 315211, China}
\address[3]{Lanzhou Center for Theoretical Physics, Key Laboratory of Theoretical Physics of Gansu Province, \\ Key Laboratory of Quantum Theory and Applications of MoE, \\ Gansu Provincial Research Center for Basic Disciplines of Quantum Physics, \\ Lanzhou University, Lanzhou 730000, China}
\address[4]{Institute of Theoretical Physics \& Research Center of Gravitation, Lanzhou University, Lanzhou 730000, China}
\address[5]{School of Physical Science and Technology, Southwest University, Chongqing 400715, China}
\address[6]{Institute of Theoretical Physics, Chinese Academy of Sciences, Beijing 100190, China}
\address[7]{School of Fundamental Physics and Mathematical Sciences, Hangzhou Institute for Advanced Study, \\ University of Chinese Academy of Sciences, Hangzhou 310024, China}
\address[8]{School of Physical Sciences, University of Chinese Academy of Sciences, Beijing 100049, China}
\address[9]{School of physics, Huazhong University of Science and Technology, Wuhan 430074, China}
\address[10]{\;International Centre for Theoretical Physics Asia-Pacific\textcolor{black}{, University of Chinese Academy of Sciences, Beijing 100190, }China}
\address[11]{\;School of Physics and Astronomy, Beijing Normal University, Beijing 100875, China}
\address[12]{\;Department of Physics, Institute of Interdisciplinary Studies, \\ Hunan Research Center of the Basic Discipline for Quantum Effects and Quantum Technologies, \\ and Synergetic Innovation Center for Quantum Effects and Applications, Hunan Normal University, Changsha, Hunan 410081, China}
\address[13]{\;CENTRA, Department of Physics, Instituto Superior Técnico, University of Lisbon, 1049-001 Lisbon, Portugal}
\address[14]{\;Institute for Theoretical Physics and Cosmology, Zhejiang University of Technology, Hangzhou, 310023, China}
\address[15]{\;United Center for Gravitational Wave Physics (UCGWP), Zhejiang University of Technology, Hangzhou, 310023, China}

\date{}

\abstract{
Gravitational waves (GWs) originating from cosmological sources offer direct insights into the physics of the primordial Universe, the fundamental nature of gravity, and the cosmic expansion of the Universe. In this review paper, we present a comprehensive overview of our recent advances in GW cosmology, supported by the national key research and development program of China, focusing on cosmological GW sources and their implications for fundamental physics and cosmology. We first discuss the generation mechanisms and characteristics of stochastic gravitational wave backgrounds generated by physical processes occurred in the early Universe, including those from inflation, phase transitions, and topological defects, and summarize current and possible future constraints from pulsar timing array and space-based detectors. Next, we explore the formation and observational prospects of primordial black holes as GW sources and their potential connection to dark matter. We then analyze how GWs are affected by large-scale structure, cosmological perturbations, and possible modifications of gravity on GW propagation, and how these effects can be used to test fundamental symmetry of gravity. Finally, we discuss the application of GW standard sirens in measuring the Hubble constant, the expansion history, and dark energy parameters, including their combination with electromagnetic observations. These topics together show how GW observations, especially with upcoming space-based detectors, such as LISA, Taiji, and Tianqin, can provide new information about the physics of the early Universe, cosmological evolution, and the nature of gravity. \Authorfootnote}

%\keywords{Gravitational wave, cosmological phase transition, primordial black holes, parity violation, Lorentz violation, standard siren, stochastic gravitational wave background}
\PACS{04.30.-w, 04.30.Nk, 98.80.-k, 98.80.Cq }
\maketitle

\begin{multicols}{2}

%%%%%%%%%%%%%%%%%%%%%%%%%%%%%%%%%%%%%%%%%%%%%%%%%%%%%%%%%%%%%%%%%%%%%%%%%%%%%%%%%%%%%%
\section{Introduction}
%%%%%%%%%%%%%%%%%%%%%%%%%%%%%%%%%%%%%%%%%%%%%%%%%%%%%%%%%%%%%%%%%%%%%%%%%%%%%%%%%%%%%%
%contributor:

The direct detection of the first gravitational waves (GWs) from the merger of a binary black hole system by the advanced laser interferometer gravitational-wave observatory (aLIGO) and Virgo has offered new opportunities to explore the Universe and fundamental physics \cite{LIGOScientific:2016emj, LIGOScientific:2016vbw, LIGOScientific:2016vlm}. Following the first detection, the LIGO-Virgo-KAGRA (LVK) scientific collaboration has reported nearly 90 GW detections in the following years \cite{LIGOScientific:2017vwq, LIGOScientific:2016sjg, LIGOScientific:2018mvr, LIGOScientific:2020ibl, KAGRA:2021vkt}, providing a unique and independent means to study violent astrophysical processes and test the nature of gravity. In addition, several pulsar timing arrays (PTA) collaborations \cite{NANOGrav:2023gor,Reardon:2023gzh,Xu:2023wog,EPTA:2023sfo}, including the North American Nanohertz Observatory for Gravitational Waves (NANOGrav)~\cite{NANOGrav:2023gor}, European Pulsar Timing Array (EPTA) \cite{EPTA:2023sfo}, Parkes Pulsar Timing Array (PPTA) \cite{Reardon:2023gzh}, and Chinese Pulsar Timing Array (CPTA) \cite{Xu:2023wog}, announced strong evidence for a stochastic gravitational wave background (SGWB) at nanohertz frequencies, opening yet another observational window into the GW spectrum.

With these achievements, the next-generation detectors, including space-based missions such as LISA \cite{Audley:2017drz}, Taiji \cite{Ruan:2018tsw, Hu:2017mde,Luo:2021qji}, and Tianqin \cite{Luo:2025ewp}, along with advanced ground-based observatories like KAGRA \cite{KAGRA:2020agh}, Voyager \cite{LIGO:2020xsf}, the Einstein Telescope (ET) \cite{Amann:2020jgo, Sathyaprakash:2012jk}, and Cosmic Explorer (CE) \cite{Hall:2020dps, Dwyer:2014fpa}, are poised to extend GW observations to unprecedented redshifts, up to $z\sim 20$. These detectors will access not only high-redshift astrophysical sources but also cosmological GW sources originating from the physical processes in the early Universe, such as inflation, phase transitions, cosmic strings, primordial black holes (PBHs), and inflation. 

The detection and characterization of such cosmological GW sources plays a crucial role in probing cosmology. They offer direct insight into the physics of the primordial Universe, enable independent measurements of the Hubble constant and cosmic expansion history using GW standard sirens, and provide constraints on models of dark energy, dark matter, and modifications to general relativity (GR). Thus, future GW detections are expected to profoundly deepen our understanding of the physics of the early Universe, the cosmic evolution, cosmological structure, and the nature of gravity. The main purpose of this paper is to provide a comprehensive overview of our recent advances on these mentioned subjects. 

Cosmological GW sources are closely related to the physical processes that happened in the early Universe, such as inflation~\cite{Bartolo:2016ami}, cosmological first-order phase transitions~\cite{Caprini:2015zlo}, and networks of topological defects~\cite{Auclair:2019wcv}, which lead to the generation of SGWBs. Such GWs travel freely in the early Universe, carrying information about the processes that produced them. The measurement of SGWBs opens a new window for exploring the physics of the early Universe or the physics of particles beyond the Standard Model (SM). Therefore, it is necessary to investigate the characteristic properties of SGWBs, which allow them to be distinguished by future GW detectors. Our review will focus on some production mechanisms of SGWBs during inflation, the electroweak phase transition, quantum chromodynamics (QCD) phase transition, and cosmic domain wall decay, and presents constraints on cosmological phase transitions from PTA observations.

% with Sec. 3
Another important cosmological source is the PBHs. PBHs are supposed to originate from the gravitational collapse of the overdense region in the primordial Universe \cite{Zeldovich:1967lct,Hawking:1971ei,Carr:1974nx}. PBHs not only offer a compelling explanation for the GW events observed by the LVK collaboration, but also represent a promising candidate for dark matter in the Universe. Both the GWs from the binary PBHs and the scalar induced GWs, inevitably associated with the formation of PBHs, provide important sources for the space-based GW detectors. In this review, we discuss mechanisms for generating PBHs and the features associated with PBHs for distinguishing PBHs from astrophysical ones. 

% with Sec. 4
On the other hand, the detection of GWs either from cosmological sources or astrophysical sources offers an unprecedented approach to explore the large-scale structure and composition of the inhomogeneous Universe and the fundamental nature of gravity. GWs originating from distant astrophysical sources propagate across cosmic distances through this inhomogeneous Universe before arriving at GW detectors, either space- and/or ground-based ones. The cumulative effects of cosmological structures and potential deviations from GR on GW propagation are expected to become significant over such vast distances. In particular, studies involving the breaking of Lorentz symmetry and parity symmetry suggest that GW observations can provide valuable insights into the fundamental symmetries of gravity, thereby advancing our understanding of its nature and potentially guiding the development of quantum gravity. We thus review recent theoretical and observational progress in utilizing GW propagation to explore the properties of cosmic large-scale structures, probe the nature of gravity, and search for possible signatures of new physics, including Lorentz and parity violations.

GWs also provide highly precise measurements of luminosity distances, making them promising standard sirens \cite{Schutz:1986gp, Holz:2005df} and an independent method for probing the evolution and composition of the Universe. To circumvent the zero-point calibration problem suffered by type Ia supernovae (SNe Ia) while leveraging their model-independent measurements on fundamental cosmological parameters, this review discusses calibrating SNe Ia using GWs from massive binary black hole (MBBH) or binary neutron star (BNS) coalescences, and summarize the progress and future outlook in GW-based cosmology.

In this review paper, we present a comprehensive overview of our recent advances in GW cosmology.  Specifically, we have conducted comprehensive analyses in the following four directions: (1) the production mechanisms and spectral features of SGWBs generated during inflation, the electroweak and QCD phase transitions, and cosmic domain wall decay and the current and prospective observational constraints on cosmological phase transitions derived from PTA data; (2) the formation scenarios and observational signatures of PBHs, including their role as potential sources for stochastic and transient GW signal; (3) the influence of cosmic large-scale structures and modified gravities on the propagation and polarizations of GWs, as well as the implications for probing the fundamental symmetries of gravity; (4) the application of GW standard sirens as independent cosmological probes, including their use in measuring the expansion history of the Universe and determining the Hubble constant.

The paper is organized as follows. Sec .~\ref{sec:cs} is devoted to the theoretical mechanisms and observational prospects for SGWBs originating from the early Universe. In Sec .~\ref{sec:pbh}, we discuss the formation, detection prospects, and cosmological implications of PBHs. Sec.~\ref{sec:tg} focuses on the use of GWs as probes of the large-scale structure and the fundamental properties of gravity, and Sec.~\ref{sec:cp} discusses the application of GW standard sirens to probe cosmology. Finally, we summarize our main conclusions and outline future directions in Sec.~\ref{sec:co}.

%%%%%%%%%%%%%%%%%%%%%%%%%%%%%%%%%%%%%%%%%%%%%%%%%%%%%%%%%%%%%%%%%%%%%%%%%%%%%%%%%%%%%%
\section{Cosmological sources \label{sec:cs}}
%%%%%%%%%%%%%%%%%%%%%%%%%%%%%%%%%%%%%%%%%%%%%%%%%%%%%%%%%%%%%%%%%%%%%%%%%%%%%%%%%%%%%%
%contributor: Zong-Kuan Guo

\subsection{GWs produced during inflation}

Inflation, a period of accelerated expansion in the early Universe, not only generated the primordial density fluctuations seeding the large-scale structure~\cite{Mukhanov:1981xt} but also produced a stochastic background of primordial GWs (PGWs)~\cite{Starobinsky:1979ty,Rubakov:1982df}. PGWs are tensor perturbations of spacetime, arising from quantum fluctuations in the metric field during inflation. PGWs induce characteristic quadrupole anisotropies in the photon field within the last scattering surface, generating a unique B-mode polarization signature in the cosmic microwave background (CMB)~\cite{Verkhodanov:2021ddu}. Consequently, precision measurements of large-scale CMB B-mode polarization are regarded as the most promising approach to detect PGWs. Furthermore, predictions for PGWs vary markedly across different inflationary models, primarily quantified by the tensor-to-scalar ratio $r$. For instance, in the single-field slow-roll inflation, $r$ is related to the energy scale of inflation by $V =  \left( 1.88 \times 10^{16} \, {\rm GeV} \right)^{4} (r / 0.1)$~\cite{Guzzetti:2016mkm,Guo:2010mm}. Recent joint analyses of BICEP/Keck and Planck 2018 data~\cite{Planck:2018jri,BICEP:2021xfz} impose an upper limit $r_{0.05} < 0.036$ (95\% CL) for a scale-invariant tensor power spectrum. Upcoming projects such as LiteBIRD~\cite{Matsumura:2013aja} aim to refine measurements of $r$, offering a decisive test to differentiate between inflationary models.

While single-field slow-roll inflation predicts a nearly scale-invariant GW spectrum, deviations from this paradigm could imprint distinct spectral signatures. Beyond the standard inflationary model, GWs can be produced during inflation through multiple channels: first-order scalar perturbations, including the inflationary perturbations~\cite{Matarrese:1997ay,Cai:2018tuh,Cai:2019jah,Cai:2019bmk,Zhou:2020kkf,Peng:2024eok,Fang:2025tgk} and perturbations of the extra field such as the curvaton~\cite{Bartolo:2007vp,Enqvist:2008be,Suyama:2011pu,Kawasaki:2013xsa,Chen:2019zza} and spectator field~\cite{Biagetti:2013kwa,Biagetti:2014asa,Fujita:2014oba,Inomata:2022ydj}, act as classical secondary sources for tensor modes, particle production during inflation, including transient bursts of scalar particles~\cite{Romano:2008rr,Barnaby:2009mc,Green:2009ds,Cook:2011hg,Carney:2012pk,Fedderke:2014ura,Ozsoy:2014sba,Goolsby-Cole:2017hod,Yu:2023ity} or tachyonic instability in gauge fields coupled to the pseudo-scalar inflaton~\cite{Barnaby:2010vf,Sorbo:2011rz,Barnaby:2011vw,Barnaby:2011qe,Anber:2012du,Ferreira:2014zia,Domcke:2016bkh,Peloso:2016gqs,Cheng:2018yyr,Ozsoy:2020kat,He:2024bno,He:2025ieo}, generates anisotropic stress sourcing GWs; and modified gravity theories (e.g., Horndeski~\cite{Kobayashi:2011nu}, Higgs G-inflation~\cite{Kamada:2010qe}) introduce deviations like altered propagation speeds ($c_T \neq 1$) or blue-tilted spectra ($n_T > 0$) while preserving energy conditions.

Our recent works explore novel mechanisms for the production of GWs during inflation. In Ref.~\cite{Peng:2021zon}, we investigated GWs produced in a single-field model with a periodic modulation superimposed on the inflaton potential. The potential is defined as~\cite{Cai:2019bmk} 
\begin{align}  
V(\phi) = \bar{V}(\phi) + \xi \cos\left(\frac{\phi}{\phi_*}\right) \Theta(\phi - \phi_e) \Theta(\phi_s - \phi),  
\end{align}  
where $\bar{V}(\phi)$ is favoured by CMB observation, and the form we have chosen is the Starobinsky potential  $\bar V(\phi) = \Lambda^4 \left[1-\exp\left(-\sqrt{2/3}\phi/M_{\rm p}\right) \right]^2$ with $\Lambda =0.0032M_{\rm p}$. $\xi \ll \Lambda^4$ governs the amplitude of the periodic structure, $\phi_*$ sets its period, and $\Theta$ functions confine the modulation to the field range $\phi_s \leq \phi \leq \phi_e$. During inflation, the oscillatory term triggers resonant amplification of the inflaton field perturbations. These enhanced perturbations, in turn, source a stochastic GW background through second-order nonlinear interactions. 

\begin{figure*}[t]
		\centering
		\includegraphics[width=0.6\textwidth]{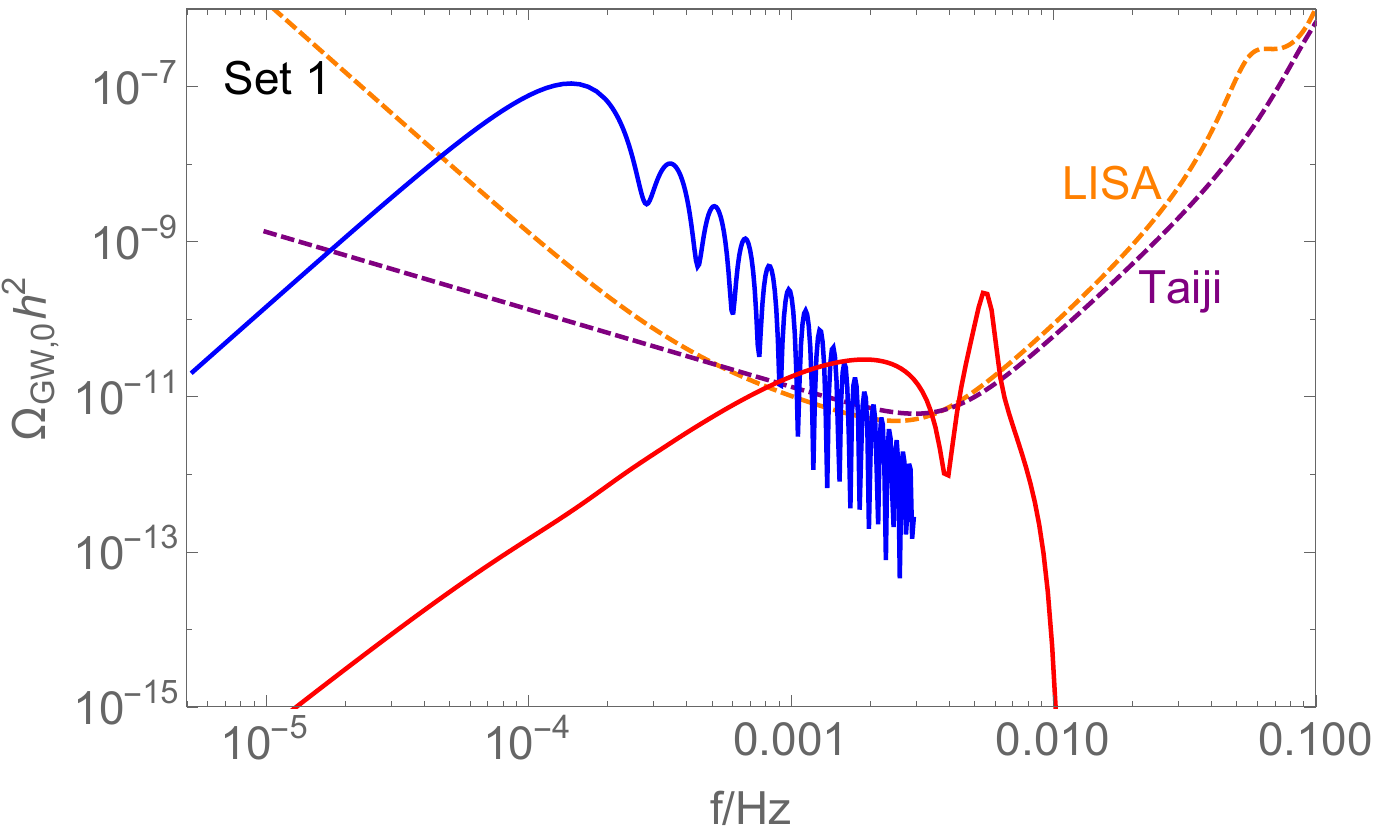}
		\caption{ The present energy spectra of total induced GWs. The blue solid line represents  GW produced from inflation and the red solid line represents the one from the radiation-dominated era . The orange and purple dashed lines represent the sensitivity curves of LISA \cite{Audley:2017drz} and Taiji \cite{Ruan:2018tsw}, respectively. Copied from Ref.~\cite{Peng:2021zon} with permission.}
		\label{peng:omega}
	\end{figure*}

We compute the energy spectra of GWs produced both during inflation and in the radiation-dominated era, and find that the peak of the energy spectrum of the former is much higher than that of the latter, but is located at a lower frequency (See Fig.~\ref{peng:omega}). Furthermore, the energy spectrum of GWs produced during inflation displays a distinctive oscillatory pattern in the ultraviolet regime. These two stochastic GW backgrounds, one resonant from inflation and the other scalar-induced post-inflation, are expected to be detected by future space-based interferometers, such as LISA and Taiji, offering complementary probes of early Universe dynamics.

\begin{figure*}
		\centering
		\includegraphics[width=0.7\textwidth]{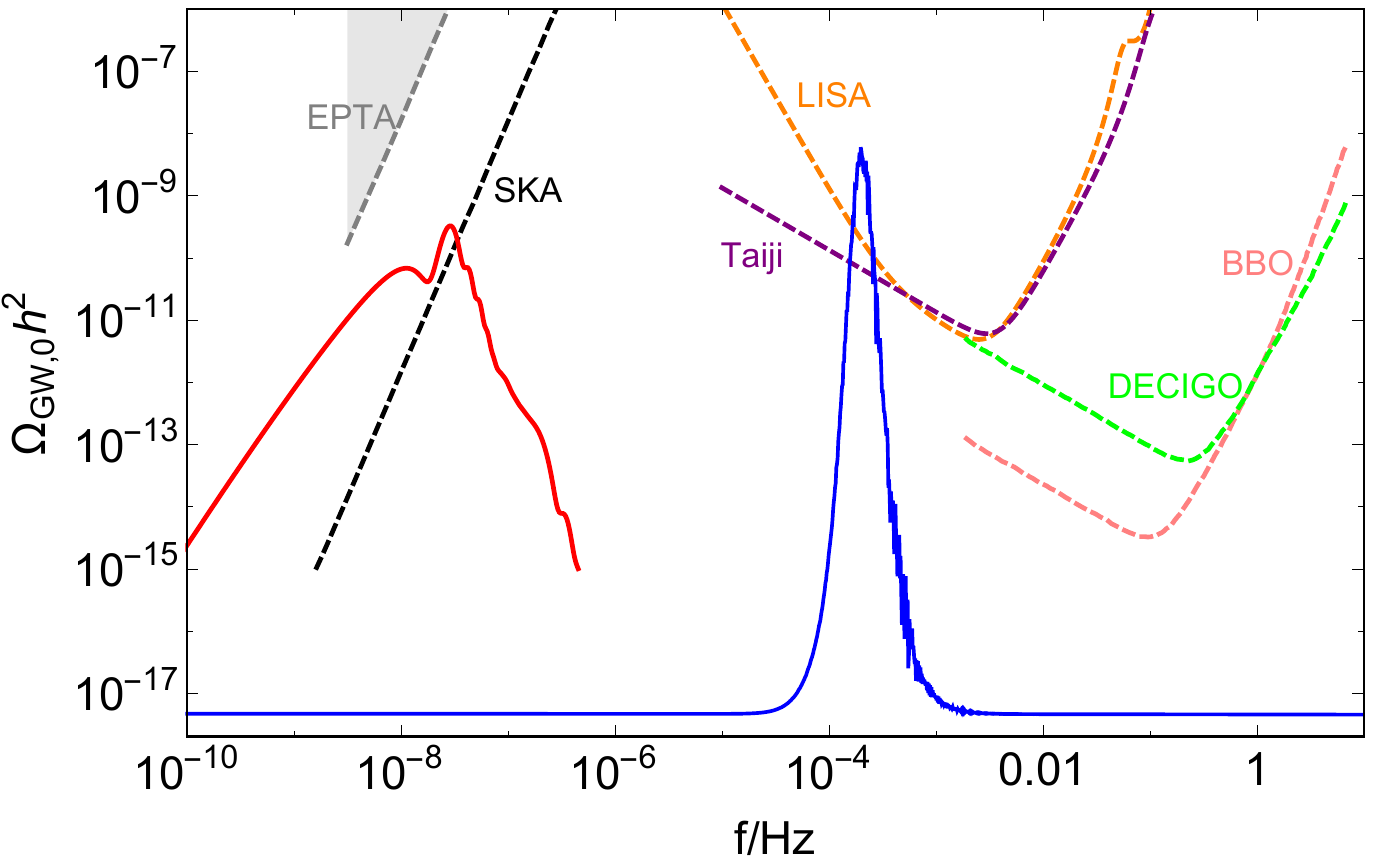}
		\caption{Present energy spectra for induced GWs and primordial GWs for Set 1, respectively.
			The red solid line represents induced GWs in the radiation-dominated era, and the blue represents PGWs generated during inflation.
			Other dashed lines are the expected sensitivity curve of the future GW projects summarized in \textcolor{black}{Ref.~}\cite{Moore:2014lga}. The shaded regions represent the existing constraints on GWs \cite{Kohri:2018awv, Lentati:2015qwp}. Copied from Ref.~\cite{Peng:2022ttg} with permission.}
		\label{peng:PVomega}
	\end{figure*}

In Ref.~\cite{Peng:2022ttg}, we investigated the generation of GWs in a two-field model embedded in dynamical Chern-Simons (dCS) gravity. The concrete action is 
\begin{eqnarray}
S[\phi,\chi] &=&\int d^4 x\,\sqrt{-g}\,\left[\frac{M_{\rm p}^2}{2} R-\frac{1}{2}(\partial \phi)^2-\frac{e^{2b(\phi)}}{2}(\partial\chi)^2 \right. \nonumber \\
 &&  ~~~~~~~~~~~~~~~~~~~~~~~ -V(\phi,\chi)+\mathcal{L}_{\rm {{dCS}}} \Bigg].
\end{eqnarray}
The model introduces a canonical inflaton field $\phi$ and a heavy spectator field $\chi$, dynamically coupled to the dCS term $ \frac{\alpha}{8}\chi R\tilde{R} $. A transition function $b(\phi)$ suppresses the kinetic energy of $\chi$ during the early phase of inflation, freezing its evolution. When the inflaton crosses a critical value $\phi_c $, the suppression is lifted, triggering oscillations of $\chi$ ($m_\chi \gg H$). These violent oscillations lead to an increase in the third derivative of $\chi$ with respect to the cosmic time, thus triggering the resonant amplification of
primordial GWs on small scales and imprinting a signature of parity violation in the power spectrum. 

Fig.~\ref{peng:PVomega} illustrates the present energy spectrum $\Omega_{\rm GW,0}h^2$ for a specific  parameter setting. The PGW spectrum (blue curve) peaks at $f \sim 10^{-4}$–$10^{-3}$ Hz within the sensitivity ranges of LISA and Taiji. For comparison, scalar-induced GWs (red curve) generated from post-inflation curvature perturbations, peak at lower frequencies $f \sim 10^{-8}$–$10^{-7}$ Hz, detectable by SKA. The relation of peak frequency between these two stochastic GW backgrounds is only dependent on the potential ratio. Varying the ratio, the total GW spectrum spans the sensitive frequency bands of various current of future observation plans.

\subsection{Electroweak phase transition}

In the early Universe, particles in the standard model obtain their masses from the vacuum expectation value of the Higgs boson during the electroweak phase transition (EWPT). In this epoch, the baryon asymmetry of the Universe could also be generated via the scenario of electroweak baryogenesis~\cite{Kuzmin:1985mm}, which requires a strongly first-order EWPT to satisfy the Sakharov conditions~\cite{Sakharov:1967dj}. However, it is well-known that the pattern of the EWPT in the SM is only a crossover according to the lattice simulation~\cite{Dine:1992vs}. Therefore, to realize the first-order EWPT, it is necessary to introduce new physics beyond the SM.

An interesting signature of the first-order phase transition is the stochastic gravitational wave from the collision of the nucleated bubbles. The GW spectrum from the phase transition can typically be characterized by the following parameters: the released latent heat $\alpha$, the inverse of the duration of phase transition $\beta$, and transition temperature $T_t$. These parameters can be determined from the shape of the scalar potential for electroweak symmetry breaking; thus, it is likely that new physics models with a first-order phase transition can be explored by GW observations. In particular, the frequency of the GW spectrum produced by the first-order EWPT is typically $10^{-3}$--$10$ Hz. Such a GW spectrum with sub-Hz frequency should be detected by the space-based interferometers in the future, such as LISA~\cite{Audley:2017drz}, Taiji~\cite{Ruan:2018tsw}, DECIGO~\cite{Seto:2001qf} etc.
 %\cite{LISA:2017pwj}, 

To investigate the patterns of the EWPT, let us explore the Higgs potential at the finite temperature. In the standard model, the Higgs potential can be parametrized as 
\begin{eqnarray}
    V_{\mathrm{eff}}(\varphi ; T) \simeq D\left(T^2-T_0^2\right) \varphi^2-E T \varphi^3+\frac{\lambda_T}{4} \varphi^4, 
\end{eqnarray}
where $D$ and $T_0^2$ are constant terms and  $\lambda_T$ is a slowly varying function of $T$. The cubic term $E=\frac{2 m_W^3+m_Z^3}{2 \pi v_0^3}$ is obtained via the thermal correction of the gauge bosons, and it plays the role of barrier of the potential. 
The barrier can be enhanced if additional scalars contribute to the thermal plasma, which can be written as
\begin{eqnarray}
    E=\frac{1}{2 \pi v_0^3}\left(2 m_W^3+m_Z^3+\text { new scalar mass }\right).
\end{eqnarray}
Usually if there are more scalars, then the cubic term $E$ would be enhanced by the new scalar contributions to the thermal plasma, and thus provides stronger first order EWPT. 
The barrier can also be enhanced through the tree level cubic term in the scalar potential
\begin{eqnarray}
    V(\phi) = \frac12 m^2 \phi^2 + \kappa \phi^2 + \lambda \phi^4.
\end{eqnarray}
This cubic term can be obtained in the singlet scalar extension case. 
Furthermore, the extended scalar model with discrete $\mathbb{Z}_2$ symmetry could realize the two-step phase transition, in which the first step is realized by the tree-level cubic term effectively~\cite{Cao:2022ocg}. 
In literature, extended scalar sectors includes scalar singlet, doublet, triplet, and quadruplet, the stop sector in the supersymmetry, and dark/hidden sectors, etc~\cite{Profumo:2007wc,Patel:2012pi}. In some cases, the fermion extensions can also trigger the first order PT~\cite{Cao:2021yau}.

These new physics models, especially the scalar extensions of the SM, could be described and parametrized in the effective field theory (EFT) framework, in which the new particles are integrated out and only the SM degrees of freedom are kept. The EFT description was adopted particularly for the higher-dimensional-operator extensions of the SM~\cite{Grojean:2004xa}, which characterize the effect on the low-energy degrees of freedom when we integrate out the heavy degrees of freedom. However, the EFT description is only valid in the presence of a clear separation of scales, which is in conflict with the relatively low scale of the new degrees of freedom so as to introduce a large correction to the SM Higgs potential~\cite{Postma:2020toi} in order to trigger a PT. Exceptions could be made for the Higgs-singlet extension with tree-level matching, though the EFT description is at most qualitative for dimension-six extension. Therefore, for the electroweak phase transition process, the EFT description for some scalar extensions is not enough since in many cases, the new light degree of freedom would contribute to the thermal plasma and thus one cannot integrate it out during the phase transition.

An intermediate strategy that lies between the specific new physics model and the EFT treatment. There could be a simplified model to illustrate features of the EWPT, the effective model description of the phase transition~\cite{Cai:2022bcf}. In this description, to capture different patterns of the EWPT and to compare the difference between new physics model and EFT description, a specific effective model description has been proposed: the general model extends the SM with an isospin N-plet scalar field, of which the light scalar case consists of a model with classical scale invariance (CSI) (model I) and a model without CSI (model II), while the heavy scalar case is simply a model with higher-dimensional operators, for example, a dimension-six operator (model III). The above cases could describe the patterns of the electroweak symmetry breaking via, for example, (1) radiative symmetry breaking, (2) Higgs mechanism, and (3) EFT description of EWSB. The effective model description already covers those scalar models with N- plet on the market, such as (1) singlet models including a real scalar singlet extension of SM, composite Higgs model like SO(6)/SO(5) model, extra dimension model like radion model, dilaton model; (2) doublet models including SUSY model like minimal supersymmetry model (MSSM), two Higgs doublet model (2HDM), minimal dark matter model; (3) triplet models including left-right model, Type-II seesaw model. In other perspective, the effective model description consists of the simplified models (effective models I and II) for the realistic models (SUSY, composite Higgs, etc.) and an EFT model (model III).
The future space-borne GW detectors would provide a promising probe for the new physics beyond the standard model that admits the first-order phase transitions. The predictions for the GW background vary sensitively among different concrete particle physics models but also share a large degeneracy in the model buildings, which could discriminate different effective models on the phase transition based on different patterns of the electroweak symmetry breaking.

\subsection{QCD phase transtion}
%arXiv:arXiv:2201.02004
%contributor: Li Li
%liliphy@itp.ac.cn

QCD is the fundamental theory describing strong interactions. Its structure is highly intricate and remains incompletely understood due to challenges in theoretical and numerical methodologies. However, QCD phase transition from the quark-gluon plasma to confined hadrons is expected to take place when the Universe was around a microsecond old after the big bang. Therefore, GWs generated during the QCD phase transition in the early Universe are an interesting topic that bridges particle physics and cosmology. Moreover, another interesting consequence is the production of PBHs which in turn can account for a fraction or the entire dark matter observed abundance. With substantial progress, current non-perturbative approaches to studying the QCD phase structure still face significant challenges. In particular, as a first-principle approach, lattice QCD has struggled with computational limitations due to the famous sign problem at finite baryon chemical potential $\mu_B$~\cite{Philipsen:2012nu}. However, in the Standard Model, lattice QCD simulations have confirmed a crossover at vanishing and small baryon chemical potentials ($\mu_B/T\le 3.5$)~\cite{Borsanyi:2021sxv}.

The cosmological QCD phase transition holds significant implications as a potential source for a stochastic GW background if it is of the first order. Until now, measurements of primordial element abundances and the CMB have imposed a strict limitation on the baryon asymmetry $\eta_B\equiv n_B/s\approx 10^{-10}$, where $n_B$ and $s$ denote the baryon number density and entropy density, respectively~\cite{dodelson2003modern}. This has led to the prevailing belief that the cosmic trajectory passes the QCD phase diagram at (extremely) small $\mu_B$ where lattice QCD reveals a crossover. Thus far, numerous QCD model buildings beyond the SM have been proposed to trigger a first-order phase transition, including hidden dark sectors~\cite{Schwaller:2015tja,Aoki:2017aws,He:2022amv,Bigazzi:2020avc}, large lepton asymmetries~\cite{Gao:2021nwz}, exotic interactions~\cite{Iso:2017uuu,Lu:2022yuc,Sagunski:2023ynd}, Peccei-Quinn symmetry breaking~\cite{DelleRose:2019pgi,VonHarling:2019rgb}. There leaves a large part of the parameter space for which the GW signal will be detectable in the next generation facilities, ranging from pulsar timing arrays to space-and ground-based interferometers, depending on the dynamical scale of each scenario. 

Although both experimental data and lattice QCD provide insights mainly within the crossover region, the possibility of the existence of a critical end point (CEP) and a region of a first-order transition has been unsettled in SM. Recently, the holographic approach was used to study the non-perturbative dynamics of QCD in terms of a gravity system with one higher dimension (see~\cite{Rougemont:2023gfz,Chen:2022goa,Jarvinen:2021jbd} for recent reviews). The model of~\cite{Cai:2022omk} showed good quantitative agreement with state-of-the-art lattice QCD~\cite{Borsanyi:2021sxv}, and exhibited consistency with measurements of heavy-ion collisions~\cite{Li:2023mpv}. It predicted the CEP located at $(T_\mathrm{CEP}=105~\text{MeV}, \mu_\mathrm{CEP}=555~\text{MeV})$.  Moreover, the theory, within the confines of the SM, suggests that around $\mu_B=1000$ MeV, not only does the QCD phase transition become first order, but also the inferred value of $\eta_B$ aligns with cosmological observations. This presents a concrete scenario for which a first-order QCD phase transition happens in the early Universe. It was then shown that this phase transiton could serve as a promising source of GW backgrounds reported recently by different observations using pulsar timing array~\cite{
NANOGrav:2023gor,Reardon:2023gzh,Xu:2023wog,EPTA:2023sfo}, and meanwhile the associated PBHs are still allowed by current observations. The first-order cosmological QCD phase transition with $\mu_B<1000$ Mev would require a ``little inflation''~\cite{Boeckel:2009ej,Boeckel:2011yj} to achieve the observed baryon asymmetry.

\subsection{Constraining cosmological phase transitions}
%arXiv:2110.03096
%contributor: Ligong Bian
%lgbycl@cqu.edu.cn

Pulsar timing array experiments detect gravitational waves in the nanohertz frequency range by precisely measuring the arrival times of radio pulses from an ensemble of highly stable millisecond pulsars within the Milky Way. Over the past decade, the three leading PTA collaborations - the Parkes Pulsar Timing Array~\cite{manchester2013parkes}, the European Pulsar Timing Array~\cite{EPTA16}, and the North American Nanohertz Observatory for Gravitational Waves~\cite{NANOGrav12.5data} - have monitored the times of arrival (ToAs) of dozens of pulsars with weekly to monthly cadence. These long-term observations provide a unique window into the stochastic gravitational-wave background in the nanohertz regime, offering critical insights into astrophysical and cosmological phenomena \cite{Lentati:2015qwp,Caprini:2018mtu,Arzoumanian:2018saf,Garcia-Bellido:2017aan,Barack:2018yly,Caprini:2010xv,Breitbach:2018ddu,Liu:2021svg,Liu:2022lvz,Zeng:2024snl}.

If a first-order phase transition (FOPT) occurs at a relatively low temperature (e.g., 
$T_\star\sim\mathcal{O}$(MeV), it is expected to generate a SGWB with a peak frequency in the range $f\sim 10^{-9}-10^{-8}$ HZ \cite{Caprini:2010xv,Bian:2020urb,Bian:2023dnv}. This frequency band lies within the sensitive region of PTA experiments \cite{foster1990constructing,hellings1983upper}. We utilize the latest PPTA dataset to search for an SGWB signature from such a cosmological FOPT. 
Pulsar Timing Array experiments measure pulse times of arrival from highly stable millisecond pulsars. The stochastic gravitational-wave background manifests through distinctive inter-pulsar correlations in the timing residuals, following theoretical predictions \cite{allen1999detecting,maggiore2000gravitational}. Ref.~\cite{Xue:2021gyq} employs a comprehensive noise model derived from rigorous single-pulsar studies \cite{goncharov2020identifying}, and demonstrate that the PPTA data can probe a significant portion of the FOPT parameter space. Therein, potential contributions from other SGWB sources have been neglected; consequently, the constraints derived on the FOPT should be regarded as conservative. 

The NANOGrav collaboration also extended its search for a gravitational-wave background to include signals from FOPTs in their 12.5-year dataset \cite{NANOGrav21fopt}. In their analysis, the team suggested that a phase transition-generated SGWB could potentially explain the observed common-power-law (CPL) process. However, as highlighted in Ref.~\cite{Boris21PPTAcrn}, the interpretation of a CPL detection may be susceptible to model misspecification. Consequently, we conservatively treat the feature identified in our data as a noise component of undetermined origin and proceed to place constraints on FOPT parameters. A more definitive characterization of the CPL process will require both extended observational timelines and an expanded pulsar array, prospects that may be realized through future data releases from the International Pulsar Timing Array \cite{IPTAdr2}.

GWs generated during a FOPT primarily arise from three mechanisms: (i) bubble collisions \cite{Caprini:2007xq, Huber:2008hg}, (ii) sound waves in the primordial plasma \cite{Hindmarsh:2017gnf, Hindmarsh:2013xza, Hindmarsh:2016lnk, Cutting:2019zws, hindmarsh2015numerical}, and (iii) magnetohydrodynamic (MHD) turbulence \cite{Caprini:2009yp}. Among these, sound waves are generally expected to dominate the GW signal in most realistic phase transition scenarios. For conservative estimates, one can focus exclusively on the sound wave contribution to the SGWB energy density. The spectral energy density from sound waves can be expressed as a function of the characteristic frequency 
$f$ \cite{Caprini:2015zlo, Hindmarsh:2017gnf, Ellis:2018mja, Ellis:2019oqb}:
  \begin{align}
	\Omega_{sw}(f)h^2 &= 2.65\times 10^{-6}v_w
	\left(\frac{H_*}{\beta}\right)\left(\frac{\kappa\alpha}{1+\alpha}\right)^2\left(\frac{100}{g_*}\right)^{1/3}\nonumber\\
	&\times	\left[(f/f_{sw})^3\left(\frac{7}{4+3(f/f_{sw})^2}\right)^{7/2}\right]\nonumber\\
	&\times \min\left[1,~(8\pi)^{1/3}v_w\left(\frac{H_*}{\beta}\right)\left(\frac{3}{4}\frac{\kappa\alpha}{1+\alpha}\right)^{-1/2} \right],\label{edensity}
	\end{align}
where, $g_\star$ represents the effective number of relativistic degrees of freedom;
$\kappa$ denotes the fraction of the latent heat converted into plasma kinetic energy \cite{Espinosa:2010hh};
$v_w
$	
  is the bubble wall velocity (fixed at 
$v_w=1
$ in our analysis);
$H_\star$ corresponds to the Hubble parameter evaluated at the phase transition temperature 
$T_\star$ ;
$\alpha$ characterizes the phase transition strength, primarily determining the GW spectrum amplitude;
$\beta$ represents the inverse duration of the phase transition (or transition rate parameter), which together with 
$T_\star$	
  determines the GW spectrum's peak frequency; and
$f_{sw}$ indicates the peak frequency of the sound wave contribution:
\begin{align}
	f_{sw} &\simeq \frac{1.9\times 10^{-5}~\mathrm{  Hz}}{v_w}\left(\frac{\beta}{H_*}\right)\left(\frac{T_*}{100\textrm{ GeV}}\right)\left(\frac{g_*}{100}\right)^{1/6}\;.\label{freq}
	\end{align}
We assign fixed values to 
the $g_\star$ based on the phase transition temperature 
$T_\star$
\begin{itemize}
\item 
$g_\star
=
100$
$T_\star >0.2$	
 GeV
\item 
$g_\star
=
10$ for 0.1 MeV$<T_\star <0.2$ MeV	

\item 
$g_\star
=
2$ for 
$T_\star <0.1$ MeV
\end{itemize}
following the standard treatment in Ref.~\cite{husdal2016effective}. The gravitational wave spectrum is completely characterized by the phase transition parameters 
$\alpha$ (transition strength) and 
$\beta$ (inverse duration), both evaluated at the nucleation temperature. Although 
$g_\star$ undergoes significant changes near the threshold temperatures, these variations have minimal impact on our model's accuracy because 
$g_\star$	
  appears with a small exponent in both Eqs.~(\ref{edensity}) and (\ref{freq}).  

The energy density of a SGWB, denoted by
$\Omega_{gw} h^2$, is related to its one-sided power spectral density $S_{gw}(f)$ through:
\begin{equation}
S_{\text{gw}}(f) = \frac{1}{12\pi^2} \frac{1}{f^5} \frac{3H_{100}^2}{2\pi^2}\Omega_{\text{gw}}(f)h^2,\label{gwb}
\end{equation}
where
$H_{100} = 100{~ {\rm km} ~ {\rm s}^{-1} ~ {\rm Mpc}^{-1}}$ is the Hubble constant and $h \approx 0.7$ is the dimensionless Hubble parameter.
The two-point correlation function for timing residuals $\delta t$ between pulsars $I$ and pulsar $J$ is given by
\begin{equation}\label{corr}
\langle \delta t_i^I \delta t_j^J \rangle = \int \mathrm{d}f, S(f),\Gamma^{IJ}(\theta_{IJ}), \cos\left[2\pi f (t_i - t_j)\right],
\end{equation}
where
$t_i$ and $t_j$ represent ToAs, $\Gamma^{IJ}$ is the HD correlation function, and $\theta_{IJ}$ denotes the angular separation between pulsars. This expression assumes an isotropic, unpolarized gravitational wave background and considers only Earth terms in the pulsar response.

For computational efficiency, the authors of Ref.~\cite{Xue:2021gyq} employ a time-frequency approach that approximates the power spectrum using a finite set of Fourier frequencies. They examine two distinct cases for the correlation analysis: 1) Full HD correlation: Includes both cross-correlations between different pulsars and autocorrelations for individual pulsars; and 2) no-auto HD correlation: Uses a modified correlation function to effectively remove pulsar autocorrelations while preserving the inter-pulsar HD correlations.

Ref.~\cite{Xue:2021gyq} displays the 95\% credible-level exclusion contours for the FOPT model parameters, considering phase transition strengths of 
$\alpha =$ 0.2, 0.5, and 1. Several key observations emerge from the analysis:
1) Dependence on 
$\alpha$: the constraints strengthen significantly with increasing 
$\alpha$, reflecting the greater sensitivity to stronger phase transitions; 2) phase transition duration: the excluded range of the transition temperature 
$T_\star$
  expands for longer phase transition durations with larger $(\beta/H_*)^{-1}$; 3) impact of  no-auto HD correlation: the excluded parameter space is notably larger when the Hellings-Downs (HD) auto-correlation is omitted. This aligns with the slight disfavoring of FOPT models in the absence of SGWB auto-correlation.
Compared to the Big Bang Nucleosynthesis (BBN) bound (
($T_*> 1 $ MeV), the analysis provides significantly stronger constraints in the intermediate temperature range 
$1\ {\rm MeV} <T_* < 100\ {\rm MeV}$.
Therefore, the analysis excludes a substantial region of the cosmological phase-transition parameter space that was previously unconstrained by observations. The PPTA data provide particularly strong constraints on phase transitions with characteristic temperatures of $\sim 1-100$~MeV and duration parameter $(\beta/H_*)^{-1}\sim 10^{-2}-10^{-1}$. These ranges are particularly relevant for both
low-scale dark sector phase transitions \cite{Ratzinger:2020koh,Bian:2020urb,Breitbach:2018ddu} and cosmological first-order QCD phase transitions \cite{Caprini:2010xv}.
These results establish novel constraints on new physics scenarios involving low-energy-scale FOPTs in the early Universe, while simultaneously demonstrating the potential of pulsar timing arrays as a unique probe of low-energy particle physics phenomena.

\subsection{GWs from cosmic domain walls}
%arXiv:2010.03225
%contributor: Jing Liu
%liujing@ucas.ac.cn
Cosmic domain walls (DWs), two-dimensional topological defects formed during symmetry-breaking phase transitions in the early Universe, serve as relics of vacuum state degeneracy~\cite{Vilenkin:1984ib,Vilenkin:1981zs}. These structures emerge when scalar fields settle into distinct minima separated by energy barriers, such as in $ Z_2 $-symmetric models where a single scalar field adopts positive and negative vacuum expectation values.  
The cosmological evolution of DW networks follows a dynamical scaling regime, where their energy density self-adjusts to maintain $ \rho_{\mathrm{DW}} \propto \sigma H $, where $\sigma$ is the DW tension, balancing Hubble expansion and inter-wall interactions, as by various numerical simulations~\cite{Garagounis:2002kt,Hindmarsh:1996xv,Li:2025gld}. The scaling behavior reveals that stable domain walls could finally dominate the Universe.  The observations of cosmic microwave background and the large scale structures also compose strict constraints on the DW tension for currently existing DWs, $\sigma\lesssim \mathrm{MeV}^{3}$~\cite{Hiramatsu:2012sc}. To address this issue, a proposed solution involves introducing explicit symmetry-breaking terms~\cite{Chang:1998tb,Sikivie:1982qv} (e.g., Planck-suppressed operators to create energy bias), inducing the domain wall decay in the early Universe. For alternative approaches to solve the DW problem, see Refs.~\cite{Dine:2023qsq,Li:2023yzq,Zhang:2023gfu,Ibe:2019yew,Chatterjee:2019rch,Barr:2014vva,Dankovsky:2024ipq,Zeng:2023jut} for instance.

The evolution and annihilation of the DW networks provide another important source of GWs~\cite{Saikawa:2017hiv}. The nonlinear evolution of GWs and the characteristic GW energy spectrum are given by precise lattice simulations~\cite{Hiramatsu:2013qaa,Notari:2025kqq,Dankovsky:2024ipq}, where transverse-traceless part of tensor metric perturbations are solved numerically with DWs providing the source term. The typical wavelength of GWs from the collapse of DWs is close to the averaged curvature radius of DWs, i.e., the Hubble radius, at the annihilation time, which is expressed by
\begin{equation}
f_{\mathrm{peak}}=\left(\dfrac{H^{2}(t_{\mathrm{ann}})}{H_{0}^{2}\Omega_{\mathrm{rad}}(t_{0})}\left(\dfrac{g_{*\mathrm{ann}}}{g_{*0}}\right)^{1/3}\right)^{-1/4}H(t_{\mathrm{ann}})\,,
\end{equation}
where $g_{*0}$ and $g_{*\mathrm{ann}}$ are the effective relativistic degrees of freedom at the present time $t_{0}$ and the annihilation time $t_{\mathrm{ann}}$, respectively. The corresponding peak value of the GW energy spectrum reads
\begin{equation}\label{eq:redshift}
\Omega_{\mathrm{GW,peak}}(t_{0})h^{2}=\Omega_{\mathrm{rad}}(t_{0})h^{2}\left(\dfrac{g_{*0}}{g_{*\mathrm{ann}}}\right)^{1/3}\Omega_{\mathrm{GW,peak}}\left(t_{\mathrm{ann}}\right)\,,\
\end{equation}
\begin{equation}
\Omega_{\mathrm{GW},\mathrm{peak}}\left(t_{\mathrm{ann}}\right)=\dfrac{ \tilde{\epsilon}_{\mathrm{GW}} \mathcal{A}^{2} \sigma^{2}}{24\pi H^{2}\left(t_{\mathrm{ann}}\right)}\,,
\end{equation}
where $\mathcal{A}\approx 0.8$ and $\tilde{\epsilon}_{\mathrm{GW}}\approx 0.7$ are constants fixed by numerical results~\cite{Hiramatsu:2013qaa}, $\Omega_{\mathrm{rad}}h^{2}=4.2\times 10^{-5}$ is the current density fraction of radiation, and $H_{0}=67.7\, \mathrm{km}\,\mathrm{s}^{-1}\,\mathrm{Mpc}^{-1}$ is the Hubble constant from the results of Planck 2018~\cite{Aghanim:2018eyx}.
Multiband GW observers will test DW models across 13 orders of magnitude in frequency~(from nHz to kHz), scanning the energy scale from $ \text{MeV} $ to $ 10^{10} \, \text{GeV} $.  The SGWB from DWs also serves as a competitive candidate to explain the recent PTA results of the nHz signal~\cite{Gouttenoire:2023ftk,Ferreira:2022zzo,Bian:2020urb,Liu:2020mru,Zhang:2023nrs}.
Beyond sourcing GWs, DWs have broader cosmological implications, such as providing the seeds of phase transitions~\cite{Blasi:2022woz,Blasi:2023rqi}, generating PBHs~\cite{Garriga:2015fdk,Deng:2016vzb,Liu:2019lul,Lu:2024ngi,Gouttenoire:2023ftk,Ge:2019ihf} and axion dark matter~\cite{Chang:2023rll,Li:2023yzq}.

Besides the characteristic energy spectrum of gravitational waves, the anisotropies of the cosmological GW backgrounds are attracting growing scientific attention. These anisotropies could be generated by the sources~\cite{Bethke:2013aba,Bethke:2013vca,Geller:2018mwu,Jenkins:2018lvb,Liu:2020mru,Luo:2025lgr,Yu:2023jrs} and the processes during propagation~\cite{Contaldi:2016koz,Bartolo:2019oiq,Bartolo:2019yeu,DallArmi:2020dar}. Due to the intrinsic limitations of GW detectors in resolving angular anisotropies, sources generating stronger anisotropic signatures are more likely to be detected in the near future~\cite{Allen:1996gp,Seto:2004np,Hotinli:2019tpc,Chu:2020qiw,Mentasti:2020yyd,Alonso:2020rar,Ding:2023xeg,Tian:2024ljg,Zhao:2024yau}. Such gravitational wave sources exhibit large-scale perturbations with amplitudes surpassing those observed in the CMB temperature fluctuations. This distinctive property finds natural theoretical realization through light fields during cosmic inflation~\cite{Bethke:2013aba,Bethke:2013vca,Geller:2018mwu,Jenkins:2018lvb,Liu:2020mru,Luo:2025lgr,Yu:2023jrs}.
The perturbations of these inflationary light fields maintain constant after horizon-crossing during inflation, which is independent of the gereration of curvature perturbations responsible for CMB temperature anisotropies. This theoretical independence allows for the possibility of significantly enhanced amplitude in the GW background anisotropies. Such a characteristic makes them particularly promising targets for future gravitational wave observatories, potentially opening new windows into the nature of inflation. 

In Ref.~\cite{Liu:2020mru}, we proposed a novel mechanism to establish a link between inflationary energy scales and the statistical properties of the SGWB anisotropies. By analyzing the imprint of large-scale perturbations of light scalar fields on the energy density distributions of DW networks, we demonstrate that the interplay between quantum diffusion during inflation and post-inflationary DW dynamics generates a unique scale-invariant angular power spectrum in the SGWB characterized by
\begin{equation}
l(l+1)C_l \approx \begin{cases} 
\displaystyle\frac{\pi}{N_{\text{peak}}}\alpha_{\text{peak}}^2, & \alpha_{\text{peak}} \gg 1\,, \\[10pt]
\displaystyle\frac{1}{N_{\text{peak}}}, & \alpha_{\text{peak}} \ll 1\,, 
\end{cases}
\label{eq:angular_spectrum}
\end{equation}
where the peak e-folding number $N_{\text{peak}} \equiv \ln(k_{\text{peak}}/H_0)$ and $\alpha_{\text{peak}} \equiv \sqrt{2}\pi\phi_i/(H_{\rm inf}\sqrt{N_{\text{peak}}})$. This mechanism bypasses the limitations of traditional probes like the tensor-to-scalar ratio $r$~\cite{Aghanim:2018eyx,Liddle:1993ch,Guo:2010mm,Planck:2018jri}, offering a complementary pathway to constrain low-scale inflationary scenarios.

The light scalar field $\phi$, responsible for the DW formation, acquires superhorizon perturbations during inflation with an amplitude $\langle\delta\phi^2\rangle^{1/2} = H_{\rm inf}/(2\pi)$. According to the the Fokker-Planck equation~\cite{Espinosa:2015qea}, these perturbations modulate the spatial distribution of DW nucleation sites through the formation probability

\begin{equation}
P(t) = \frac{1}{2}\operatorname{erfc}\left(\frac{\sqrt{2}\pi\phi_i}{H_{\rm inf}\sqrt{N(t)}}\right)\,,
\label{eq:probability}
\end{equation}
where the efolding number $N(t)=H_{\mathrm{inf}}(t-t_{i})$ assuming $H_{\mathrm{inf}}$ is almost a constant during inflation, and $t_{i}$ satisfies $a(t_{0})H_{0}=a(t_{i})H_{\mathrm{inf}}$.
The resulting DW energy density contrast is determined by the first-order expansion
\begin{equation}
\delta P(t, \boldsymbol{x}) = \frac{1}{2}\operatorname{erfc}\left[\alpha(t)\left(1+\frac{\delta\phi(\boldsymbol{x})}{\phi_i}\right)\right] - \frac{1}{2}\operatorname{erfc}[\alpha(t)]\,,
\label{eq:deltaP}
\end{equation}
leading to the fractional energy density contrast
\begin{equation}
\frac{\delta\rho_{\rm DW}}{\rho_{\rm DW}} = c_1\delta\phi,\quad c_1 = \frac{2\alpha_{\text{peak}}}{\sqrt{\pi}\phi_i}\frac{e^{-\alpha_{\text{peak}}^2}}{\operatorname{erfc}(\alpha_{\text{peak}})}\,.
\label{eq:density_contrast}
\end{equation}

The resultant SGWB anisotropies exhibit a distinctive scale-invariant angular power spectrum $l(l+1)C_l \sim \mathcal{O}(10^{-2})$, significantly larger than the CMB temperature anisotropies.  This enhancement arises from the nonlinear dependence encoded in Eq.~\eqref{eq:probability}, amplified by the error function's sensitivity to perturbations. Intuitively, the amplitude of vacuum fluctuations during inflation, $H_{\mathrm{inf}}/(2\pi)$, only needs to reach the scale of the vacuum expectation value to produce strong anisotropies in the observable SGWB. This requirement is significantly weaker than that for primordial tensor perturbations to generate observable B-mode polarization in CMB observations. Such a signature remains robust against post-inflationary thermalization effects, as shown by the frozen superhorizon perturbations until DW formation.

%%%%%%%%%%%%%%%%%%%%%%%%%%%%%%%%%%%%%%%%%%%%%%%%%%%%%%%%%%%%%%%%%%%%%%%%%%%%%%%%%%%%%%
\section{Primordial black holes \label{sec:pbh}}
%%%%%%%%%%%%%%%%%%%%%%%%%%%%%%%%%%%%%%%%%%%%%%%%%%%%%%%%%%%%%%%%%%%%%%%%%%%%%%%%%%%%%%
%contributor: Qing-Guo Huang

\subsection{Log-dependent slope of scalar induced gravitational waves in the infrared regions} 
Although the energy spectrum of scalar-induced gravitational waves (SIGWs), $\Omega_{\mathrm{GW}}$ depends on the shape of the primordial curvature power spectrum, the infrared (IR) behavior of $\Omega_{\mathrm{GW}}$ exhibits unique scalings. Analytical derivations for understanding of this phenomenon were studied in \textcolor{black}{Refs.}~\cite{Yuan:2019wwo,Li:2024lxx} for Gaussian perturbations, the scaling of $\Omega_{\mathrm{GW}}$ in the IR region is given by 
\begin{equation}
\Omega_{\mathrm{GW}}(k) \propto
\begin{cases}
k^{2\alpha}, & \alpha < 3/2, \\
k^3\ln^3\left({k_{\star}^2 / k^2}\right), & \alpha = 3/2, \\
k^n\ln^2\left({k_{\star}^2 / k^2}\right) , & \alpha > 3/2,
\end{cases}
\end{equation}
where the IR region refers to $k\ll k_*$ with $k_*$ to be the peak of $\Omega_{\mathrm{GW}}$, $k_\star$ is a pivot scale and $\alpha$ denotes the slope of the primordial power spectrum in the infrared region, such that $\mathcal{P}_\zeta$ grows as $\mathcal{P}_\zeta(k)\sim k^\alpha$. The value of $n$ lies in the range $2\le n \le 3$, depending on the width of the primordial power spectrum, $\Delta \equiv (k_+-k_-)/k_*$ where $k_+$ and $k_-$ denote the upper and lower cut-off of the power spectrum. For example, the scales at which the power drops to $1\%$ of its peak value. For narrow power spectrum $\Delta \ll 1$, the value of $n$ is $n\simeq 2$ in the region $\Delta\ll k/k_*\ll1$ and $n\simeq 3$ in the region $k/k_* \ll \Delta$. For a wide spectrum $\Delta \gtrsim 1$, the scaling in the entire IR region corresponds to $n\simeq 3$. 

For the most relevant case, $\alpha\ge 3/2$, the IR behavior of $\Omega_{\mathrm{GW}}$ exhibits a unique log-dependent scaling, providing a smoking gun for distinguishing SIGWs from other signals. This log-dependent infrared scaling has also been widely reproduced in studies of SIGWs generated in various models \cite{Fu:2019vqc,Lin:2020goi,Domenech:2020kqm,Yuan:2020iwf,Kohri:2020qqd,Atal:2021jyo,Adshead:2021hnm,Bastero-Gil:2021fac,Wu:2021zta,Solbi:2021rse,Teimoori:2021pte,Kozaczuk:2021wcl,Rezazadeh:2021clf,Heydari:2021qsr,Balaji:2022rsy,Li:2022avp,Balaji:2022dbi,Spanos:2022euu,Heydari:2023xts,Domenech:2023dxx,Ashrafzadeh:2023ndt,NANOGrav:2023hvm,Heydari:2023rmq,Domenech:2024kmh,Solbi:2024zhl,Heydari:2024bxj,Fei:2024vfu,LISACosmologyWorkingGroup:2024hsc,Ashrafzadeh:2024oll,Kohri:2024qpd,Iovino:2024crh,Luo:2025lgr,Domenech:2025bvr}

\begin{figure*}
\centering
\includegraphics[width =0.65\textwidth]{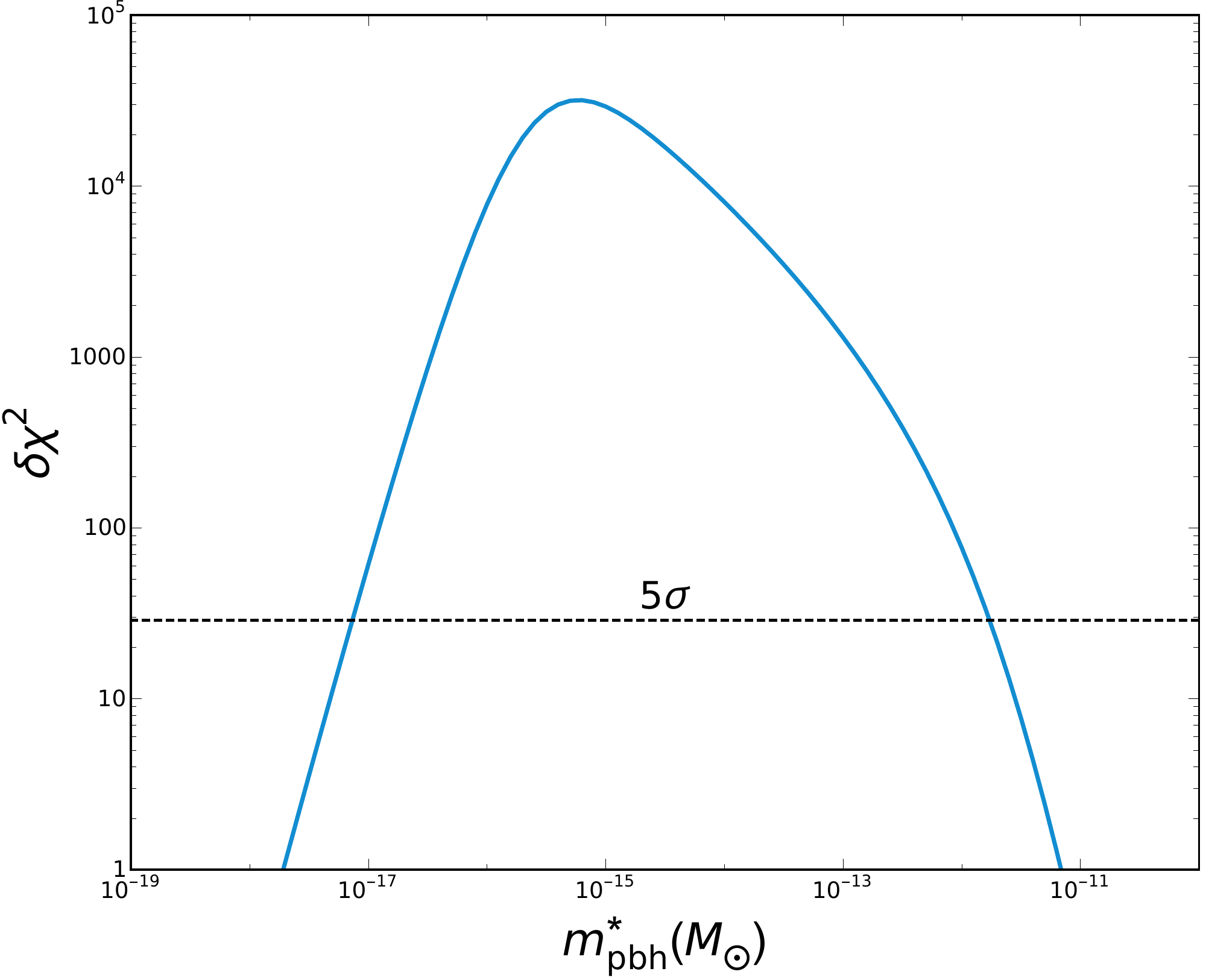}
    \caption{\label{NGWchi}
     The plot of $\delta\chi^2$ with the peak mass of PBHs ($m_{\mathrm{pbh}}^*/M_\odot \approx2.3\times10^{18}\left({2\pi H_{0}}/{k_*}\right)^{2}$) generated by the log-normal power spectrum with $\sigma_*=0.5$. The dashed line corresponds to $\delta\chi^2=28.74$. \textcolor{black}{Copied from Ref.~\cite{Yuan:2019wwo} with permission.}}
\end{figure*}	

Although the log-dependent behavior is simply treated as power-law in some work, it is important to consider the logarithm correction in observations. To illustrate this point, \cite{Yuan:2019wwo} considered two SIGW templates: one given by the exact numeric result, and the other is given by
\begin{equation}
\Omega^{\mathrm{fid}}_{\mathrm{GW}}(k) =
\begin{cases}
\Omega_{\mathrm{GW}}(0.1k_\star)\left( \dfrac{k}{0.1k_\star} \right)^3, & k < 0.1k_\star, \\
\Omega_{\mathrm{GW}}(k), & k \geq 0.1k_\star.
\end{cases}
\end{equation}
which follow a power-law scaling $\Omega_{\mathrm{GW}} \propto k^3$ in the infrared region. The distinguishability between these two templates by can be quantified by computing the $\delta\chi^2$ statistic:
\begin{equation}
    \delta\chi^2 \simeq T \int_0^\infty df \left( \frac{\Omega_{\mathrm{GW}}(f) - \Omega_{\mathrm{fid}}(f)}{\Omega_{\mathrm{fid}}(f) + \Omega_n(f)} \right)^2,
\end{equation}
where $T$ is the observation time (taken as $1$ year), $\Omega_n(f)$ is the noise power spectrum of the GW detector.

The resulting $\delta\chi^2$ values, as shown in Fig.~\ref{NGWchi} of~\cite{Yuan:2019wwo}, indicate that LISA can distinguish the log-dependent template from the power-law scaling model at more than $5\sigma$ confidence level over a wide range of PBH masses. This suggests that future GW experiments could be able to confirm SIGW detection based on its unique scaling features and distinguish SIGWs from other signals.

\begin{figure*}
\centering
\includegraphics[width =0.7\textwidth]{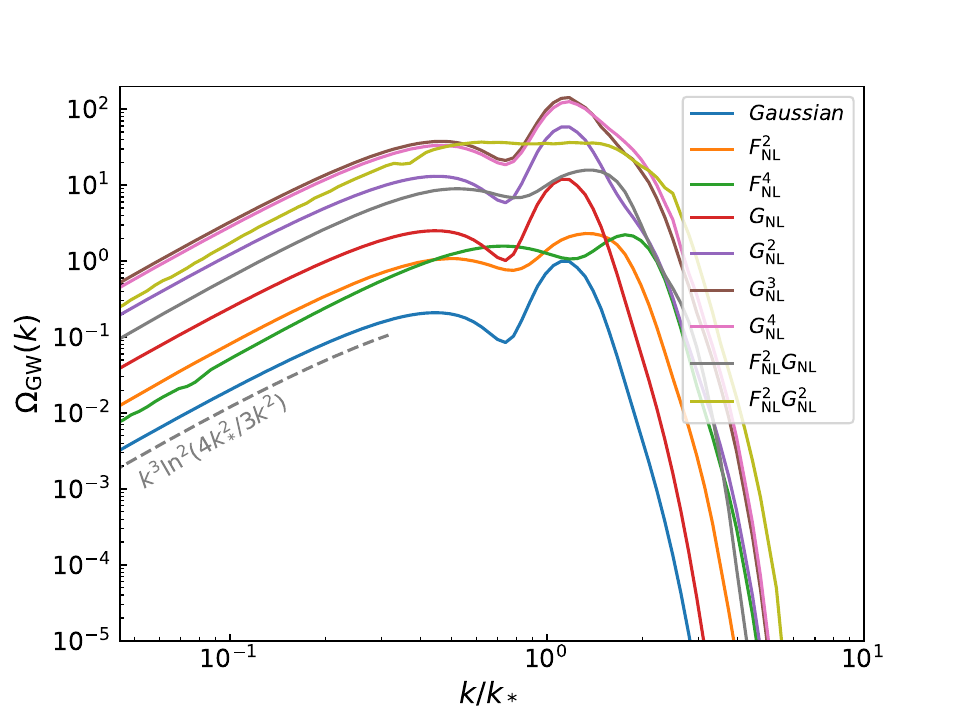}
    \caption{\label{SIGW-Ogwng}  
    The non-Gaussian components of the SIGWs energy spectrum generated by a log-normal power spectrum with $\sigma_*=0.2$. The dashed line corresponds to $\Omega_{\rm GW}(k) \propto k^3 \ln^2 \left( \frac{4k_*^2}{3k^2} \right)$. \textcolor{black}{Copied from Ref.~\cite{Yuan:2023ofl} with permission.} }
\end{figure*}	

The analysis of the infrared scaling was further extended to the case of non-Gaussian primordial curvature perturbations in \textcolor{black}{Ref.}~\cite{Yuan:2023ofl}. The authors considered contributions from the local type non-Gaussianities up to third-order, namely
\begin{equation}    \zeta\left(\zeta_{g}\right)=\zeta_{g}+F_{\mathrm{NL}}\left(\zeta_{g}^{2}-\left\langle\zeta_{g}^{2}\right\rangle\right)+G_{\mathrm{NL}} \zeta_{g}^{3},
\end{equation}
where $\zeta_g$ is the Gaussian curvature perturbations, $F_{\mathrm{NL}}$ and $G_{\mathrm{NL}}$ are the non-Gaussian parameters. Since SIGWs are sourced at second order by scalar perturbations, their energy spectrum $\Omega_{\rm GW}(k)$ is determined by the four-point correlation function of $\zeta^4$. Substituting the above local-type expansion into the four-point correlation function leads to a series of contributions from different orders. These include not only the Gaussian term ($\zeta_g^4$), but also a variety of non-Gaussian terms whose coefficients are given by:
\begin{itemize}
    \item $F_{\rm NL}^2$, $F_{\rm NL}^4$: contributions from quadratic non-Gaussianity,
    \item $G_{\rm NL}$, $G_{\rm NL}^2$, $G_{\rm NL}^3$, $G_{\rm NL}^4$: contributions from cubic non-Gaussianity,
    \item Mixed terms $F_{\rm NL}^2 G_{\rm NL}$ and $F_{\rm NL}^2 G_{\rm NL}^2$.
\end{itemize}
\textcolor{black}{Ref.~}\cite{Yuan:2023ofl} has demonstrated that in the infrared region, $\Omega_{\mathrm{GW}}$ can also yield log-dependent scaling. A plot of non-Gaussian terms is shown in Fig.~\ref{SIGW-Ogwng} for a log-normal power spectrum,
\begin{equation}
    \mathcal{P}_g(k) = \frac{A}{\sqrt{2\pi\sigma_*^2}} \exp\left( -\frac{\ln^2(k/k_*)}{2\sigma_*^2} \right).
\end{equation}
where $\sigma_*$ denotes the dimensionless width of the spectrum, $A$ is related to the amplitude and $k_*$ corresponds to the peak of the spectrum. For $\sigma_*=0.2$, it can be seen that all these non-Gaussian terms shows the same infrared scaling, namely
\begin{equation}
\Omega_{\rm GW}(k) \propto k^3 \ln^2 \left( \frac{4k_*^2}{3k^2} \right).
\end{equation}

Overall, these works establish the log-dependent IR slope as a robust and universal feature of SIGWs generated by a wide class of inflation models. This feature offers a promising observational signature to distinguish SIGWs from other stochastic gravitational wave backgrounds.

\subsection{Growth of power spectrum for generating PBHs}

It is well known that in the standard slow-roll inflation the amplitude of the power spectrum of the curvature perturbations $\mathcal{R}$ can be expressed as 
\begin{eqnarray}
   \mathcal{P_R}= \frac{H^2}{8\pi^2\epsilon c_s} 
\end{eqnarray} at the moment when the mode exits the horizon during inflation, where $H$ is the Hubble parameter which is approximately constant during inflation, $\epsilon$  is  the slow-roll parameter and $c_s$ is the sound speed of the curvature perturbations. So,  a natural way to  enhance  the curvature perturbations is to decrease  the rolling speed of the inflaton which is   proportional to $\epsilon$ or to suppress the   sound speed.  

The reduction in the inflaton rolling speed can be realized in the ultra-slow-roll inflation~\cite{Yokoyama:1998pt,Choudhury:2013woa,Germani:2017bcs,Ezquiaga:2017fvi,Di:2017ndc, Fu:2019ttf,Chen:2022dqr,Gao:2018pvq,Xu:2019bdp,Liu:2020oqe},  in which the slow-roll parameter $\eta$, which is defined to be $\eta=\frac{\dot{\epsilon}}{\epsilon H}$ and satisfies the condition $|\eta|\ll 1$ in the slow-roll inflation, equals approximately $-6$, where an overdot denotes a derivative with respect to cosmic time $t$. During  inflation, the evolution of curvature perturbations satisfies the Sasaki-Mukhanov equation. Its solution for modes outside the Hubble horizon during the slow-roll inflation consists of a constant term and a decaying one, which leads to a nearly scale-invariant spectrum.   During the transition of $\eta$ from $\eta\simeq 0$ to $-6$, the term, which is decaying with time before the transition, becomes ``growing". This growing term  gradually  dominates  if the ultraslow-roll inflation lasts a sufficiently long time and it results in the enhancement of the curvature perturbations to meet the requirement of formation of a sizable amount of PBHs. The power spectrum of the curvature perturbations displays a $k^4$ growth and  has a dip preceding the $k^4$ dependence since the constant and growing terms cancel each other~\cite{Byrnes:2018txb,Carrilho:2019oqg, Liu:2020oqe}.

If the sound speed,  which equals  $1$ in the canonical scalar field inflation model, decreases during inflation, the curvature perturbations can also be amplified~\cite{Ballesteros:2018wlw,Kamenshchik:2018sig,Gorji:2021isn,Romano:2020gtn,Ballesteros:2021fsp}. We have found that   the solution of the curvature perturbations does not contain growing terms and still has a constant component and a decay part. However,  the constant component contains a $k$-independent term and two $k$-dependent terms, which leads to the growth of the power spectrum at the small scale. These characters are different from that in the case of the transition from the slow-roll inflation to the ultraslow-roll one, where there is an appearance of the growing term~\cite{Zhai:2022mpi}. We also have obtained that the power spectrum of the curvature perturbations has a $k^2$ growth and there is no dip in the power spectrum since no term cancels the constant one, which is different from the case of the ultraslow-roll inflation~\cite{Zhai:2023azx}.

When both the sound speed of the curvature perturbations and the inflaton’s rolling speed are suppressed during inflation, the power spectrum can display a $k^6$ growth under certain conditions in the cases of the simultaneous changes of the sound speed and the slow-roll parameter $\eta$ and the change of the sound speed  preceding that of the slow-roll parameter $\eta$~\cite{Zhai:2023azx}. The $k^6$ growth is the steepest growth of the power spectrum reported so far.

%\cite{Chen:2024gqn}

\subsection{PBHs generated by the non-minimal spectator field}
The action for an inflation field, $\phi$, and a non-minimal spectator field, $\chi$, can be written as
\begin{align}
S=\int \mathrm{d}^4 x\sqrt{-g} \Bigg[&-\frac{1}{2}(\partial \phi)^{2}-V(\phi)-\frac{1}{2} f(\phi)^{2}(\partial \chi)^{2}-\frac{1}{2} m^{2} \chi^{2}\Bigg],
\label{actionpc}    
\end{align}
where $f(\phi)$ is the  non-minimal coupling function between $\phi$ and $\chi$. Expanding the action to second-order in momentum space, the equation of motion for the perturbations $\delta\phi$ and $\delta\chi$ can be obtained as follows
\begin{align}
\label{sigma1}
\delta\phi_k''+2{\cal H} \delta\phi_k'+\left[k^2+a^2V_{,\phi\phi}-(f_{,\phi}^2+ff_{,\phi\phi})\chi'^2\right]\delta\phi_k\nonumber\\
-2ff_{,\phi}\chi'\delta\chi_k'=0,\\
\delta\chi_k''+2\left({\cal H}+{f'\over f}\right)\delta\chi_k'+\left(k^2+{m^2a^2\over f^2}\right)\delta\chi_k\nonumber\\
+{2\over a^2f^2}{\mathrm{d}\over \mathrm{d}\tau}\left(a^2ff_{,\phi}\chi'\delta\phi_k\right)=0.    
\end{align}

\begin{figure*}
\centering
\includegraphics[width=0.6\textwidth]{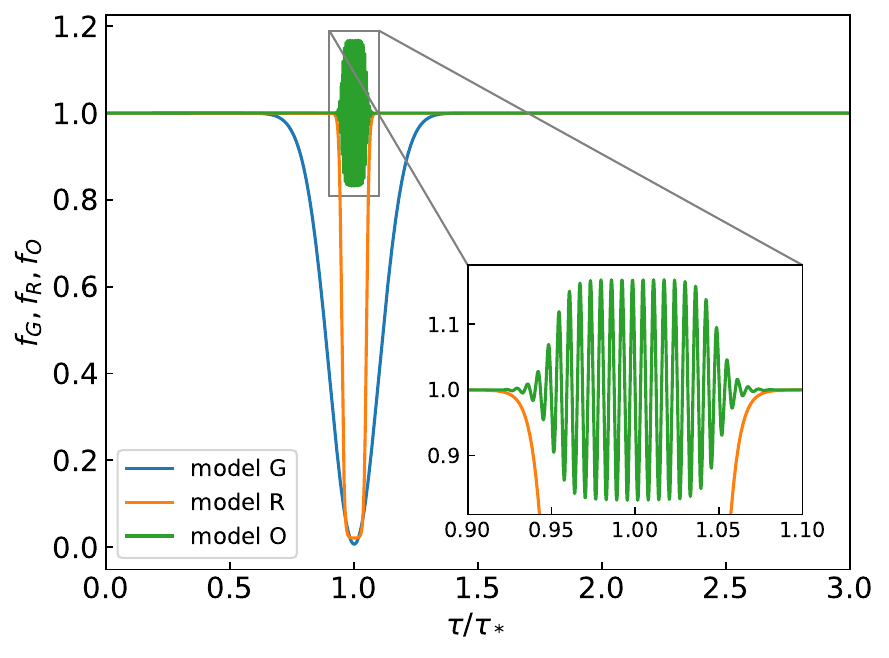}
\caption{\label{Non-minimal-fplot} 
The non-minimal coupling functions for model G, model R and model O. Here we set $\Delta=0.1$, $\Lambda=0.01$, $\xi=0.001$ and the values of $A_\text{G}$, $A_\text{R}$ and $A_\text{O}$ are chosen to let PBHs make up all of the DM.
}
\end{figure*}

\begin{figure*}[htbp!]
\centering
\includegraphics[width=0.45\textwidth]{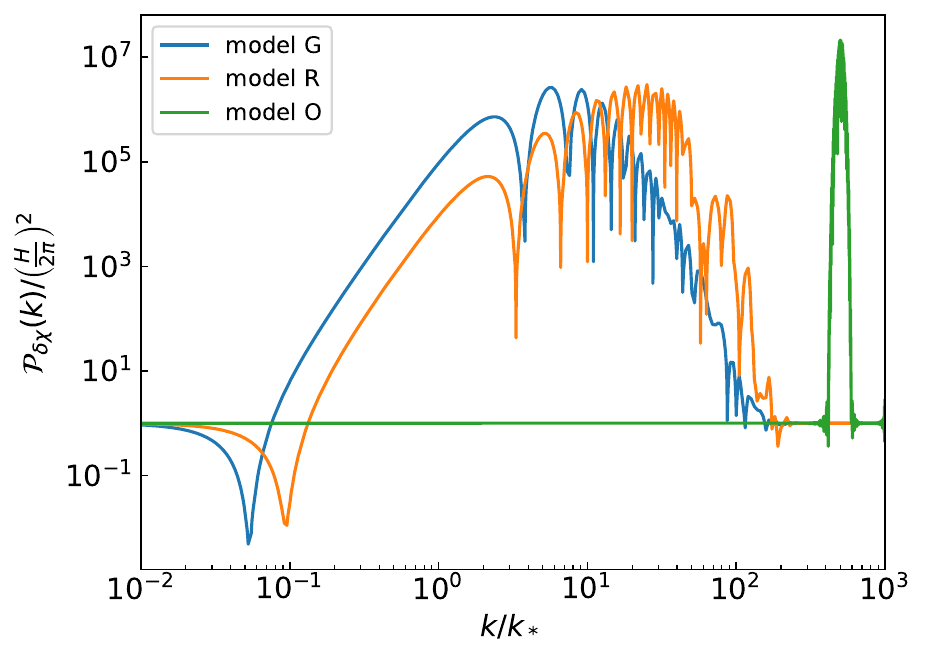}
\includegraphics[width =0.45\textwidth]{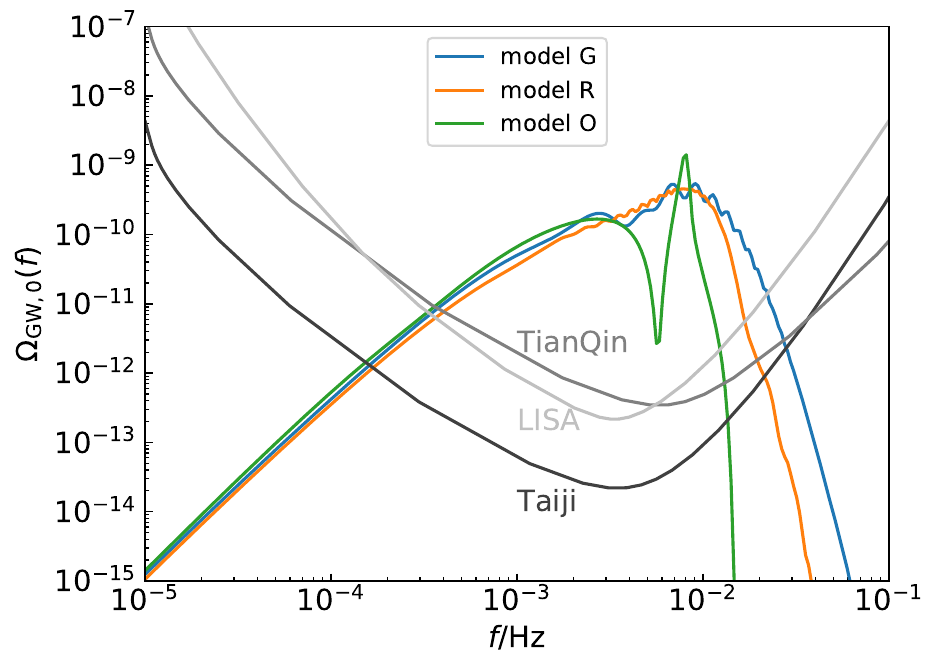}
\caption{\label{non-minimal-plot} 
The parameters of the coupling function is fixed to be $\Delta=0.1$, $\Lambda=0.01$ and $\xi=0.001$. The amplitudes $A_\text{G}$, $A_\text{R}$ and $A_\text{O}$ are chosen for PBHs making up all of the CDM, see \cite{Meng:2022low} for details.
\textit{Left panel}: The power spectra of $\delta \chi$ for model G, model R and model O. 
\textit{Right panel}: The resulting SIGW energy spectrum for the three models, together with power-law integrated sensitivity curves for Taiji \cite{Hu:2017mde}, TianQin \cite{TianQin:2015yph} and LISA \cite{Audley:2017drz} respectively, assuming a four year detection. The reference scale $\phi_*$ is chosen to generate PBHs with masses $\simeq 10^{-12}M_{\odot}$.
}
\end{figure*}
Introduce a canonical variable $\varphi_k$ which is related to $\delta\chi_k$ by $\delta \chi_k=\varphi_k/(af)$. The equation of motion for $\varphi_k$ reads
\begin{align}
\label{sigma}
\varphi_k''+\left[k^2-\frac{(af)''}{af}+\frac{m^{2}a^2}{f^{2}}\right]\varphi_k+\frac{2}{af}{\mathrm{d}\over \mathrm{d}\tau}\left(a^2ff_{,\phi} \chi'\delta \phi_k\right)=0.    
\end{align}
Neglecting the effective mass of the spectator field and the above equation can be simplified to 
\begin{align}
\label{sigmam}
\varphi_k''+\left[k^2-\frac{(af)''}{af}\right]\varphi_k=0.  
\end{align}

The idea of introducing a coupling function with a sharp dip to enhance the curvature power spectrum was studied in \textcolor{black}{Ref.}~\cite{Pi:2021dft}, where $f(\phi)$ is dramatically varying at $\phi=\phi_*$ and $f(\phi)=1$ elsewhere. Such a idea was then generalized in \textcolor{black}{Ref.}~\cite{Meng:2022low} (labeled as model G, model R and model O) such that
\begin{equation}
f_\text{G}(\phi)=1-A_\text{G} \exp\left[{-{(\phi-\phi_*)^2\over 2\Delta_\phi^2}}\right],    
\end{equation}
\begin{equation}
f_\text{R}(\phi)=1-{A_\text{R}\over 2}\left[\text{Tanh}{\phi-(\phi_*-\Delta_\phi/2)\over \Lambda_\phi}-\text{Tanh}{\phi-(\phi_*+\Delta_\phi/2)\over \Lambda_\phi}\right], 
\end{equation}
\begin{align}
f_\text{O}(\phi)=1-&{A_\text{O}\over 2}\left[\text{Tanh}{\phi-(\phi_*-\Delta_\phi/2)\over \Lambda_\phi}-\text{Tanh}{\phi-(\phi_*+\Delta_\phi/2)\over \Lambda_\phi}\right]\nonumber\\
&\times\sin {\phi-\phi_*\over \xi_\phi},     
\end{align}
where the amplitudes of three models are characterized by $A_\text{G}$, $A_\text{R}$ and $A_\text{O}$ respectively and $\phi_*$ corresponds to the field value at a reference conformal time $\tau_*$. The plot of the coupling functions are shown in Fig.~\ref{Non-minimal-fplot}.

Taking the Bunch-Davies vacuum condition in the sub-horizon limit, the power spectrum of the spectator field can be obtained numerically, which is defined as
\begin{equation}
    \lim _{k \tau \to 0^{-}}\left\langle\delta \chi_{\boldsymbol{k}}(\tau) \delta \chi_{\boldsymbol{k}^{\prime}}(\tau)\right\rangle=(2 \pi)^{3} \delta^{(3)}\left(\boldsymbol{k}+\boldsymbol{k}^{\prime}\right) \frac{2 \pi^{2}}{k^{3}} \mathcal{P}_{\delta \chi}(k).
\end{equation}
The power spectrum and the resulting SIGW energy spectrum are shown in Fig.~\ref{non-minimal-plot}. 

In the above non-minimal spectator field, the curvature perturbation is mainly generated from the perturbation of the spectator field. 
Importantly, when the reheating hypersurface is approximately straight or when the inflaton decay rate depends linearly on the spectator field, the curvature perturbation sourced from the spectator field remains nearly Gaussian \cite{Huang:2009vk}. In the case of the curvaton scenario with a quadratic potential, the non-Gaussianity parameter $F_{\rm NL}$ becomes small when the curvaton dominates the energy density at the time of its decay \cite{Sasaki:2006kq}, which is required by the perturbativity condition \cite{Meng:2022ixx}. Therefore, under such setups, the curvature perturbation can achieve a large amplitude suitable for PBH formation, while maintaining suppressed local-type non-Gaussianity.

\subsection{Testing PBH with multiband gravitational-wave observations}

The inflationary paradigm predicts that primordial curvature perturbations generated in the early Universe can leave imprints on various cosmological observables. While these perturbations are tightly constrained on large scales by CMB measurements, their properties on smaller scales are less certain~\cite{Mesinger:2005ah, Bringmann:2011ut, Chluba:2012we}. If the small-scale curvature perturbations have a large amplitude, they can give rise to two distinct types of SGWBs that may be detectable across multiple frequency bands by current and future GW experiments.

The first type of SGWB, known as SIGWs, is generated by the nonlinear coupling of primordial curvature perturbations to tensor perturbations during the radiation-dominated era \cite{Ananda:2006af,Baumann:2007zm,Cai:2018dig,Kohri:2018awv}. The second type arises from the coalescence of binary PBHs that form when sufficiently large curvature perturbations re-enter the horizon \cite{Zeldovich:1967lct, Hawking:1971ei, Carr:1974nx}. These two SGWBs typically have different characteristic frequencies and spectral shapes, encoding valuable information about the primordial perturbations and the early Universe.

In Ref.~\cite{Liu:2021jnw}, we uncover a remarkable relation between the peak frequencies of the SIGW and PBH binary GW spectra as 
\begin{equation}    
\begin{aligned}
\frac{\nu_{\mathrm{I}}^2}{\nu_{\mathrm{B}}} & =H_0 \Omega_{r, 0}^{1 / 2}\left[\frac{1}{2}\left(\frac{g_{* s}\left(T_*\right)}{g_{* s}\left(T_0\right)}\right)^{-\frac{2}{3}}\left(\frac{g_{* r}\left(T_*\right)}{g_{* r}\left(T_0\right)}\right)^{\frac{1}{2}}\right]\left[\frac{C_{\mathrm{I}}(\Theta)^2}{C_{\mathrm{B}}(\Theta)}\right] \\
& \equiv H_0 \Omega_{r, 0}^{1 / 2} G\left(k_*\right) C(\Theta) \equiv H_0 \Omega_{r, 0}^{1 / 2} Y\left(k_*, \Theta\right),
\end{aligned}
\end{equation}
where $\nu_{\mathrm{I}}$ and $\nu_{\mathrm{B}}$ denote the peak frequencies of SIGWs and PBH merger GWs, respectively, $H_0$ represents the Hubble constant, and $\Omega_{r, 0}$ is the present-day radiation energy density fraction. The parameters $k_*$ and $\Theta$ characterize the position and shape of the unimodal power spectrum of primordial curvature perturbations, respectively.

\begin{figure*}
\centering
\includegraphics[width=0.8\textwidth]{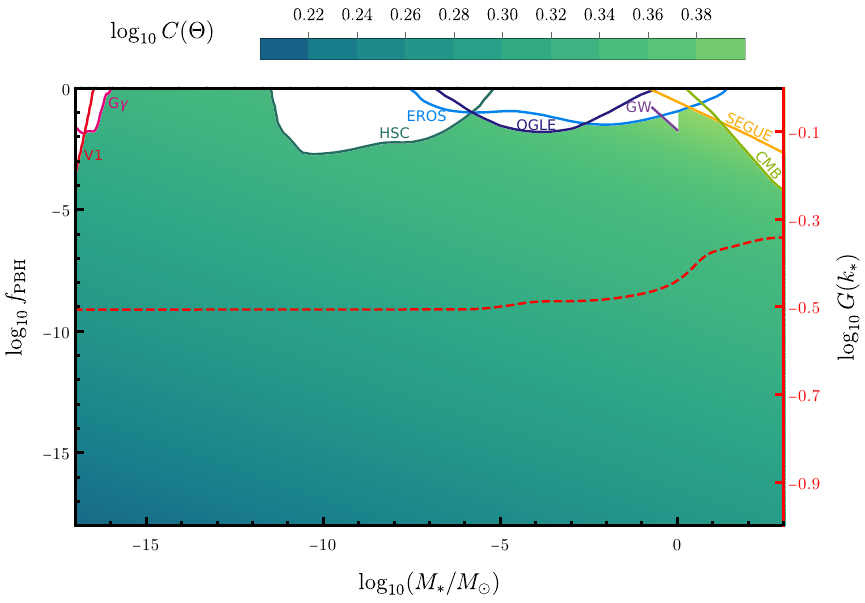}
\caption{\label{fig:lgC_lgG_delta} 
Plot of $C(\Theta)$ and $G(k_{*})$ in the $M_{*}-f_{\mathrm{PBH}}$ plane. The colorized region denotes $\log_{10}C(\Theta)$, and the red dashed line indicates $\log_{10}G(k_{*})$. Various observational constraints on the PBH abundance are shown as excluded regions, derived from the galactic 511 keV line from Hawking radiation (G$\gamma$ \cite{DeRocco:2019fjq, Laha:2019ssq, Dasgupta:2019cae, Laha:2020ivk}), Voyager 1 measurements (V1 \cite{Boudaud:2018hqb}), gravitational lensing events (HSC \cite{Niikura:2017zjd}, EROS \cite{EROS-2:2006ryy}, OGLE \cite{Niikura:2019kqi}), LIGO/Virgo observations (O2 \cite{LIGOScientific:2019kan}, O3a \cite{Hutsi:2020sol}), dynamical effects (SEGUE \cite{Koushiappas:2017chw}), and cosmic microwave background (CMB~\cite{Poulin:2017bwe}). Copied from Ref.~\cite{Liu:2021jnw} with permission.
}
\end{figure*}

\begin{figure*}[htpb]
    \centering
    \includegraphics[width=0.6\textwidth]{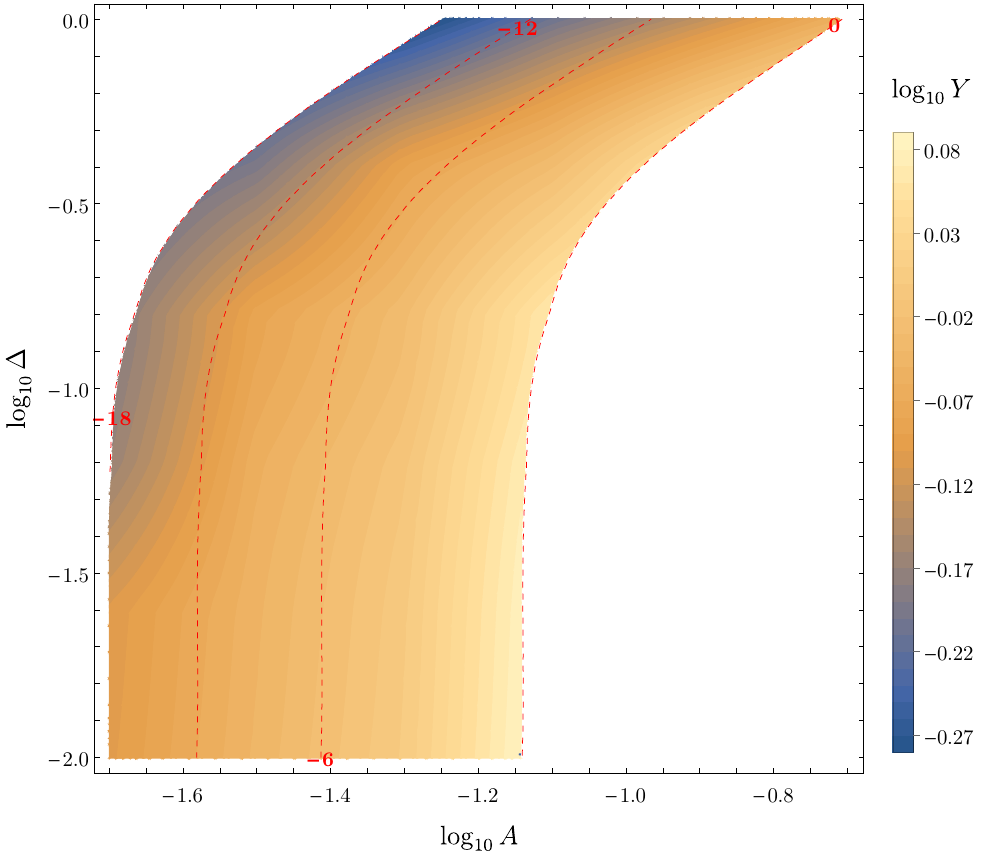}
    \caption[]{The numerical value of $Y$ in the $A-\Delta$ plane with $M_*=10 M_\odot$. The red dashed lines represent the contour lines of $\log_{10}f_{\mathrm{PBH}}$. Copied from Ref.~\cite{Liu:2021jnw} with permission.}
    \label{fig:lgY}
\end{figure*}

The position parameter $k_*$ affects the peak frequency relation through the factor $G(k_{*})$, which depends on the thermal history of the Universe. Interestingly, $G(k_{*})$ remains nearly constant \cite{Saikawa:2018rcs}. To investigate the influence of the shape parameters $\Theta$ on the peak frequency relation, we first consider the delta spectrum $\mathcal{P}_{\mathcal{R}}(k)=A\ \delta( \ln(k/k_{*}) )$, where $\Theta =\{A\}$. Although the parameter $A$ affects the PBH abundance $f_{\mathrm{PBH}}$ exponentially, it only affects $C(\Theta)$ polynomially. Consequently, $C(\Theta)$ remains nearly constant even when $f_{\mathrm{PBH}}$ varies over a wide range. Fig.~\ref{fig:lgC_lgG_delta} numerically demonstrates the values of $\log_{10}G(k_{*})$ and $\log_{10}C(\Theta)$, confirming this expectation. For the lognormal case, $\mathcal{P}_{\mathcal{R}} (k) = \frac{A}{\sqrt{2\pi}\Delta} \exp \left( -\frac{\ln^{2}(k/k_{*})}{2\Delta^{2}} \right)$, Fig.~\ref{fig:lgY} reveals that $Y(k_{*},\Theta) \sim \mathcal{O}(1)$ regardless of whether the spectrum of curvature perturbations exhibits a narrow peak $(\Delta \ll 1)$ or a broad peak $(\Delta \sim 1)$.

This peak frequency relation has several important implications. First, it provides a smoking-gun criterion for the existence of PBHs. If future multiband gravitational-wave observations detect two different gravitational-wave signals with peak frequencies satisfying this relation, it would confirm the existence of PBHs. Second, the relation is nearly constant over a large parameter space, even when the PBH mass varies by 20 orders of magnitude, making it a robust probe for PBHs.

Furthermore, we propose a new method to measure the Hubble constant using this peak frequency relation. By utilizing multiband gravitational-wave observations, we can constrain $H_0$ independently of the cosmic distance ladder, potentially providing valuable insights into the current tension between measurements of H0 from different methods.

Our work highlights the power of joint analysis of different frequency bands of the SGWB. The peak frequency relation between SIGWs and GWs from PBH mergers offers a novel way to confirm the existence of PBHs and measure the Hubble constant, demonstrating the potential of multiband gravitational-wave observations in exploring fundamental questions in astrophysics and cosmology.

%%%%%%%%%%%%%%%%%%%%%%%%%%%%%%%%%%%%%%%%%%%%%%%%%%%%%%%%%%%%%%%%%%%%%%%%%%%%%%%%%%%%%%
\section{Testing gravity in GW propagation and polarizations\label{sec:tg}}
%%%%%%%%%%%%%%%%%%%%%%%%%%%%%%%%%%%%%%%%%%%%%%%%%%%%%%%%%%%%%%%%%%%%%%%%%%%%%%%%%%%%%%
%contributor: Jin Qiao, Tao Zhu, Yu-Qi Dong, Yu-Xiao Liu

GW originating from cosmological sources must propagate through vast cosmic distances before being detected by GW detectors, inevitably encoding crucial information about the structure, composition, and evolution of the Universe. Analogous to the CMB, GWs traversing an inhomogeneous Universe can experience observable effects in their frequency, chirp mass, and luminosity distance due to the integrated Sachs-Wolfe effect \cite{Laguna:2009re, Fier:2021fbt}. Consequently, analyzing gravitational-wave data for such effects allows us to probe the inhomogeneities of spacetime and to place constraints on the properties of dark matter, dark energy, and the large-scale structure of the Universe. Furthermore, various modified gravity theories beyond GR predict alterations in the propagation of GWs, with such effects accumulating over cosmological distances \cite{Zhu:2023rrx}. In particular, studies involving breakings of Lorentz symmetry and parity symmetry suggest that GW observations can provide valuable insights into the fundamental symmetries of gravity, thereby advancing our understanding of its nature and potentially guiding the development of quantum gravity \cite{Zhu:2023rrx, Zhang:2025kcw}. In addition, GWs also exhibit different polarization modes in a large number of modified theories of gravity. The detection of polarization modes of GWs provides a powerful tool for testing various modified gravity theories. 

\subsection{GW propagation with cosmological perturbations}

GWs generated by cosmological sources propagate over vast cosmic distances before being detected by GW observatories, thereby carrying valuable information not only about their origins, but also about the details of cosmological expansion and the inhomogeneities of the Universe. These detections open a completely new window to explore the Universe by using GWs, as so far our understanding of the Universe almost all comes from observations of electromagnetic waves only (possibly with the important exceptions of cosmic rays and neutrinos). Similar to CMB, when GWs propagate through the inhomogeneous structure of the Universe, they experience a gravitational integrated Sachs-Wolf (iSW) effects, a analogue of the electromagnetic integrated Sachs-Wolf effects in cosmology \cite{Sachs:1967er}, which results additional corrections to the phase, frequency, and amplitude of the GW waveforms \cite{Laguna:2009re, Bertacca:2017vod, Mukherjee:2019wcg, Mukherjee:2019wfw, Cusin:2019rmt, Fier:2021fbt, Bonvin:2016qxr, Isaacson:1968hbi}. Testing or constraining these observational effects using GW detections can also help us to further understand the inhomogeneity of cosmic spacetime \cite{Mukherjee:2019wcg, Mukherjee:2019wfw, Mukherjee:2020hyn, Mukherjee:2020mha,Bonvin:2016qxr}, the properties of dark matter and dark energy \cite{Bertacca:2017vod, Garoffolo:2020vtd}, and the nature of gravity \cite{LISACosmologyWorkingGroup:2019mwx, Garoffolo:2019mna, Liu:2020mab,Luo:2024vls,Fier:2025huc, Tasinato:2021wol, Dalang:2019rke, Ezquiaga:2020dao, Dalang:2020eaj, Mukherjee:2020mha}.

In most of these studies, only the iSW effect of cosmological scalar perturbations on GWs has been considered, and the influence of the cosmological tensor perturbations has been neglected. The effect of cosmological tensor perturbations is also important. PGWs are one of the most important targets in the next stage (Stage IV) of CMB observations \cite{Abazajian:2019tiv}. Recently, Ref.~\cite{Fier:2021fbt} studied in detail the iSW effects due to both the cosmological scalar and tensor perturbations on the propagation of GWs. This study is the first to systematically and simultaneously investigate the effects of cosmological scalar and tensor perturbations on the propagation of GWs. It establishes a complete theoretical system for studying the propagation of GWs in nonuniform cosmological backgrounds on cosmological scales. In the following subsubsections, we present a brief summary of Ref.~\cite{Fier:2021fbt}. 

\subsubsection{The high-frequency approximation}

To describe the GWs propagating through the inhomogeneous Universe from cosmic distances to observers properly, we begin with the introduction of three fundamental scales: the typical wavelength of GWs ($\lambda$), the scale of cosmological perturbations ($L_{c}$), and the size of the observable Universe ($L$). For GWs detectable by current and future detectors, the relationship $\lambda\ll L_c \ll L$ is valid. This inequality is of utmost importance as it allows researchers to approximate these GWs as high-frequency GWs, a crucial step in simplifying the theoretical analysis. In addition to this, conditions $h_{\mu\nu}\ll 1$ and $\epsilon\ll\epsilon_c\ll1$ (where $g_{\mu\nu}=\gamma_{\mu\nu}+ h_{\mu\nu}$ with $\gamma_{\mu\nu}$ denoting the background) are essential. The former ensures that the backreaction of GWs on the background spacetime is negligible, while the latter validates the use of linearized Einstein field equations. These approximations and conditions form the foundation for the entire research framework.

Under the high-frequency approximation, the spatial, traceless, and Lorenz gauge conditions can be imposed simultaneously \cite{Isaacson:1968hbi}. By initially imposing the spatial gauge ($\chi_{0\mu}=0$), one can derive the linearized Einstein field equations for $\chi_{\mu\nu}\equiv h_{\mu\nu}-\frac{1}{2}\gamma_{\mu\nu}h$. This process involves expanding the relevant tensors in the Einstein field equations and averaging over a suitable scale ${l}\; (\lambda\ll l\ll L_c)$ to separate the GWs from the background. The derived equations are then combined with the geometric optics approximation. By representing $\chi_{\alpha\beta} = Re(A_{\alpha\beta}e^{i\phi/\epsilon})= Re(e_{\alpha\beta}Ae^{i\phi/\epsilon})$ and considering the Lorenz and traceless gauges, one then can deduce important properties of GWs, such as their propagation along null geodesics and the parallel transport of the polarization bivector. This provides a comprehensive approach to studying GWs in an inhomogeneous Universe, allowing one to make predictions and interpretations that are consistent with both theoretical expectations and observational data.

\subsubsection{Propagation properties of gravitational waves}

When studying their propagation in an inhomogeneous Universe with the high-frequency approximation and geometric optics approximation, several remarkable properties of GWs have been identified. Even in the presence of cosmological scalar and tensor perturbations, GWs still propagate along null geodesics. This means that they follow the shortest paths in spacetime, similar to how light travels in a vacuum. Mathematically, this is expressed by equations such as $\kappa^{\lambda}\nabla_{\mu}\kappa_{\lambda} = \kappa^{\lambda}\nabla_{\lambda}\kappa_{\mu}$, where $\kappa_{\mu}=\nabla_{mu}\Phi$ and $\Phi$ is related to the phase of the GWs. This property holds as long as the geometric optics approximation for high-frequency GWs is valid.

Another important feature of GWs is that their polarization bivector 
$e_{\alpha\beta}$ is parallel transported along the null geodesics. This indicates that the direction of GW polarization remains consistent during the propagation process, which is described by the equation $\kappa^{\mu}\nabla_{\mu}e_{\alpha\beta}=0$ \cite{2018grav.book.....M}. In addition, the current of gravitons moving along the null geodesics, defined as $J^{\mu} \equiv A^2\kappa^{\mu}$, is conserved, that is, $\nabla_{\mu}J^{\mu} =0$, where $A$ is the amplitude of the GWs. These propagation characteristics of GWs are not only of great significance for understanding the fundamental nature of gravitational radiation but also play a crucial role in various astrophysical and cosmological studies. They provide valuable insights into the structure of the Universe and the behavior of matter and energy on large scales.

\subsubsection{Gravitational iSW effects}

Building upon the understanding of GW propagation, the gravitational integrated Sachs-Wolfe effect emerges as a key factor in deciphering the behavior of GWs in an inhomogeneous Universe \cite{Sachs:1967er}. This effect comes into play when considering the influence of cosmological scalar and tensor perturbations on GWs.

This effect is derived based on the fact that GWs, similar to electromagnetic radiation, propagate along null geodesics in an inhomogeneous Universe.
When both cosmological scalar and tensor perturbations are present, the iSW effect of GWs can be calculated. First, a conformal metric $\tilde{\gamma}_{\mu\nu}$ is introduced, which simplifies the calculations. Because the zeroth-order spacetime becomes the Minkowski spacetime, the null geodesics are straight lines in this framework. The effects of scalar and tensor perturbations are manifested through the perturbations of the null geodesics. For example, the time-component of the wave-vector perturbation $\tilde{k}^{(1)0}$ is affected by scalar and tensor perturbations, and its expression contains terms related to the scalar perturbations $\phi$ and $\psi$,  and the tensor perturbation $H_{ij}$, and the integrated effects $I^{(s)}_{iSW}$ and $I^{(t)}_{iSW}$. The former is the gravitational iSW effect caused by cosmological scalar perturbations, given by $I^{(s)}_{iSW}=\int^{\lambda}_{\lambda_e}\partial_{\tau}(\phi+\psi)d\lambda'$, while the latter is a new term representing the gravitational integrated effect caused by cosmological tensor perturbations, $I^{(t)}_{iSW}=n^kn^l\int^{\lambda}_{\lambda_e}\partial_{\tau}H_{kl}d\lambda'$.

These perturbations affect the GW phase $\varphi$, and the phase change $\delta\varphi$ is given by 
\begin{eqnarray}
\delta\varphi=\varphi-\varphi_e = \int^{\lambda}_{\lambda_e}\partial_{\tau}(\phi+\psi)d\lambda'-\frac{1}{2}n^kn^l\int^{\lambda}_{\lambda_e}\partial_{\tau}H_{kl}d\lambda'.
\end{eqnarray} 
In addition, they also affect the frequency ratio $\frac{\omega_{r}}{\omega_e}$ of received and emitted GWs, as well as the luminosity distance $D_L$ and chirp mass $M$ of a binary system. The modified luminosity distance and chirp mass are given by $D_L = \frac{d_L}{1-\Upsilon-\xi}$ and $M_r\equiv (\frac{1+z}{1-\Upsilon})M_e$ respectively, where $\Upsilon$ and $\xi$ are functions of scalar and tensor perturbations and their expressions are given in Ref.~\cite{Fier:2021fbt}. With these corrected quantities, the modified waveform of a binary system is given by
\begin{eqnarray}
\tilde h=\frac{{\cal M}_r}{D_L}(\pi f_r {\cal M}_r)^{2/3} e^{i(\varphi_e+\delta\varphi)}.
\end{eqnarray}
Overall, the gravitational iSW effect provides valuable information for understanding the nature of cosmic perturbations and GW sources. The applications of these general formulas to other studies are immediate, including the gravitational analogue of the electromagnetic Faraday rotations \cite{Hou:2019wdg, Feng:2020phf, Wang:1991nf, wanganzhong_book}, and their detections by the space- and ground-based detectors.  It would be also very important to extend such studies to include the relations between the GWs and their sources, high-order corrections to the geometrical optics approximations, and to modified theories of gravity.

\subsection{Constraints on Lorentz and parity violation from GWTC-3 data through a universal paramatrizations}

The detection of GWs has not only confirmed a key prediction of GR but also provided a novel experimental tool for testing alternative theories of gravity and exploring the fundamental nature of gravity \cite{LIGOScientific:2019fpa, LIGOScientific:2020tif, LIGOScientific:2021sio}. Using GW data from detectors such as LIGO, Virgo, and KAGRA, we can impose constraints on various modified theories of gravity, thereby advancing the development of gravitational theories. The use of GW detection to investigate the fundamental nature of gravity is both timely and highly relevant.

Symmetry and symmetry breaking have played a pivotal role in the development of modern physics. As early as 1954, parity symmetry breaking was discovered in weak interactions \cite{Lee:1956qn, Wu:1957my}, and Lorentz symmetry has been validated to extraordinary precision in the context of particle physics experiments within the standard model \cite{Mattingly:2005re, Kostelecky:2008ts}. In GR, which remains the most successful theory of gravity, Lorentz symmetry and parity symmetry are two fundamental symmetries. However, when quantum effects or the quantization of gravity are considered, it is widely believed that these symmetries may be broken. Consequently, numerous modified theories of gravity have been proposed, incorporating violations of Lorentz and parity symmetries in gravitational interactions. Since the breaking of Lorentz and parity symmetries can significantly alter the propagation of GWs, the detection of GWs provides a unique experimental platform to test or constrain such fundamental symmetry violations. This has enabled a lot of tests of parity/Lorentz symmetries by GW signals detected by LIGO-Virgo-KAGRA Collaboration \cite{LIGOScientific:2019fpa, LIGOScientific:2020tif, LIGOScientific:2021sio, Wang:2020cub, Wu:2021ndf, Gong:2021jgg, Zhao:2022pun, Wang:2021gqm, ONeal-Ault:2021uwu, Haegel:2022ymk, Wang:2025fhw, Ma:2024kkz, Ng:2023jjt, Zhu:2023rrx, Gong:2023ffb, Qiao:2022mln, Zhu:2022uoq, Niu:2022yhr, Wang:2021ctl, Okounkova:2021xjv, Yamada:2020zvt, Wang:2020pgu, Wang:2017igw}. 

To describe the possible parity- and Lorentz-violating effects in GWs, recently we adopt a systematic parametric framework for characterizing possible derivations of GW propagation and modified waveform with parity-and Lorentz-violating effects \cite{Zhu:2023rrx, Zhao:2019xmm}. This parametric framework has also been used in the study of parity- and Lorentz-violating effects in PGWs \cite{Li:2024fxy, Li:2022xww}. See references \cite{Nishizawa:2017nef, Ezquiaga:2021ler, Tahura:2018zuq, Saltas:2014dha} for other parameterized frameworks. In the following subsections, we present a biref review of the parameterized framework in \textcolor{black}{Refs.}~\cite{Zhu:2023rrx, Zhao:2019xmm}, the corresponding modified waveforms, and the constraints on parity and Lorentz violations from GWTC-3 data.

\subsubsection{Parametrization}

The parameterized framework in \textcolor{black}{Refs.}~\cite{Zhu:2023rrx, Zhao:2019xmm} is based on the parametrized equation of motion for the circular polarization modes of GWs,
\begin{eqnarray}
h''_A+(2+\bar{\nu}+\nu_A)\mathcal{H}h'_A+(1+\bar{\mu}+\mu)k^2h_A=0,
\end{eqnarray}
where $A = {\rm R}$ or ${\rm L}$ represents the right- and left-handed circular polarization modes, respectively. Here, a prime denotes differentiation with respect to the conformal time $\tau$, $\mathcal{H} = a'/a$ is the conformal Hubble parameter with $a$ being the scale factor of the Universe, and $k$ is the wavenumber. The new effects from theories beyond GR are characterized by four parameters: $\bar{\nu}$, $\bar{\mu}$, $\nu_A$ and $\mu_A$. For frequency-independent effects, a non-zero $\bar{\mu}$ modifies the GW speed, and its constraint can be obtained by comparing with the arrival time of photons from the associated electromagnetic counterpart. A non-zero $\bar{\nu}$ modifies the friction term, changing the GW damping rate, and its constraint can be derived from multi-messenger measurements of luminosity distances.

The parity-violating effects, the parameters $\nu_A$ and $\mu_A$ correspond to amplitude and velocity birefringences of GWs respectively. They are often frequency-dependent and can be further parametrized as $\mathcal{H}{\nu}_A=\left[\alpha_{{\nu}}(\tau)(k/aM_{\rm PV})^{\beta_{\nu}}\right]'$ and ${\mu}_A=\alpha_{{\mu}}(\tau)(k/aM_{\rm PV})^{\beta_{\mu}}$, where where $\beta_\nu$, $\beta_\mu$ are arbitrary numbers, $\alpha_\nu$, $\alpha_\mu$ are arbitrary functions of time, and $M_{\rm PV}$ denotes the energy scale of the parity violation. Through the stationary-phase approximation (SPA), the amplitude and phase corrections to the GR-based waveform due to parity-violating effects can be calculated as \cite{Zhu:2023rrx, Zhao:2019xmm}
\begin{eqnarray}
\bar{h}_A(f) = \bar{h}^{GR}_{A}e^{\rho_A\delta h_1}e^{i(\rho_A\delta\Psi_1)},
\end{eqnarray}
where where $ \tilde h_A^{\rm GR} $ is the corresponding GR-waveform, and its explicit form can be found in the previous works \cite{Zhao:2019xmm}, $\delta h_1$ and $\delta\Psi_1$ represent the amplitude and phase corrections due to the parity-violating effects which are related to parameters $\nu_A$ and $\mu_A$ respectively.

For Lorentz-violating effects, $\bar{\nu}$ and $\bar{\mu}$ correspond to Lorentz-violating damping and dispersions. They are also frequency-dependent, parametrized as $\mathcal{H}\bar{\nu}=\left[\alpha_{\bar{\nu}}(\tau)(k/aM_{\rm LV})^{\beta_{\bar{\nu}}}\right]'$ and $\bar{\mu}=\alpha_{\bar{\mu}}(\tau)(k/aM_{\rm LV})^{\beta_{\bar{\mu}}}$, where $\beta_{\bar \nu}$, $\beta_{\bar \mu}$ are arbitrary numbers, $\alpha_{\bar \nu}$, $\alpha_{\bar \mu}$ are arbitrary functions of time, and $M_{\rm LV}$ denotes the energy scale of Lorentz violation. The modified waveform with Lorentz-violating effects is given by
\begin{eqnarray}
\tilde{h}_A(f)=\tilde{h}^{GR}_A(f)e^{\delta h_2}d^{i\delta\Psi_2},
\end{eqnarray}
where $\delta h_2$ and $\delta\Psi_2$ are the amplitude and phase corrections due to the Lorentz-violating effects which are related to parameters $\bar{\nu}$ and $\bar{\mu}$ respectively. 

The above parametrizations and calculations are crucial for analyzing the effects of parity and Lorentz violations on GWs and obtaining relevant constraints from GW data. It is shown in Refs.~\cite{Zhu:2023rrx, Zhao:2019xmm} that this parametrization provides a general framework for studying the GW propagation of possible modifications caused by various modified gravitational theories. The specific forms of the parameters ${\cal H}\bar \nu$, $\bar \mu$, ${\cal H}\nu_A$, and $\mu_A$ in various modified gravity theories are summarized in Table I of Ref.~\cite{Zhu:2023rrx}.

\subsubsection{Constraints on parity and Lorentz violations}

With the modified waveforms given in the above subsubsection, one can now analyze the constraints on parity and Lorentz violations using the GWTC-3 data.
Bayesian inference was employed to analyze the publicly available data of binary black hole merger events in GWTC-3, aiming to constrain the parity and Lorentz violations in gravity. Different values of $\beta_{\bar{\nu}}$, $\beta_{\bar{\mu}}$, $\beta_{{\nu}}$, and $\beta_{{\mu}}$ were considered separately to test the parity- and Lorentz-violating waveforms.

\begin{figure*}
\centering
\includegraphics[width=5.7cm]{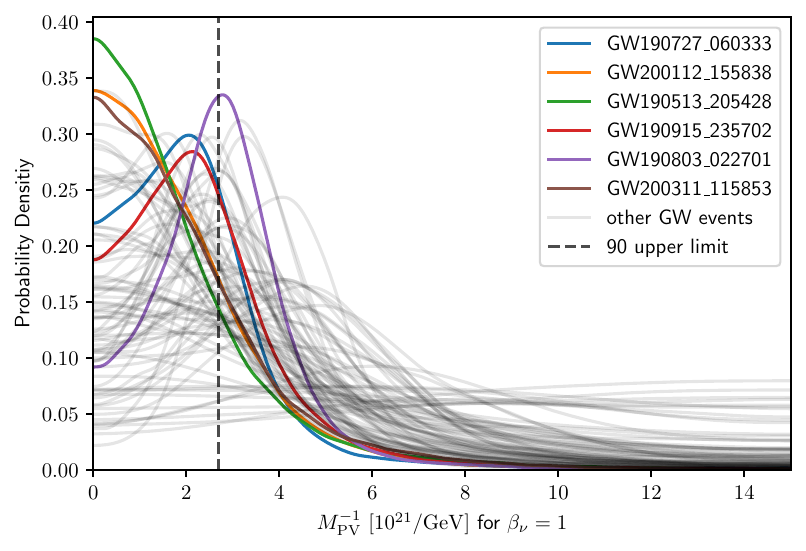}
\includegraphics[width=5.7cm]{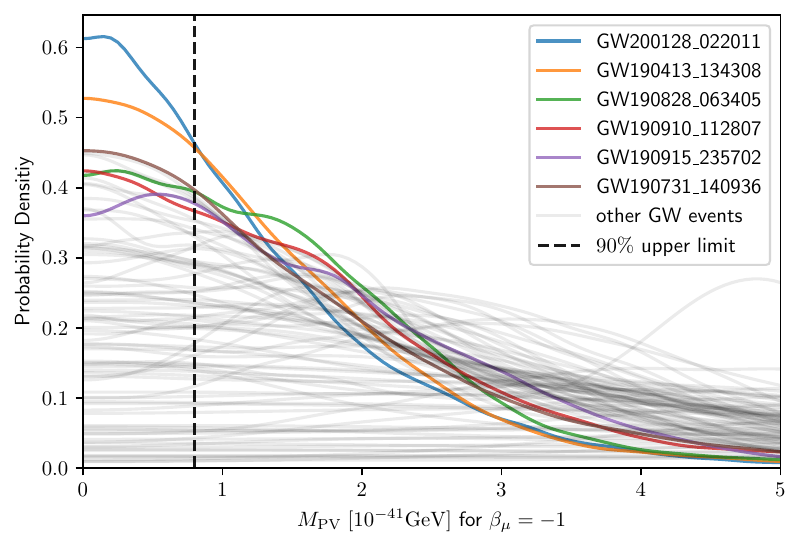}
\includegraphics[width=5.7cm]{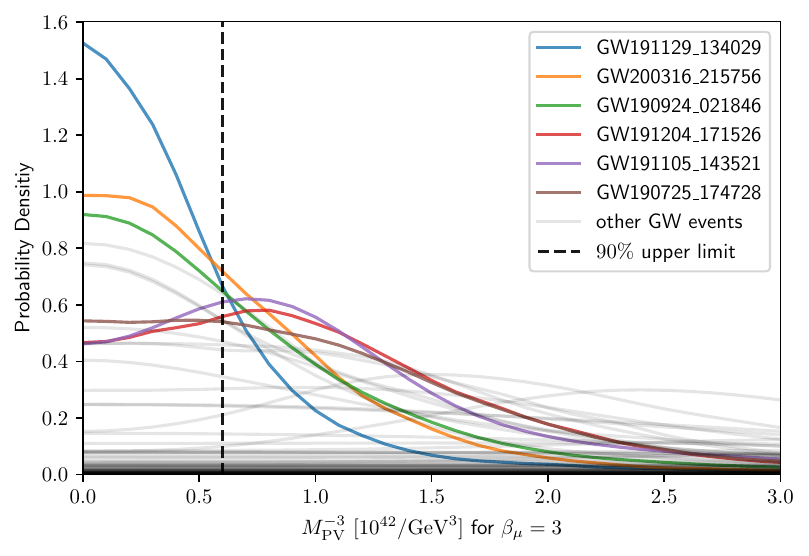}
\includegraphics[width=5.7cm]{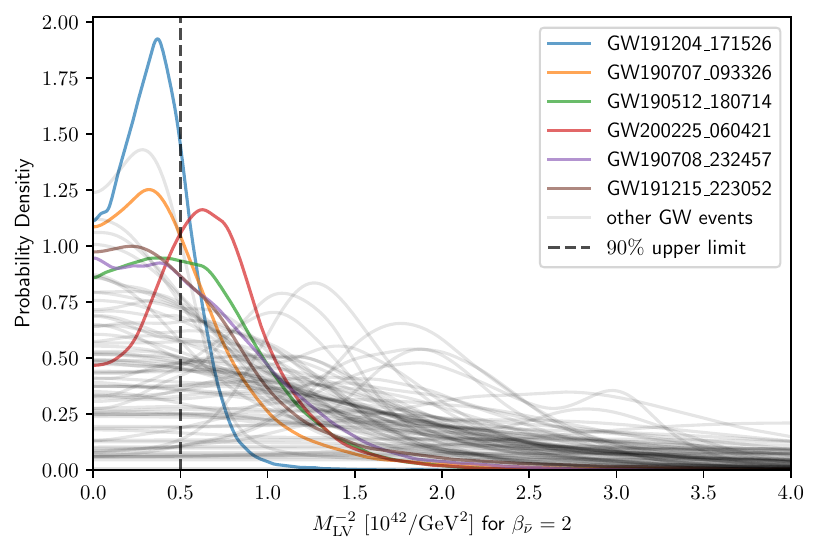}
\includegraphics[width=5.7cm]{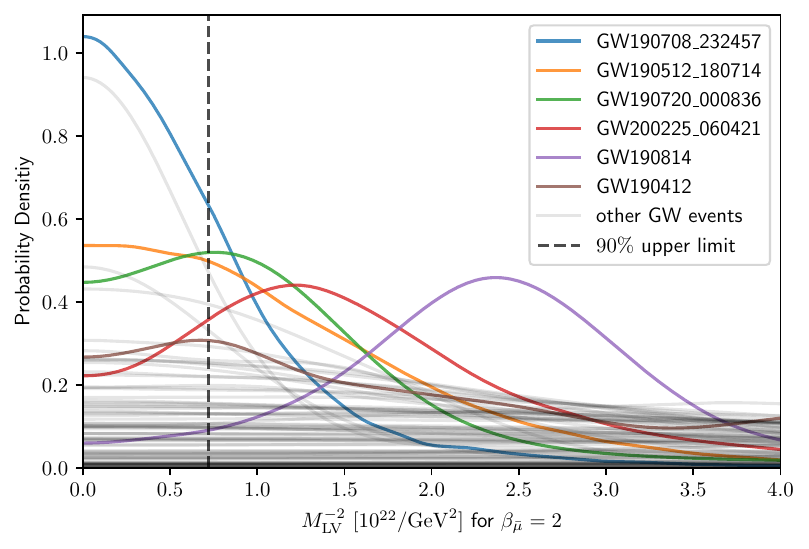}
\includegraphics[width=5.7cm]{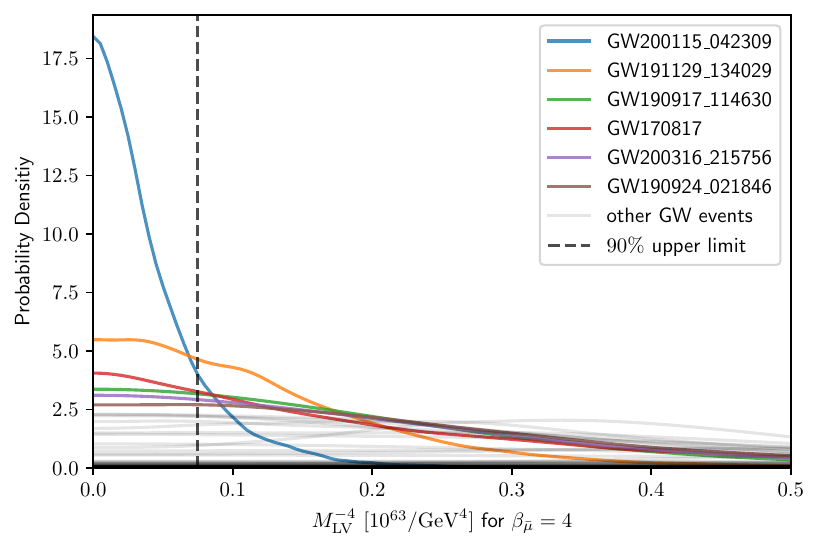}
\caption{The posterior distributions for $M_{\rm PV}^{-\beta_\nu}$ with $\beta_\nu=1$,  $M_{\rm PV}^{-\beta_\mu}$ with $\beta_\mu=-1,3$,  $M_{\rm LV}^{-\beta_{\bar \nu}}$ with $\beta_\mu= 2$, and $M_{\rm LV}^{-\beta_{\bar \mu}}$ with $\beta_{\bar \mu}= 2, 4$ from selected GW events in the GWTC-3. The legend indicates the events that give the tightest constraints. The vertical dash line in each figure denotes the 90\% upper limits from the combined result. Copied from Ref.~\cite{Zhu:2023rrx} with permission.} \label{posterior_tests}
\end{figure*}

For the parity-violating effects, the amplitude birefringence with $\beta_{\nu}=1$ and velocity birefringence with $\beta_{\mu}=-1,3$ were investigated. The test of velocity birefringence with $\beta_\mu=1$ was explored in \textcolor{black}{Refs.}~\cite{Zhao:2022pun, Wang:2021gqm}. Regarding the Lorentz-violating effects, the frequency-dependent damping with $\beta_{\bar\nu}=2$ and nonlinear dispersion relations with $\beta_{\bar\mu}=2,4$ were explored. Most of the GW events analyzed in each test did not show significant signatures of parity and Lorentz violations. From the marginal posterior distributions of relevant parameters and the redshift of the analyzed GW events, posterior distributions of $M^{-\beta_{\nu}}_{\rm PV}$, $M^{-\beta_{\mu}}_{\rm PV}$, $M^{-\beta_{\bar\nu}}_{\rm PV}$, and $M^{-\beta_{\bar\mu}}_{\rm PV}$ were obtained. By multiplying the posterior distributions of individual events, combined constraints for each analysis were derived, presented as $90\%$ upper limits. In Fig.~\ref{posterior_tests}, we display the marginalized posterior distributions of $M_{\rm PV}^{-\beta_\nu}$ with $\beta_\nu=1$,  $M_{\rm PV}^{- \beta_\mu}$ with $\beta_\mu=-1,3$,  $M_{\rm LV}^{-\beta_{\bar \nu}}$ with $\beta_{\bar \nu}= 2$, and $M_{\rm LV}^{-\beta_{\bar \mu}}$ with $\beta_{\bar \mu}= 2, 4$ from selected GW events in the GWTC-3.

\begin{table*}[t]
\footnotesize
%\centering
\caption{Results from the Bayesian analysis of the parity- and Lorentz-violating waveforms with GW events in GWTC-3. The table shows 90\%-credible upper bounds on $M_{\rm PV}$ for $\beta_{\nu}=-1$ (for velocity birefringence) and lower bounds on $M_{\rm PV}$ and $M_{\rm LV}$ for other cases. We also include bounds for several cases derived from existing tests with GWTC-1/GWTC-2/GWTC-3 in Refs.~\cite{Gong:2021jgg, Wu:2021ndf, LIGOScientific:2019fpa, LIGOScientific:2020tif, LIGOScientific:2021sio, Wang:2021gqm} for comparison.  The bounds on $M_{\rm LV}$ from the results of the parameters $A_4$ in \textcolor{black}{Refs.}~\cite{LIGOScientific:2019fpa, LIGOScientific:2020tif, LIGOScientific:2021sio} are derived via $M_{\rm LV}^{-2} = \hbar^2 A_4$ with $\hbar$ being the reduced Planck constant. Note that in deriving the bounds of $M_{\rm PV}$ and $M_{\rm PV}$, we do not transform their priors to be uniform, except for the case of $M_{\rm PV}$ with $\beta_\mu=-1$. Table is copied from Ref.~\cite{Zhu:2023rrx} with permission.}
    \vspace{.2cm}
    \label{tab}
\tabcolsep 11pt 
\begin{tabular*}{\textwidth}{ccccccc}
\toprule
     & \multicolumn{3}{c}{$M_{\rm PV}$ [GeV]} &\multicolumn{3}{c}{$M_{\rm LV}$ [GeV]} \\
\cline{2-4}  \cline{5-7}
  & $\beta_\nu=1$ & $\beta_\mu=-1$ & $\beta_\mu=3$ & $\beta_{\bar \nu}=2$ & $\beta_{\bar \mu}=2$ & $\beta_{\bar \mu}=4$     \\  
  \hline
GWTC-1  &    $1.0\times10^{-22}$ \cite{Wang:2021gqm} &      -   &  - &  - &  $0.8\times10^{-11}$ \cite{LIGOScientific:2019fpa}& - \\  
 GWTC-2  &    -   &     $6.5\times10^{-42}$ \cite{Wu:2021ndf}  &  $1.0\times10^{-14}$ \cite{Gong:2021jgg}  & - & $1.3\times10^{-11}$ \cite{LIGOScientific:2020tif} & $2.4\times10^{-16}$ \cite{Gong:2021jgg} \\
GWTC-3  &    - &     -  & - & - & $1.8\times10^{-11}$ \cite{LIGOScientific:2021sio} &- \\  
This work  &    $4.0\times10^{-22}$  &     $8.0\times10^{-42}$   & $1.2\times10^{-14}$ & $1.4\times10^{-21}$ & $1.2\times10^{-11}$ & $3.4\times10^{-16}$  \\   
\bottomrule
\end{tabular*}
\end{table*}

Table~\ref{tab} summarizes the results from the Bayesian analysis of the parity- and Lorentz-violating waveforms with GW events in GWTC-3. It shows $90\%$-credible upper bounds on $M_{PV}$ for $\beta_{\nu}=-1$ (for velocity birefringence) and lower bounds on $M_{PV}$ and $M_{LV}$ for other cases. For example, for $\beta_{\nu=1}$, the bound on $M_{PV}$ is $4.0\times 10^{-22}\rm GeV$, which improves the previous result by a factor of $4.0$. For $\beta_{\mu}=3$, the bound on $M_{PV}$ is $1.2\time10^{-14} \rm GeV$, with a $1.2$-fold improvement. For $\beta_{\bar{\mu}}=4$, the bound on $M_{LV}$ is $3.4\times 10^{16}\rm GeV$, showing a $1.4$-fold improvement compared to previous studies. The bounds on $M_{PV}$ for $\beta_{\mu}=-1$ and $M_{LV}$ for $\beta_{\bar\mu}=2$ are compatible with those obtained in previous works from different waveform templates and methods. Additionally, this study obtained the first bound on $M_{LV}$ for $\beta_{\bar\nu}=2$, representing the constraint on the Lorentz-violating damping effect in GWs.

However, in certain tests, a few GW events favored non-GR values for parity-/Lorentz-violating parameters, which are inconsistent with GR. These results might be due to limitations of existing waveform approximants, such as systematic errors during the merger phase of the waveform, or the neglect of physical effects like eccentricity in current waveform approximants. Therefore, these events were excluded from the analysis. In general, these constraints represent significant progress in understanding the parity and Lorentz symmetries in the context of GWs and provide important references for further studies in this field.

\subsection{Constraining parity and Lorentz violations with future ground and space-based GW detectors}

The future ground- and space-based GW detectors offer unprecedented opportunities to test GR with greater precision. To further explore the great potential, we recently investigate the capability of future ground-based GW detectors, the Einstein Telescope and the Cosmic Explorer, and space-based GW detectors, LISA, Taiji, and TianQin, for constraining parity and Lorentz violations in gravity \cite{Zhang:2025kcw, Zhang:2024rel}. The research continues to utilize the previously established parametric framework characterizing possible derivations of GW propagation and modified waveform with parity- and Lorentz-violating effects \cite{Zhu:2023rrx, Zhao:2019xmm}. Note that constraining parity and Lorentz violations for several specific models from future detectors have been also carried in refs.~\cite{Wang:2020cub, Lin:2024pkr, Califano:2023aji, Hu:2020rub, Zhang:2024rel}.

Specifically, typical GW signals from compact binary systems are injected into the above-mentioned detectors. Subsequently, the Bayesian inference method is applied to analyze the modified waveforms that incorporate the effects of parity and Lorentz violations, thereby obtaining crucial constraint information on the energy scales ( $M_{\rm PV}$  and $M_{\rm LV}$) closely related to parity and Lorentz violations. Through this research approach, it is possible to conduct a comparative assessment of the performance of different detector networks and gain in-depth insights into the sensitivity of these detectors when exploring fundamental physical phenomena beyond the framework of GR.

In addition, a comprehensive and in-depth analysis of the data obtained during the research process helps to more precisely understand the roles played by parity and Lorentz symmetries in gravitational interactions. This not only deepens our understanding of the nature of gravity but also significantly contributes to the sustainable development of GW astronomy, driving continuous new research progress in this field.

\subsubsection{Constraints from the third-generation ground-based GW detectors}

We injected several typical GW signals such as GW150914-like, GW170817-like, and GW190521-like signals are injected into ET and CE with the help of the BILBY software package. For different types of binary systems such as binary black holes and binary neutron stars, corresponding waveform templates are adopted (for example, IMRPhenomPv2 is used for binary black hole events, and IMRPhenomPv2 NRTidal is used for binary neutron star events).

\begin{figure*}
\centering
\includegraphics[width=4.2cm]{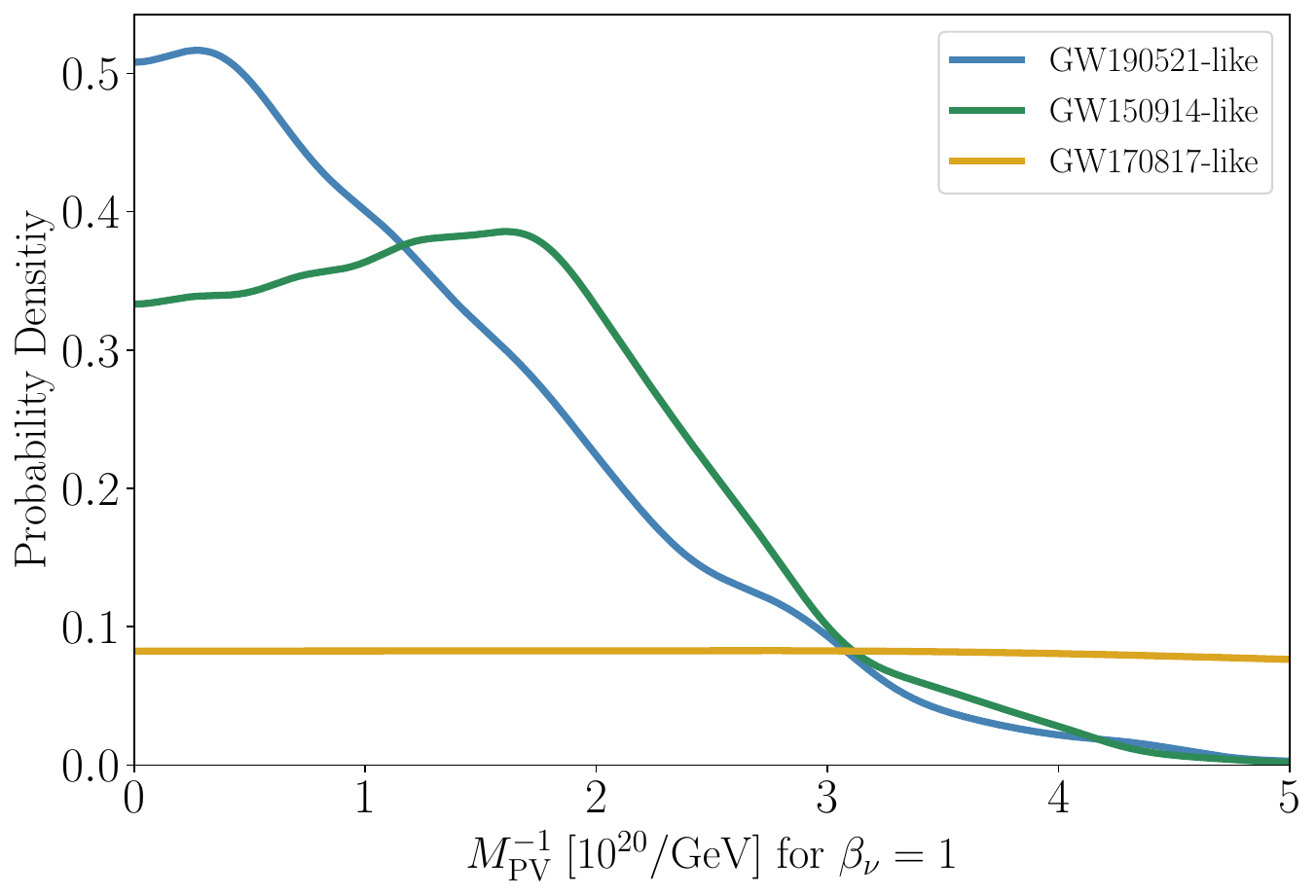}
\includegraphics[width=4.2cm]{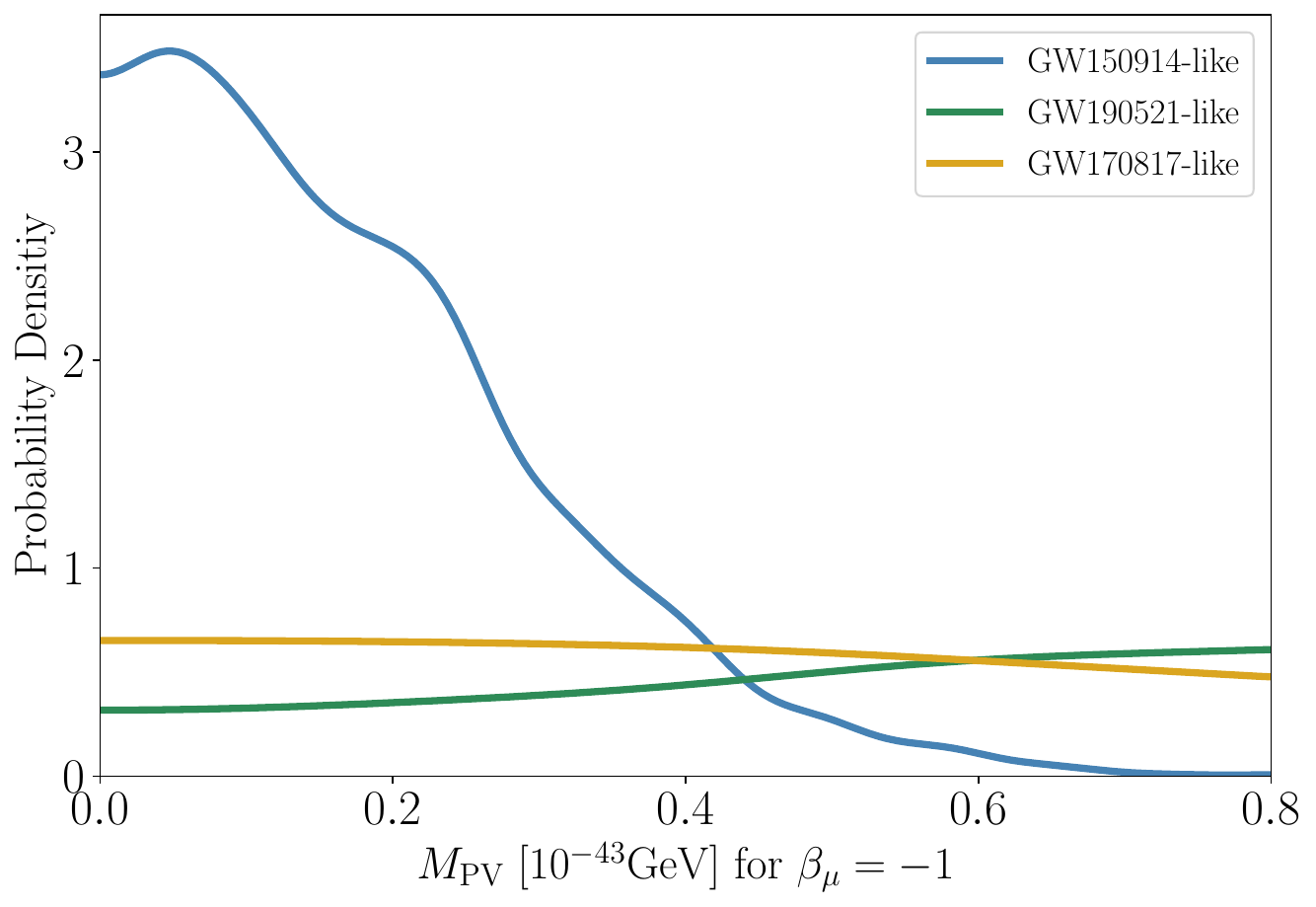}
\includegraphics[width=4.2cm]{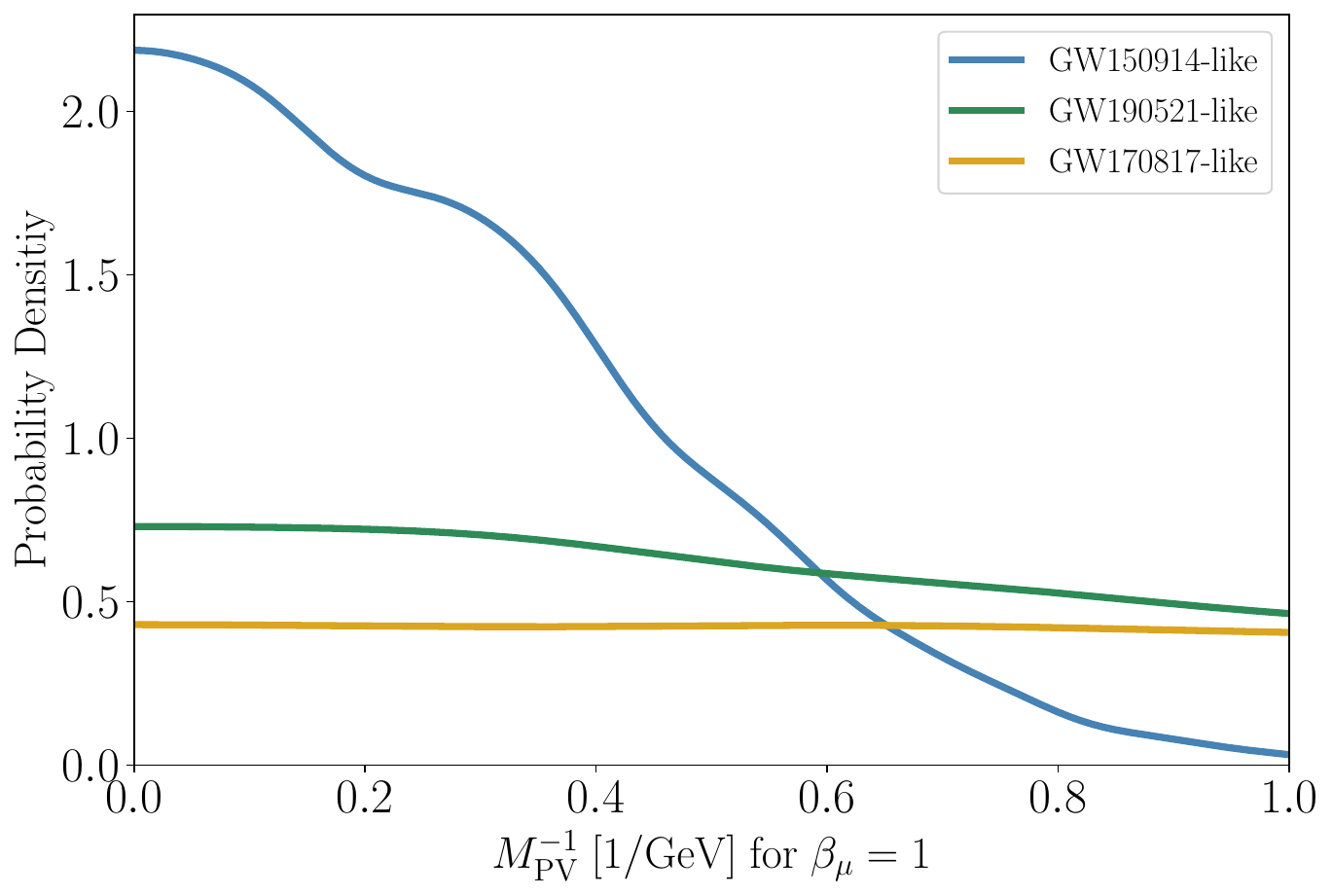}
\includegraphics[width=4.2cm]{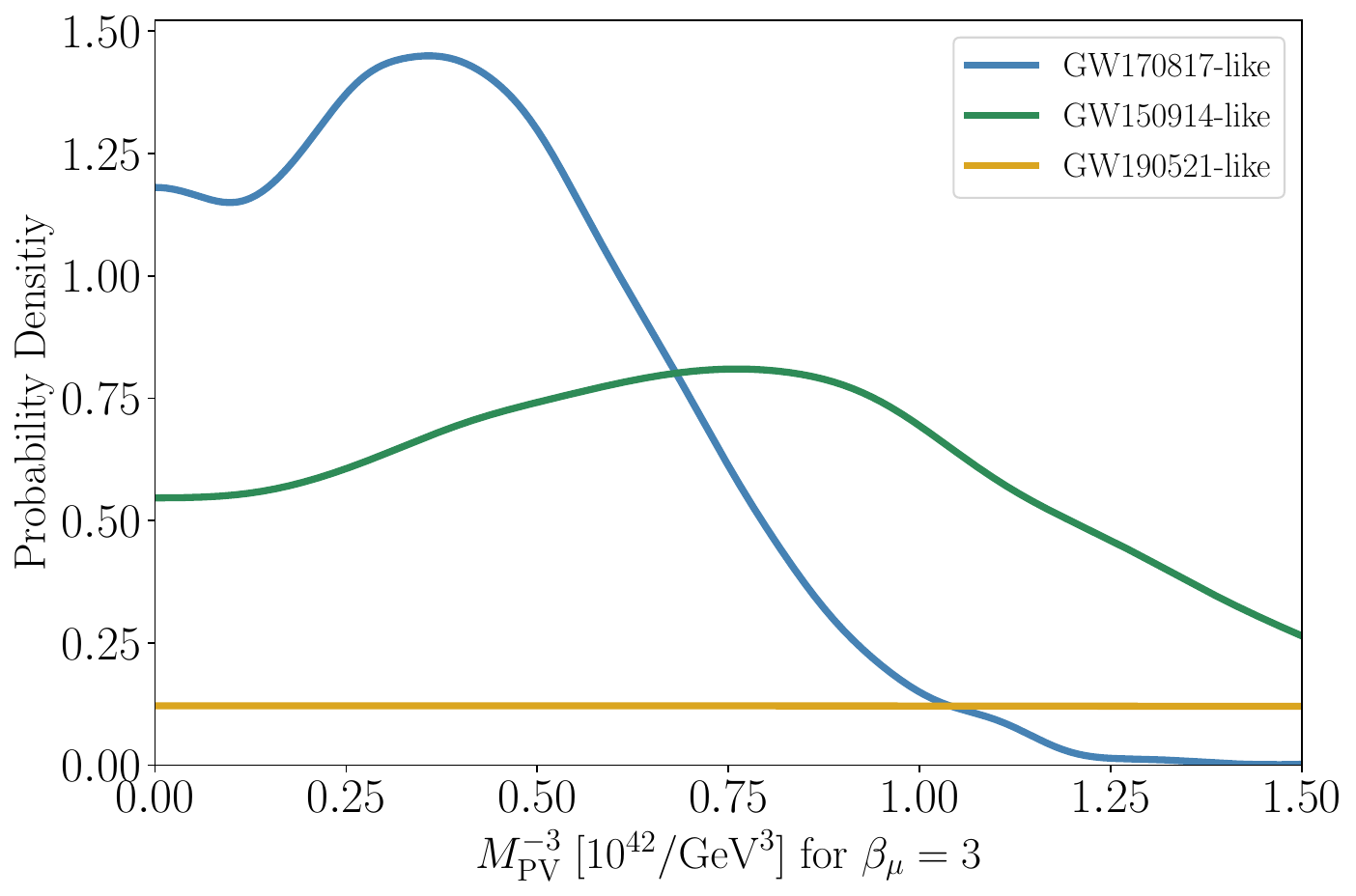}
\includegraphics[width=4.2cm]{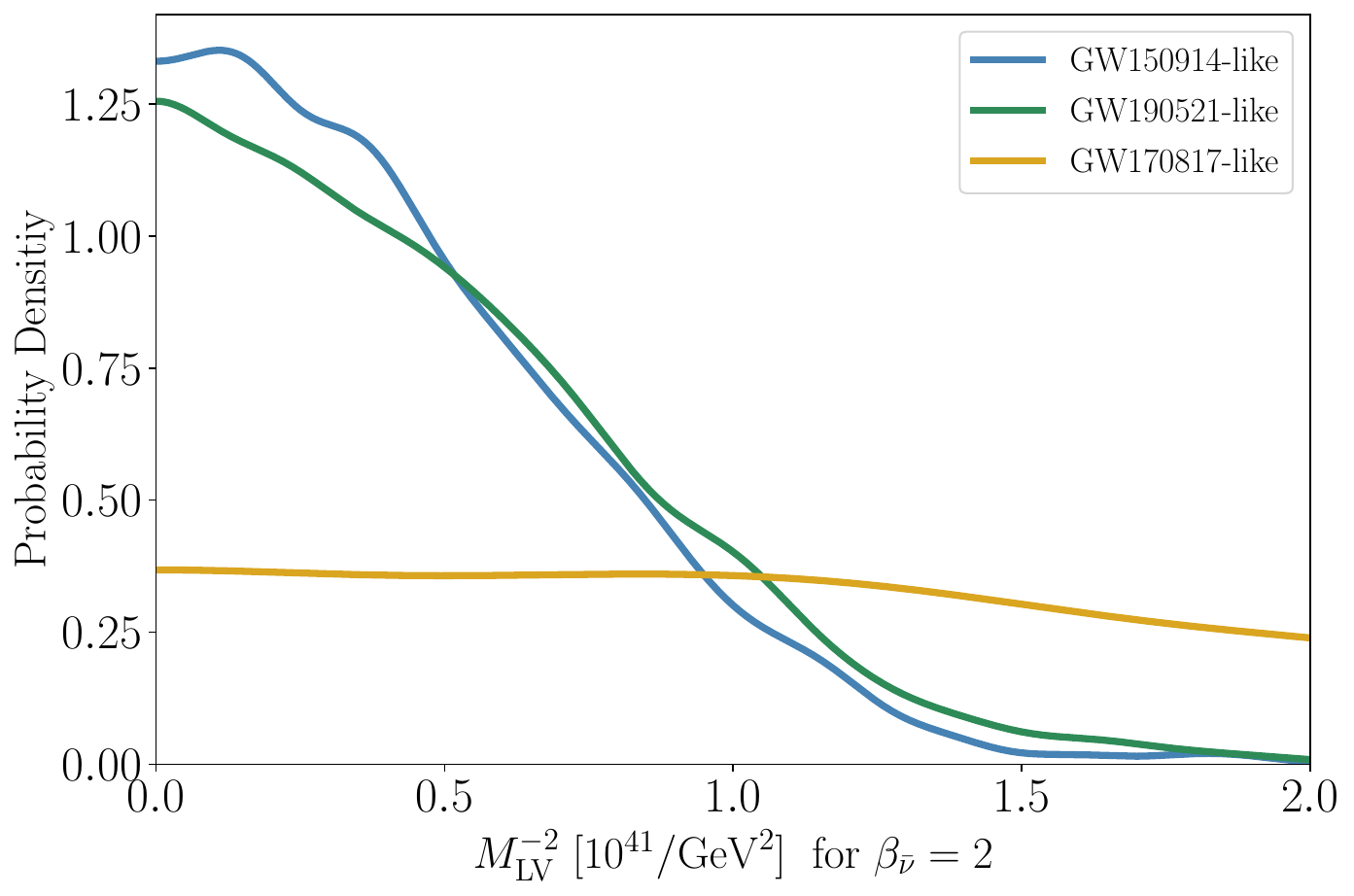}
\includegraphics[width=4.2cm]{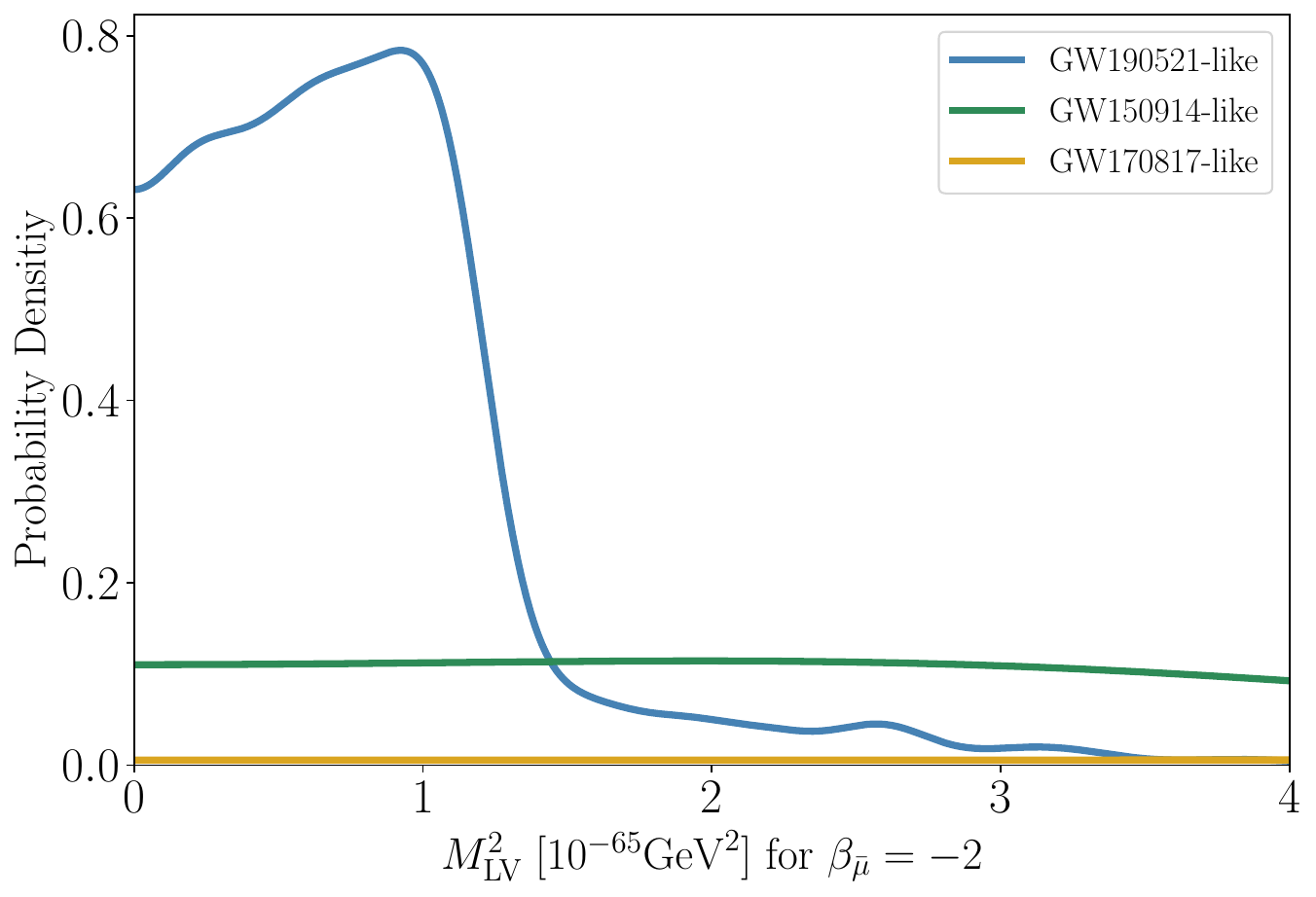}
\includegraphics[width=4.2cm]{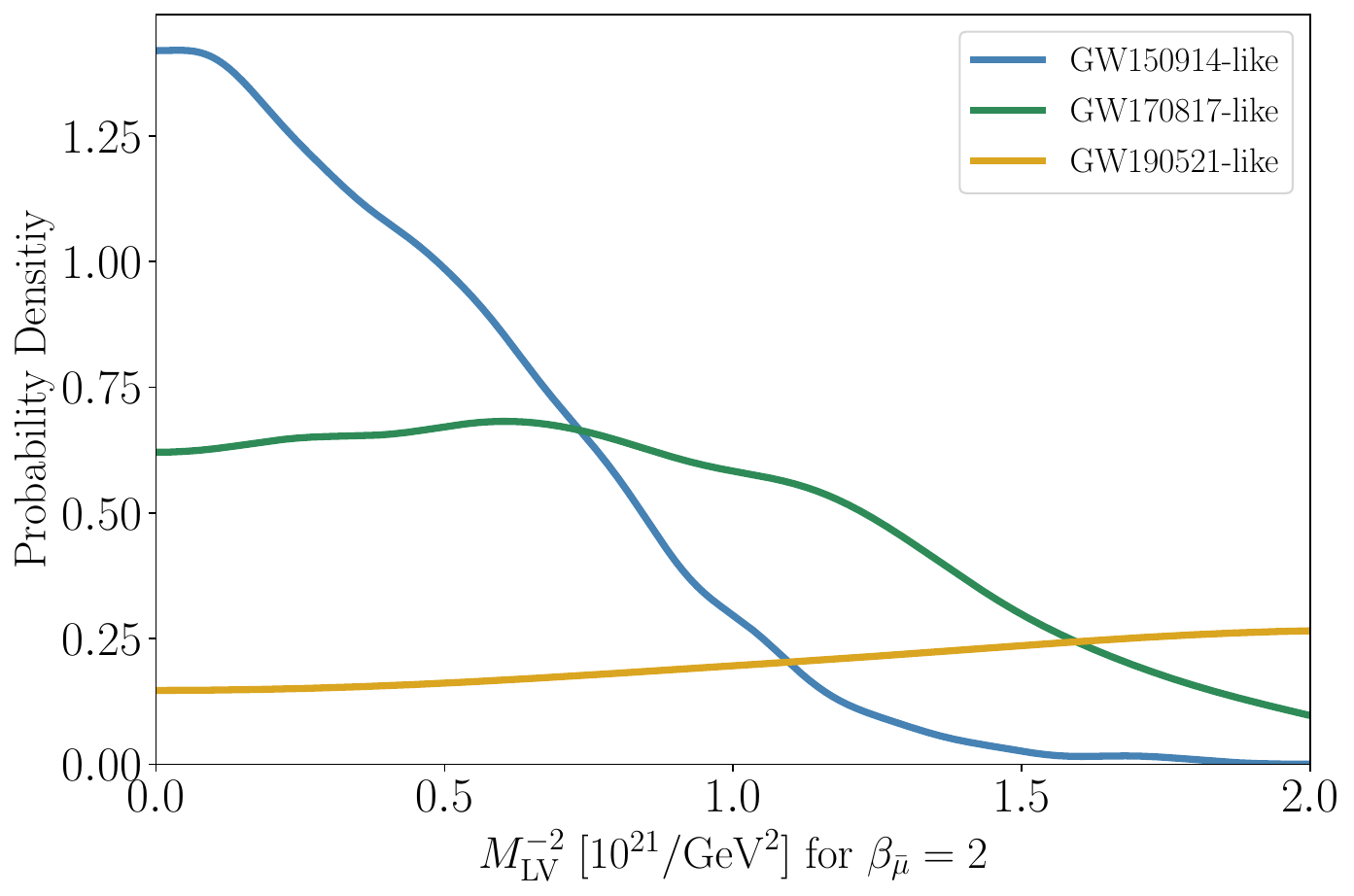}
\includegraphics[width=4.2cm]{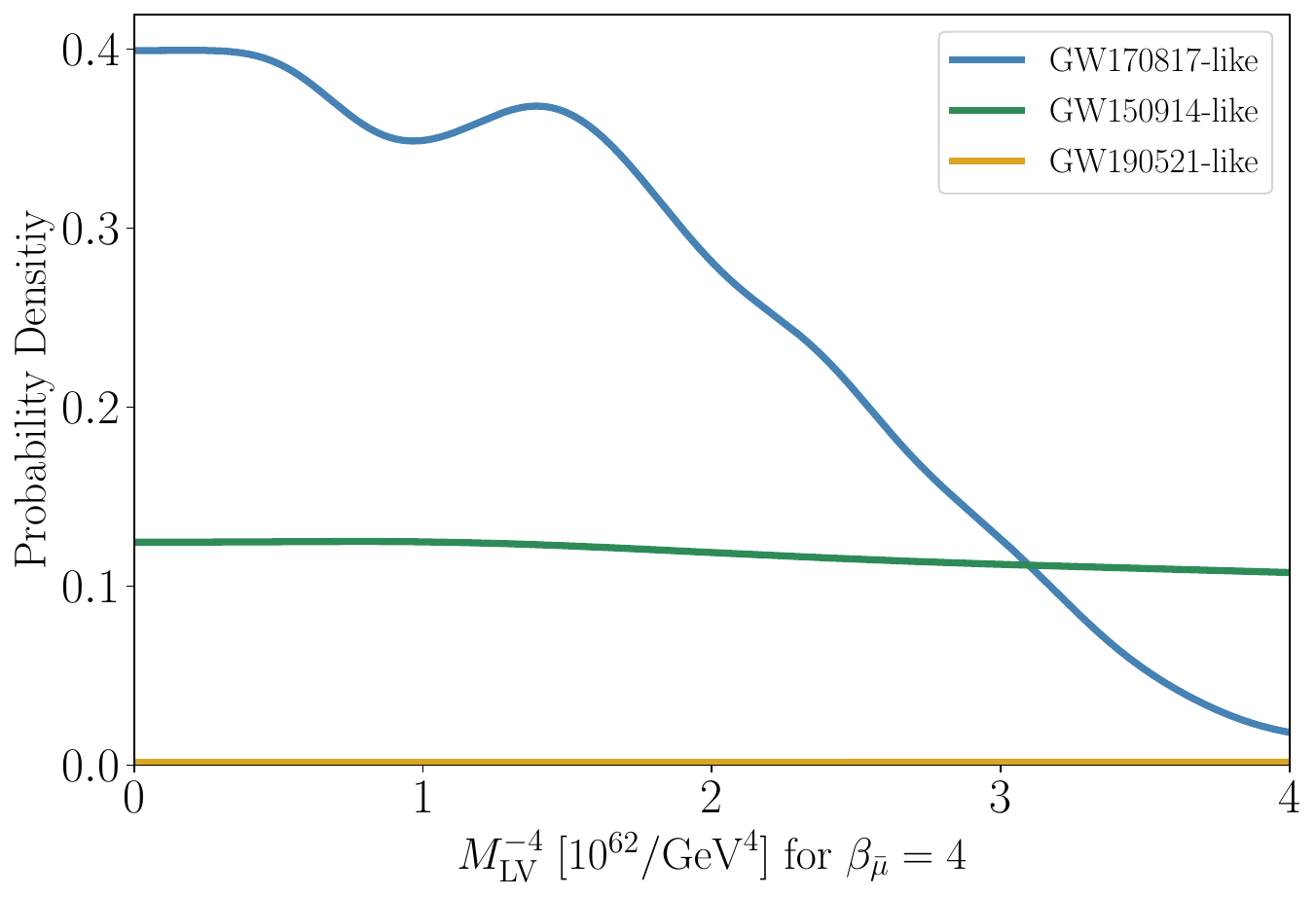}
\caption{The posterior distributions for $M_{\rm PV}^{-\beta_{\nu}}$ with $\beta_{\nu}=1$, $M_{\rm PV}^{-\beta_{\mu}}$ with $\beta_{\mu}=-1, 1, 3$, $M_{\rm LV}^{-\beta_{\bar \nu}}$ with $\beta_{\bar{\nu}}=2$, and $M_{\rm LV}^{-\beta_{\bar \mu}}$ with $\beta_{\bar{\mu}}=-2, 2, 4$ of three injected GW events GW150914-like, GW170817-like, and GW190521-like given by the ground-based GW detectors ET and CE.  Copied from Ref.~\cite{Zhang:2025kcw} with permission.
\label{fig:mpv and mlv from CE and ET}}
\end{figure*}

After processing the simulated signals using the Bayesian analysis framework on the modified waveform with parity- and Lorentz-violating effects, we find that joint observations from the ET and CE exhibit significantly enhanced capability to constrain the energy scales associated with parity and Lorentz violations ($M_{\rm PV}$ and $M_{\rm LV}$) compared to the current LIGO-Virgo-KAGRA detectors. The strength of these constraints varies considerably for different GW events, depending on the source mass, frequency, and $\beta$ parameter values. These results are presented in Fig.~\ref{fig:mpv and mlv from CE and ET}, and Table~\ref{tab:results of ground detectors}. We also present the previous results \cite{Zhu:2023rrx, Wang:2025fhw, Wang:2021gqm} from analysis with the GW data of LVK for comparison, which are also presented in Table.~\ref{tab:results of ground detectors}. Specifically, for lower-mass events (e.g., GW170817-like), the tightest constraints on $M_{\rm PV}$ and $M_{\rm LV}$ are achieved when $\beta_{\mu}$ and $\beta_{\bar{\mu}}$ are large and positive. In contrast, for higher-mass events (e.g., GW190521-like and GW150914-like), stronger constraints arise when $\beta_{\mu}$ and $\beta_{\bar{\mu}}$ are smaller.

\begin{table*}[t]
\footnotesize
%\centering
\caption{Results from the Bayesian analysis of the parity- and Lorentz-violating waveforms on GW events in LVK and injected signals to be detected by ET+CE. The upper half of the table shows the constraints from a single GW event and the joint constraint from more than 80 GW events in the set of GWTC-3. Copied from Ref.~\cite{Zhang:2025kcw} with permission.}
   \vspace{.2cm}
    \label{tab:results of ground detectors}
\tabcolsep 14pt 
\begin{tabular*}{\textwidth}{ccccccccc}
\toprule
     & \multicolumn{4}{c}{$M_{\rm PV}$ [GeV]} &\multicolumn{4}{c}{$M_{\rm LV}$ [GeV]} \\
\cline{2-5}  \cline{6-9}
  & $\beta_\nu=1$ & $\beta_\mu=-1$ & $\beta_\mu=1$ & $\beta_\mu=3$ & $\beta_{\bar \nu}=2$ & $\beta_{\bar \mu}=-2$ &$\beta_{\bar \mu}=2$ & $\beta_{\bar \mu}=4$    \\
  & $[10^{-23}]$ & $[10^{-43}]$ & $[10^{-3}]$ & $[10^{-15}]$ & $[10^{-22}]$ & $[10^{-33}]$ &$[10^{-12}]$ & $[10^{-17}]$    \\
 \hline 
GW170817 LVK    & $3.1$   & $13000$    & $8.4$ 
& $7.5$   &  ---     &  $1100$  & $2.1$     & $21.1$  \\
GW150914 LVK    & $16$    & $750$    & $7.1$  
&$1.9$  & $6.4$    &  $150$  & $3.1$     & $4.9$   \\
GW190521 LVK    & ---    & $480$   & --- 
& ---   & ---    & $68$  & ---  & ---  \\
GWTC-3 combined & $40$    & $80$      & $50$   
& $12$    & $14$     & $83.1$ 
& $12$    & $34$ \\
\hline
GW170817 ET+CE & $65$   & $2.0$   & $390$  & $11$    &  $17$   & $51.5$  & $25.2$  & $24$ \\
GW150914 ET+CE & $364$  & $0.36$    & $1720$  & $8.9$    &  $32.8$  & $11.8$ & $33.4$ & $17.7$ \\
GW190521 ET+CE & $375$  & $1.82$   & $547$  & $4.5$   &  $31.1$  & $3.7$  & $15.2 $ & $8.3$ \\  
\bottomrule
\end{tabular*}
\end{table*}

For comparison, Table~\ref{tab:results of ground detectors} lists previous analysis results for events GW150914, GW170817, and GW190521 from LVK GW data. The table also presents combined constraints on $M_{\rm PV}$ and $M_{\rm LV}$ derived from about 90 events in GWTC-3 (see Ref.~\cite{Zhu:2023rrx}). For individual events, constraints on $M_{\rm PV}$ with $\beta_{\nu}=1$ and $\beta_{\mu}=-1, 1, 3$, and on $M_{\rm LV}$ with $\beta_{\bar \nu}=2$ and $\beta_{\bar \mu}=-2, 2, 4$, improve upon those in Ref.~\cite{Zhu:2023rrx} by roughly 1 to 3 orders of magnitude.

\subsubsection{Constraints from space-based GW detectors}

\begin{figure*}
\centering
\includegraphics[width=4.2cm]{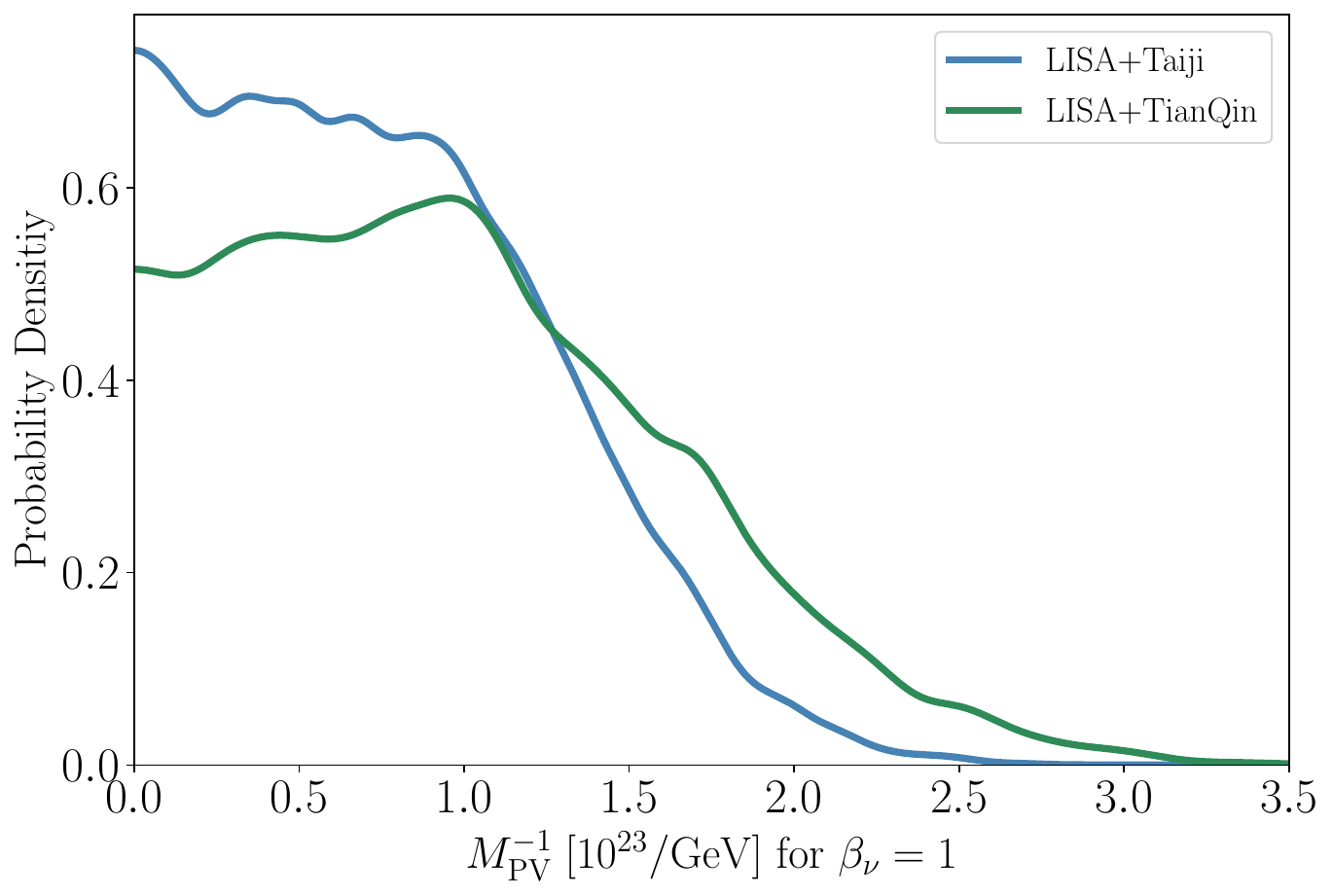}
\includegraphics[width=4.2cm]{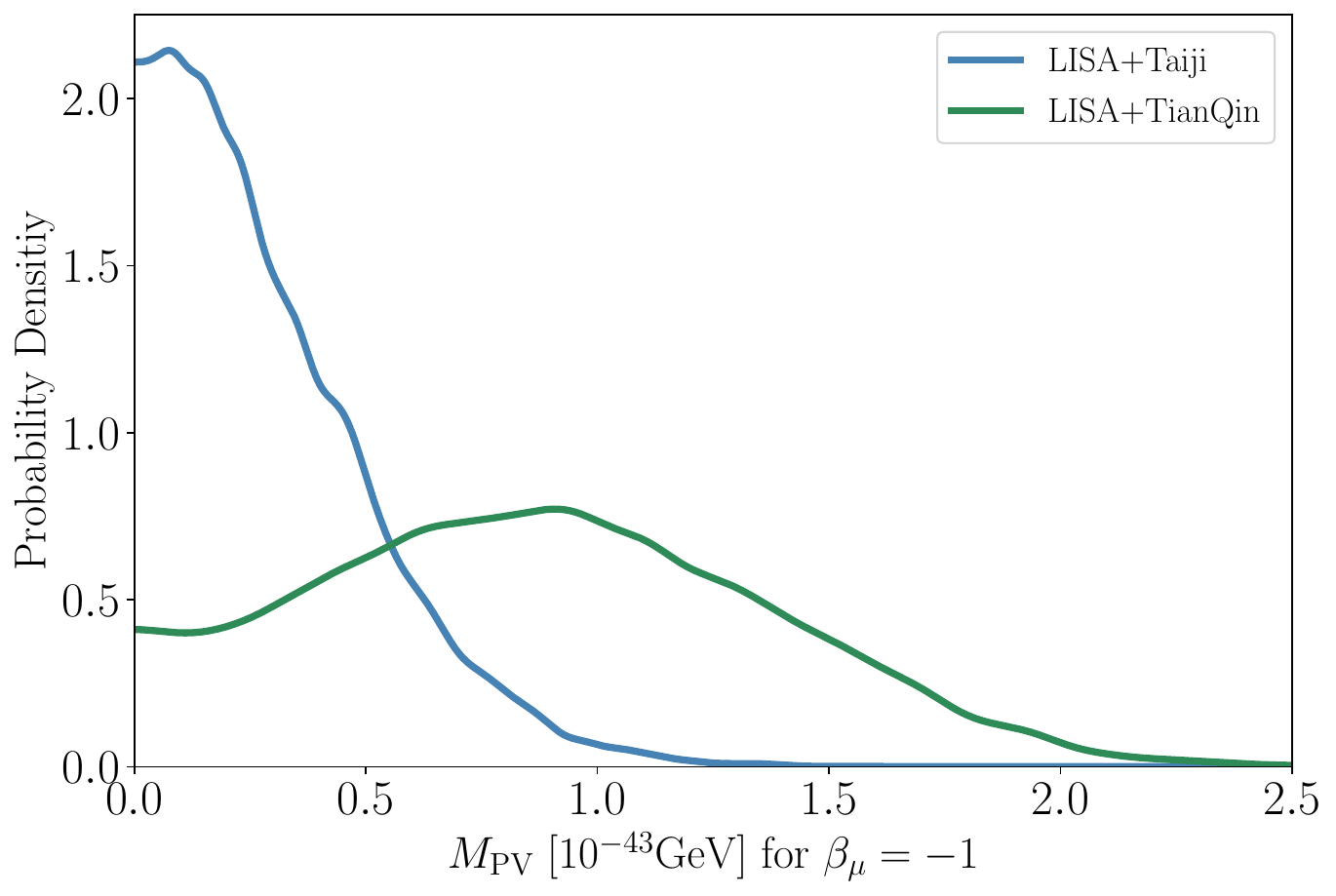}
\includegraphics[width=4.2cm]{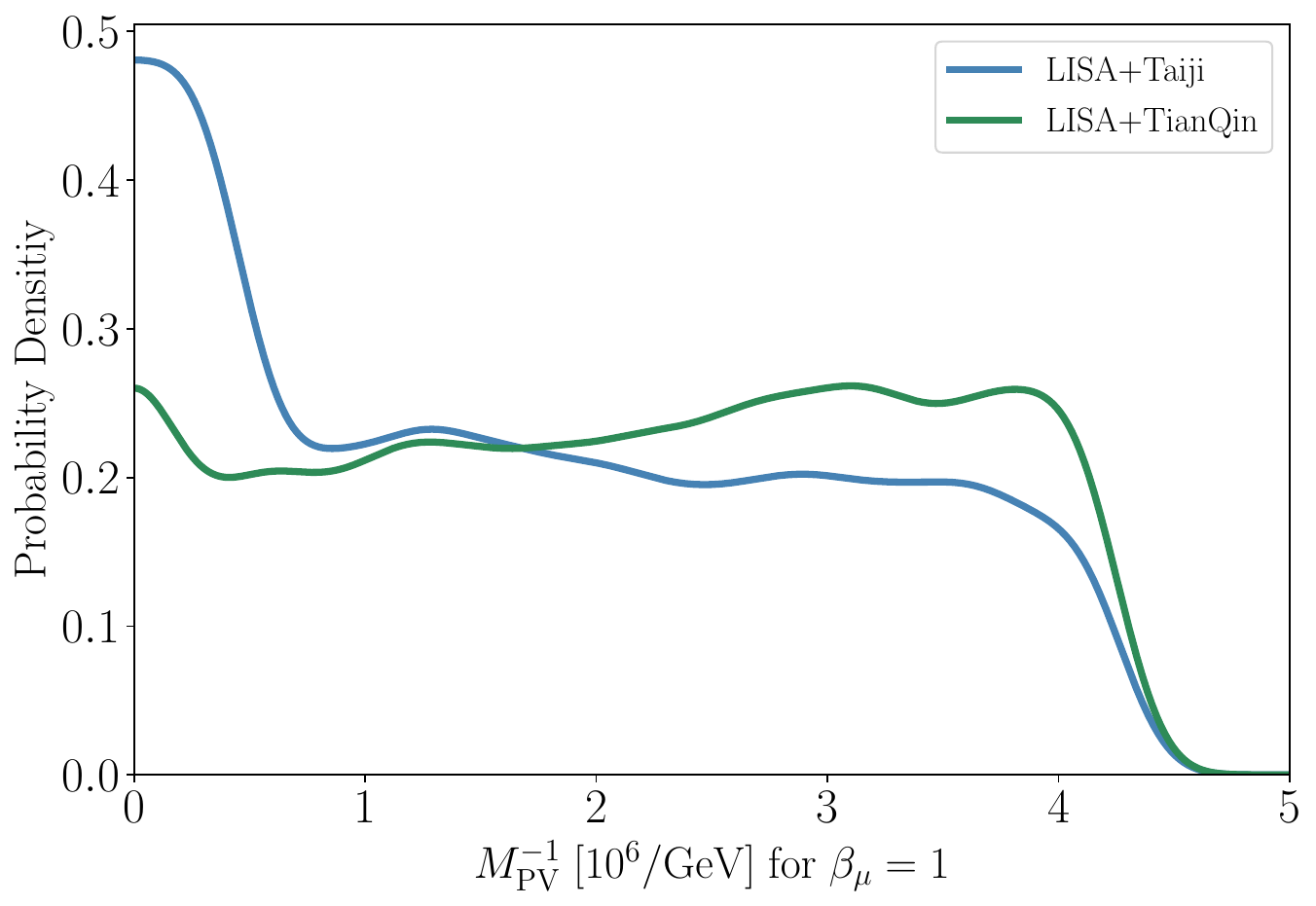}
\includegraphics[width=4.2cm]{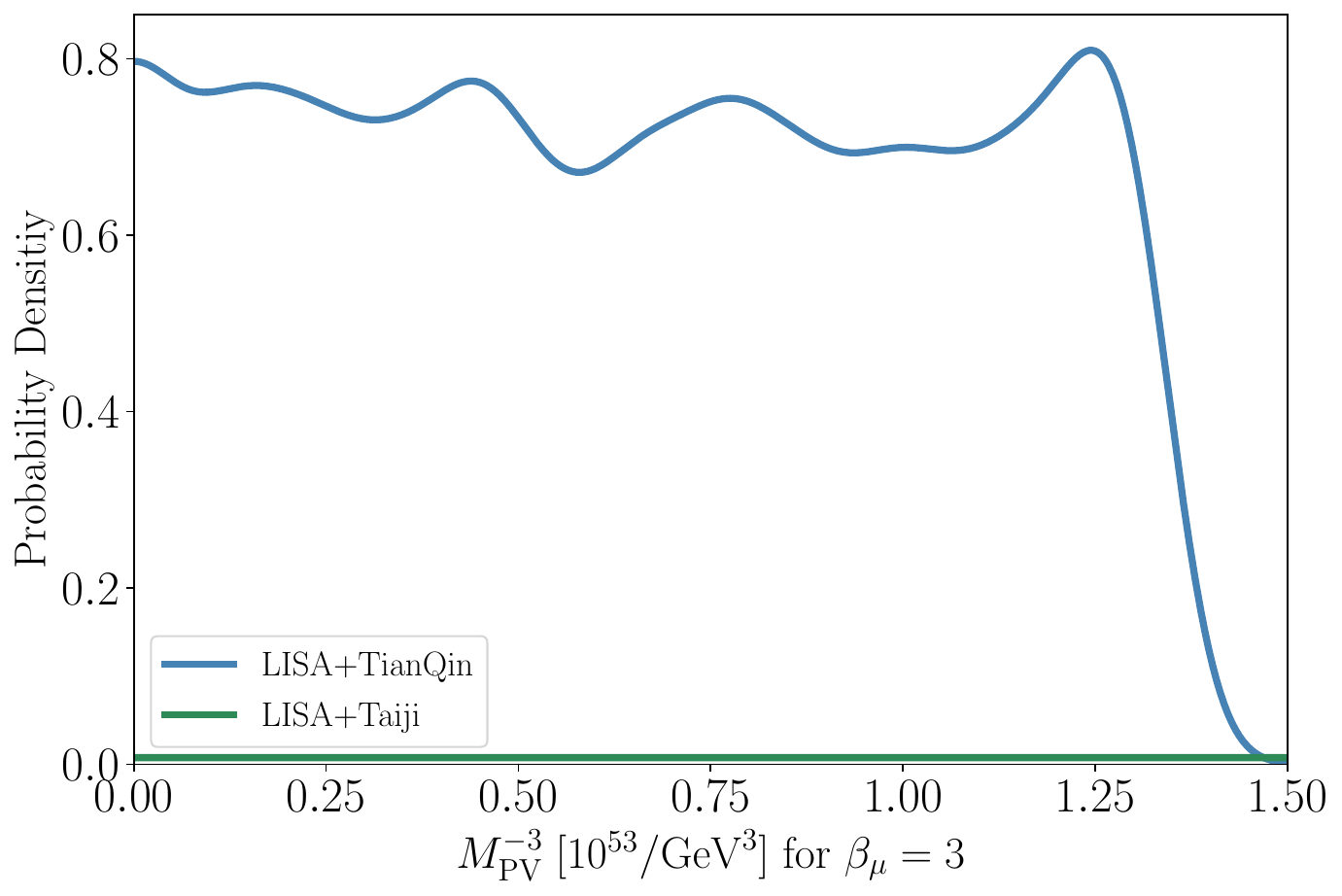}
\includegraphics[width=4.2cm]{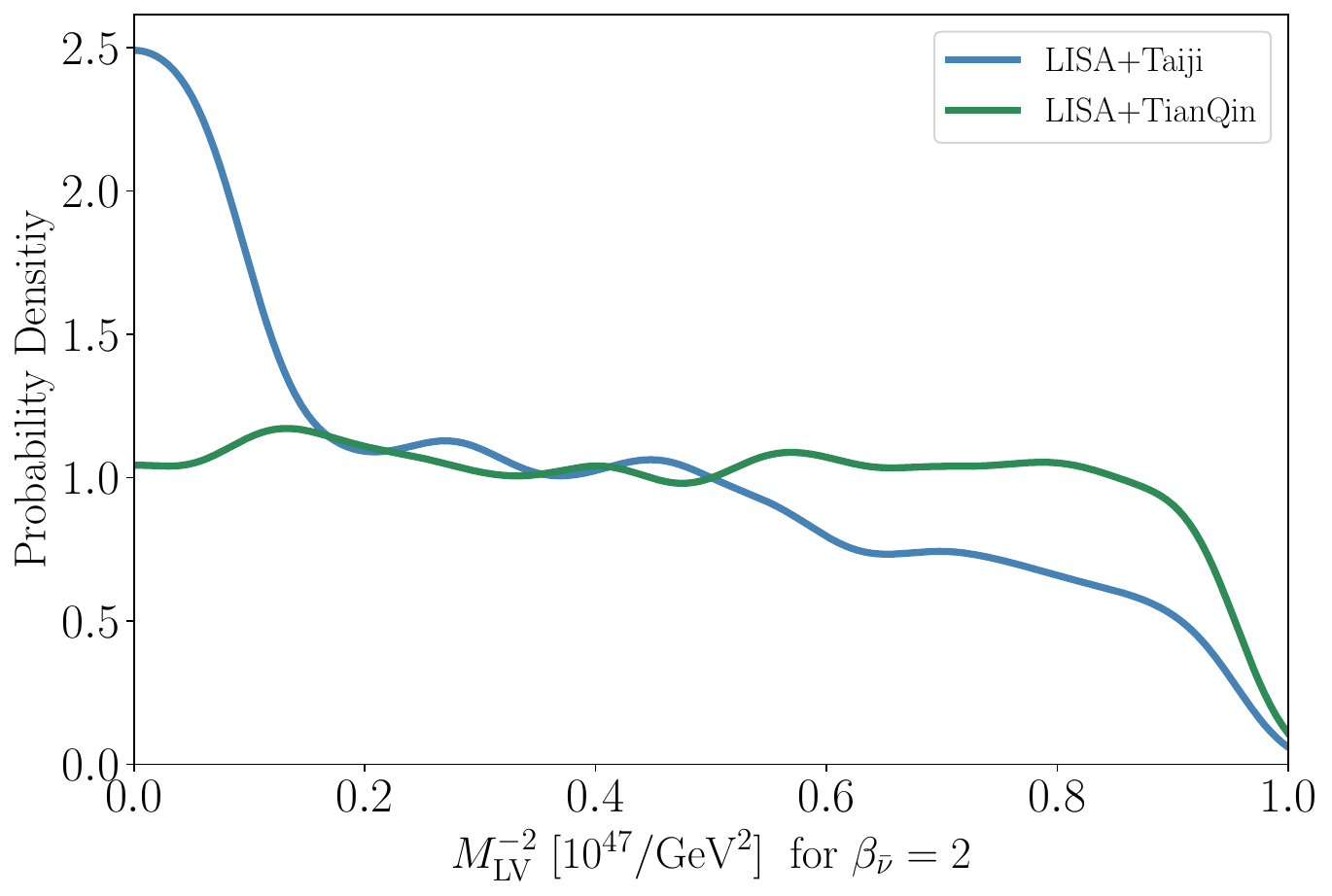}
\includegraphics[width=4.2cm]{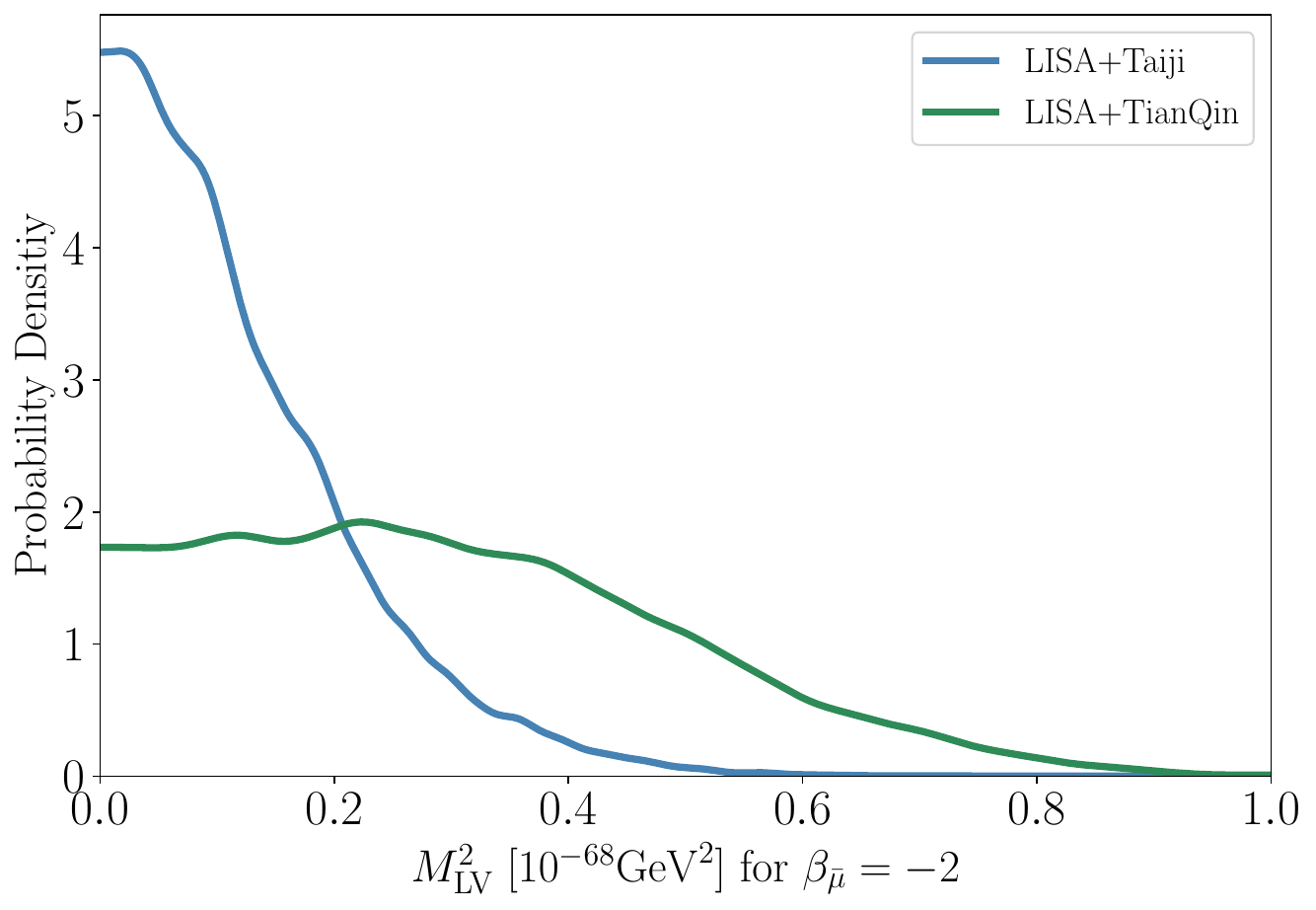}
\includegraphics[width=4.2cm]{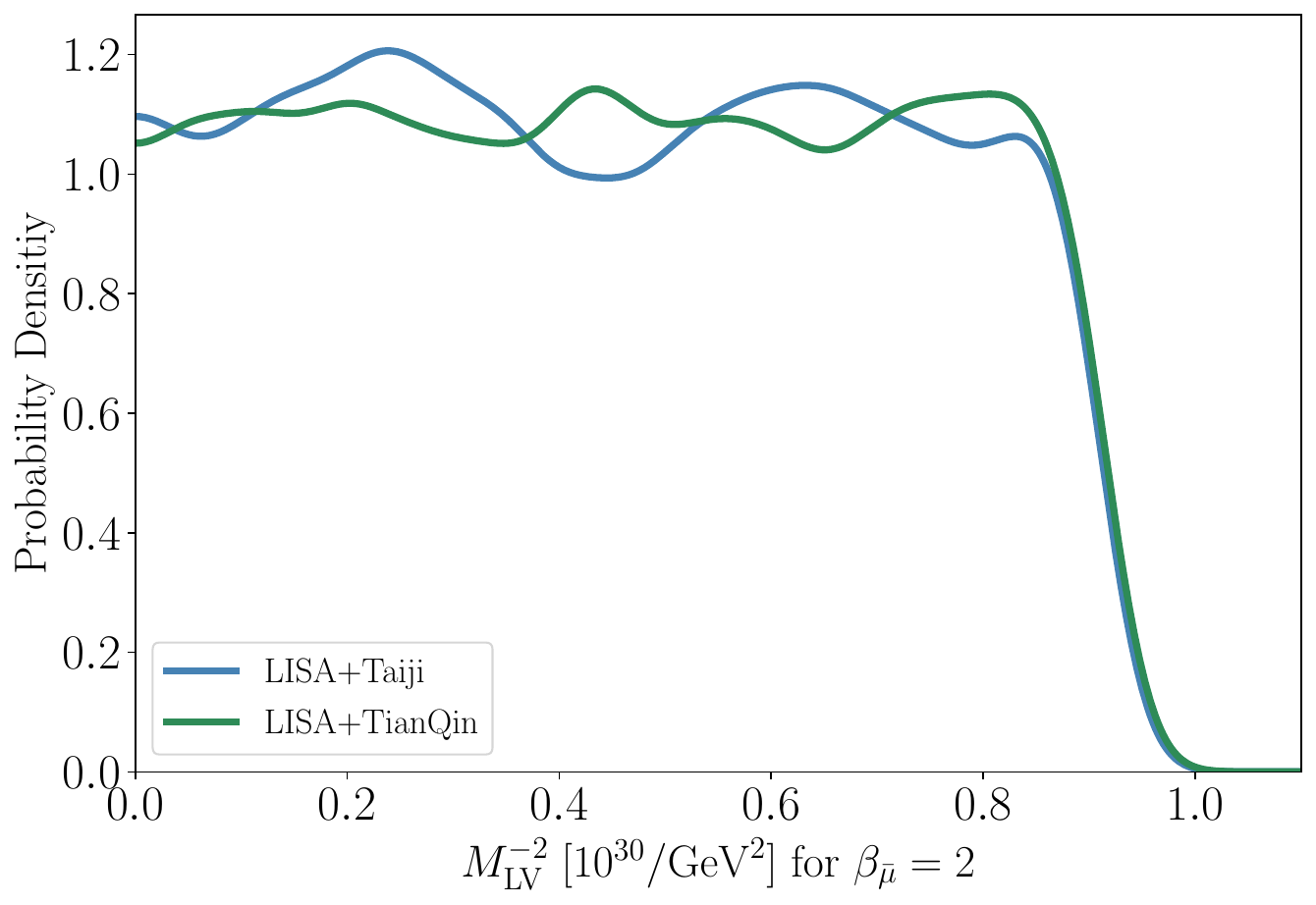}
\includegraphics[width=4.2cm]{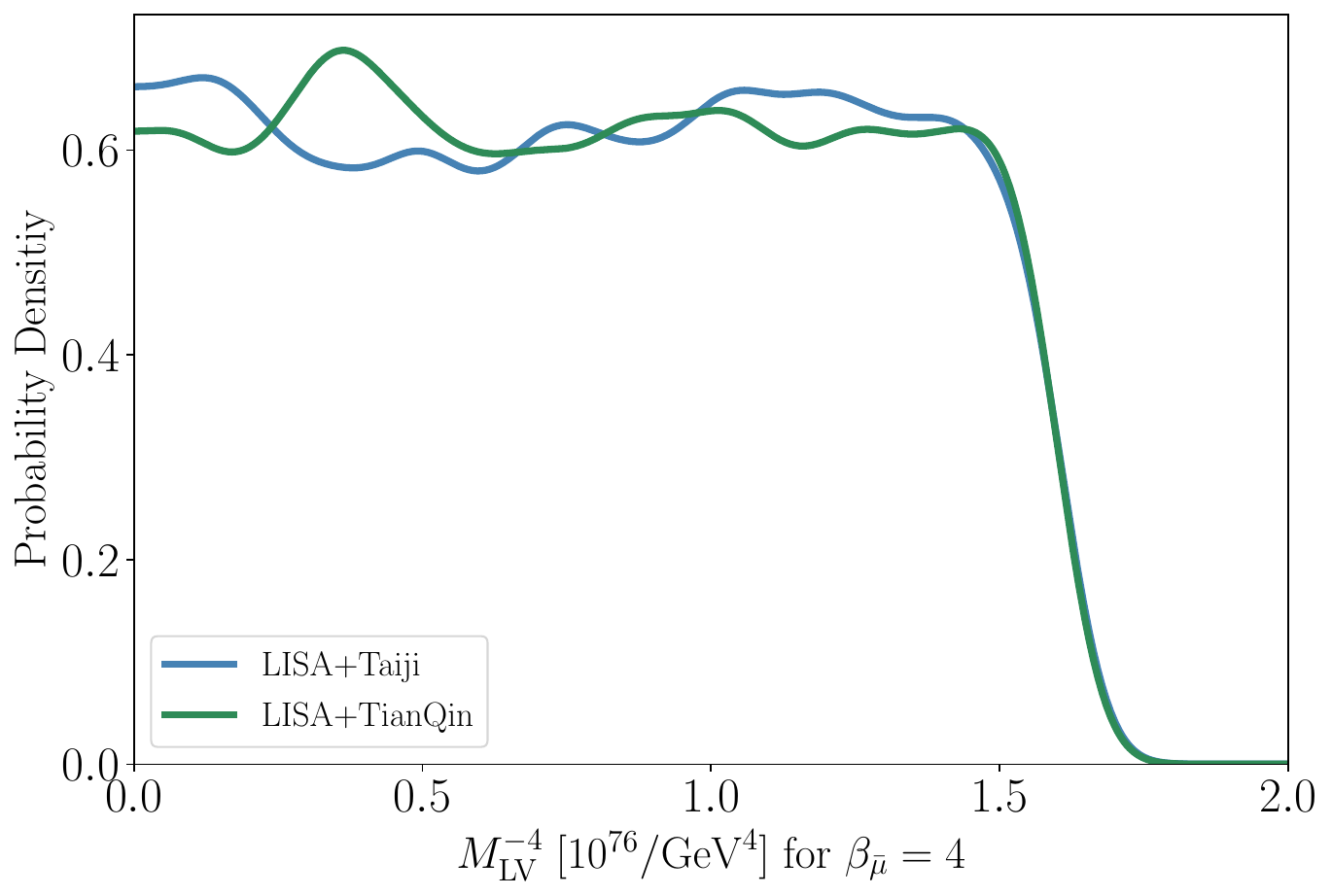}
\caption{The posterior distributions for $M_{\rm PV}^{-\beta_{\nu}}$ with $\beta_{\nu}=1$, $M_{\rm PV}^{-\beta_{\mu}}$ with $\beta_{\mu}=-1, 1, 3$, $M_{\rm LV}^{-\beta_{\bar \nu}}$ with $\beta_{\bar{\nu}}=2$, and $M_{\rm LV}^{-\beta_{\bar \mu}}$ with $\beta_{\bar{\mu}}=-2, 2, 4$ of the three simulated events for the space-based GW detectors. Copied from Ref.~\cite{Zhang:2025kcw} with permission.
\label{fig:mpv and mlv from LTT}}
\end{figure*}

We then evaluate the capabilities of two detector networks, LISA + Taiji and LISA + TianQin, in constraining parity- and Lorentz-violating effects. The BILBY software package is used to simulate the GW signals generated by the merger of supermassive binary black hole systems. The simulated signals include Event 1 ($M=2\times10^4M_{\odot}$, $z=1$), Event 2 ($M=2\times10^5M_{\odot}$, $z=5$), and Event 3 ($M=2\times10^6M_{\odot}$, $z=10$). Results are presented in Fig.~\ref{fig:mpv and mlv from LTT}, and Table~\ref{tab:results of space detectors}. 

The results show that for the cases where $\beta_{\nu}$, $\beta_{\mu}$, $\beta_{\bar{\nu}}$ and $\beta_{\bar{\mu}}$ are positive, the constraint capabilities of space-based detectors on $M_{\rm PV}$ and $M_{\rm LV}$ are 3-4 orders of magnitude weaker than those of ground-based detectors. This is mainly because the amplitude and phase corrections in these cases are more relevant to high-frequency GWs, while ground-based detectors have obvious advantages in the high-frequency band. However, in some specific cases, space-based detectors show irreplaceable unique values. For example, when $\beta_{\mu}=-1$, the constraints from space-based detectors on $M_{\rm PV}$ is better than that of LVK and comparable to that of ET + CE; when $\beta_{\bar{\mu}}=-2$, the constraint from space-based detectors on $M_{\rm LV}$ is approximately three orders of magnitude tighter than that of ground-based detectors, and this constraint can also be transformed into an upper limit on the mass of the graviton, that is, $m_g \lesssim 10^{-35}\;{\rm GeV}$, which is highly consistent with previous analysis.

%\begin{table*}
\begin{table*}[t]
\footnotesize
\caption{Results from the Bayesian analysis of the parity- and Lorentz-violating waveforms on injected GW signals to be detected by LISA+Taiji and LISA+TianQin. The upper half of the table shows the results of the constraint from a single GW event by a combination of LISA and Taiji. The lower half of the table shows the constraints from the combination of LISA and TianQin. The table shows 90\%-credible upper bounds on $M_{\rm PV}$ for $\beta_{\mu}=-1$ (for velocity birefringence) and $M_{\rm LV}$ for $\beta_{\bar{\mu}}=-2$. In the other cases, the results show the lower bounds on $M_{\rm PV}$ and $M_{\rm LV}$.  Copied from Ref.~\cite{Zhang:2025kcw} with permission.}
    \vspace{.2cm}
\tabcolsep 13.5pt 
\begin{tabular*}{\textwidth}{ccccccccc}
\toprule
    \label{tab:results of space detectors}
     & \multicolumn{4}{c}{$M_{\rm PV}$ [GeV]} &\multicolumn{4}{c}{$M_{\rm LV}$ [GeV]} \\
\cline{2-5}  \cline{6-9}
  & $\beta_\nu=1$ & $\beta_\mu=-1$ & $\beta_\mu=1$ & $\beta_\mu=3$ & $\beta_{\bar \nu}=2$ & $\beta_{\bar \mu}=-2$ &$\beta_{\bar \mu}=2$ & $\beta_{\bar \mu}=4$    \\
  & $[10^{-25}]$ & $[10^{-44}]$ & $[10^{-8}]$ & $[10^{-19}]$ & $[10^{-25}]$ & $[10^{-35}]$ &$[10^{-16}]$ & $[10^{-20}]$    \\
\hline
event1 LISA+Taiji    & 4.2  & 26    & 4.4   & 1.4   &  8.8   & 8.9  & 2.2 & 2.9   \\
event2 LISA+Taiji    & 37   & 23    & 22    & 3.4   &  20    & 9.0  & 6.8 & 5.3   \\
event3 LISA+Taiji    & 67   & 6.2   & 27    & 4.4   &  36    & 5.1  & 11  & 9.1   \\
\hline
event1 LISA+TianQin  & 7.3  & 34    & 4.9   & 4.3   & 8.7    & 12   & 2.2  & 2.9   \\
event2 LISA+TianQin  & 40   & 21    & 17    & 12    & 24     & 11   & 6.8 & 5.4   \\
event3 LISA+TianQin  & 53   & 16    & 26    & 20    & 34     & 7.6  & 11 & 9.1   \\
\bottomrule
\end{tabular*}
\end{table*}

\subsection{Polarization modes of GWs}\label{section41}

Similar to electromagnetic waves, which have two linear polarization modes, GWs  also exhibit different polarization modes. The detection of polarization modes of GWs provides a powerful tool for testing various modified gravity theories. This is because, even at a qualitative level, different theories make distinct predictions regarding the number of independent GW polarization modes and the dispersion relations of each mode.

The polarization modes of GWs are defined by the relative motion modes of test particles. As long as the matter fields in the action couple only to the metric, the relative motion of test particles satisfies the geodesic deviation equation \cite{2018grav.book.....M}:
\begin{eqnarray}
	\label{equation of geodesic deviation}
	\frac{d^{2}\eta_{i}}{dt^{2}}=-R_{i0j0}(t)\eta^{j}.
\end{eqnarray}
Here, $\eta_{i}$ represents the relative displacement between two neighboring test particles, and $R_{i0j0}$ denotes the relevant components of the linearized Riemann curvature tensor, where the indices $(i,j)$ range over the three spatial indices. Noting that $R_{i0j0}$ is symmetric with respect to the indices $(i,j)$, we can conclude that, there exist at most six independent GW polarization modes \cite{Eardley:1973zuo,Dong:2021jtd}:
\begin{eqnarray}
	\label{P1-P6}
	R_{i0j0}\coloneqq\begin{pmatrix}
		P_{4}+P_{6} & P_{5} & P_{2}\\
		P_{5}       & -P_{4}+P_{6}  & P_{3}\\
		P_{2}       &  P_{3}   &   P_{1}
	\end{pmatrix}.
\end{eqnarray}

The names of these six polarization modes, along with the relative motion of test particles under each mode.
%, are illustrated in Fig. \ref{fig1}. 
In GR, there are only the $+$ and $\times$ modes propagating at the speed of light, whereas in general modified gravity theories, additional polarization modes typically arise. Currently, GW polarization modes are being probed or planned for detection by ground-based detectors \cite{VIRGO:2014yos,KAGRA:2017tir,Abac:2025saz,Reitze:2019iox}, space-based detectors \cite{Audley:2017drz, Luo:2021qji, Hu:2017mde,Luo:2025ewp}, and pulsar timing arrays \cite{Xu:2023wog, Jenet:2009hk, B.W.Stappers_2006, Manchester_2013,ChandraJoshi:2022etw,Chen:2021wdo}. On the other hand, to effectively use polarization measurements for testing gravity theories, it is also essential to theoretically analyze the polarization properties of GWs in various gravity theories.

%%\begin{figure*}
%	\centering
%%	\setlength{\abovecaptionskip}{-0.8cm}
%	\includegraphics[width=0.8\textwidth]{fig1.pdf}
%	\caption{The six polarization modes of GWs \cite{Eardley:1973zuo}. The GW propagates in the $+z$ direction.} 
%	\label{fig1}
%\end{figure*}

\subsubsection{General analysis of polarization modes}

Introducing higher-order derivative terms and additional fields are two common approaches to modify GR. In Refs. \cite{Dong:2023bgt,Dong:2024zal}, we developed a model-independent framework for analyzing GW polarization modes in a generalized manner. Using this framework, we investigated the polarization modes of GWs in the most general pure metric theory \cite{Dong:2023bgt}, the most general scalar-tensor theory \cite{Dong:2023bgt}, and the most general vector-tensor theory \cite{Dong:2024zal}. The first two theories considered in these works can accommodate arbitrary higher-order derivative terms, while the last theory is restricted to cases that yield second-order field equations.

For the most general pure metric theory and the most general scalar-tensor theory, the polarization modes of GWs depend on the parameter spaces, and in the most general case, all six polarization modes may be present. The dispersion relations of these modes take the form $k^2=m^2$, where the mass $m$ can be either zero or nonzero. For these two theories, the polarization modes of GWs also follow the following universal properties:
\begin{itemize}
	\item The existence of the vector-$x$ and vector-$y$ modes with mass $m$ in the theory implies the presence of the $+$ and $\times$ modes with the same mass $m$. The existence of the $+$ and $\times$ modes with mass $m\neq0$ in the theory implies the presence of the vector-$x$ and vector-$y$ modes with the same mass $m$. 
	\item If the $+$ and $\times$ modes in the theory can only propagate at the speed of light, then the vector-$x$ and vector-$y$ modes do not exist.
	\item If the $+$ and $\times$ modes in the theory can only propagate at the speed of light, then the breathing and longitudinal modes can only appear in a mixed mode form. For any mixed mode, the amplitude ratio of the longitudinal mode to the breathing mode must be $m^{2}/\omega^{2}$, where $m$ is the mass of this mode, and $\omega$ is the frequency. (When $m=0$, it degenerates into a breathing mode.)
	\item If the amplitude ratio of the longitudinal mode to the breathing mode of a mixed mode is not $m^{2}/\omega^{2}$, then the theory must have a field equation higher than the second derivative and must have massive $+$, $\times$, vector-$x$, and vector-$y$ modes.
\end{itemize}

For the most general vector-tensor theory, the polarization modes of GWs also depend on the parameter spaces and can at most exhibit all six modes. However, vector-tensor theories generally do not satisfy the above four universal properties. For example, even under the condition that only the $+$ and $\times$ modes propagate at the speed of light, it is still possible for the vector-$x$ and vector-$y$ modes to exist. The detailed relationship between the polarization modes of GWs and the parameter spaces in these theories can be found in Refs. \cite{Dong:2023bgt,Dong:2024zal}.

In addition, gravitational quantum field theory (GQFT) predicts the existence of five independent GW polarization modes, in contrast to the two tensor modes allowed in GR~\cite{Gao:2024juf}. The GQFT is a general theory of the standard model which provides a unified framework for both  particle physics and cosmology and enables to understand inflationary Universe, dark matter and dark energy \cite{Wu:2025abi}. It has been demonstrated that future space-based detectors such as Taiji, LISA, and TianQin possess the sensitivity required to detect the spin-0 GW polarization mode predicted by GQFT, which is distinguishable from that arising in scalar-tensor theories such as the Brans-Dicke model~\cite{Xu:2025yrn}. Further details regarding the polarization modes in GQFT and their detection prospects with upcoming GW observatories are discussed in Ref.~\cite{Gao:2024juf}.

\subsubsection{Non-metricity and new polarization modes}

\begin{figure}[H]
	\centering
	\includegraphics[width=0.5\textwidth]{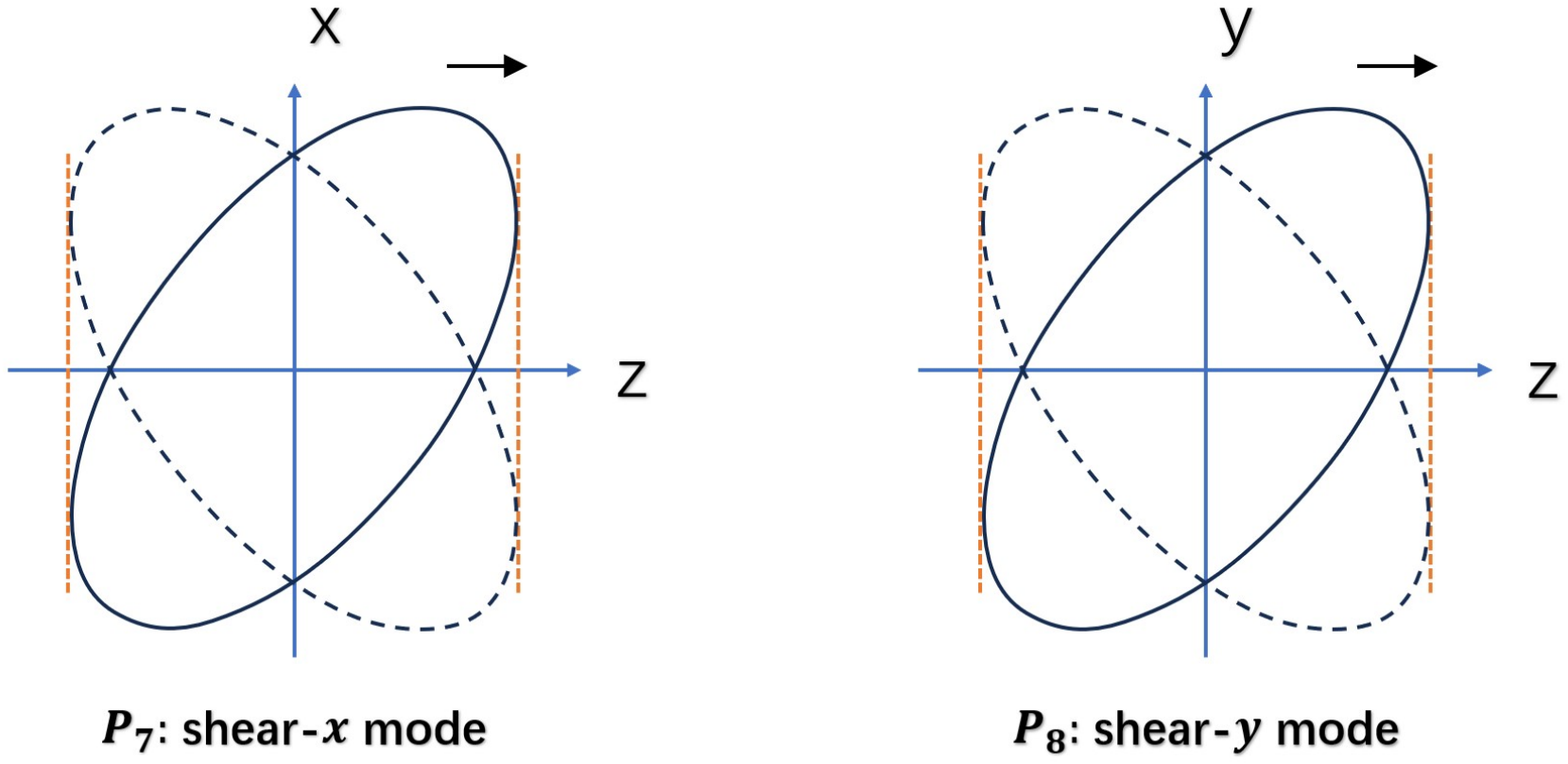}
	\caption{The two new shear modes \cite{Dong:2025ddi}.} 
	\label{fig2}
\end{figure}

Another common approach to modifying GR is to consider non-Riemannian geometry. This approach typically introduces the hypermomentum tensor into the theory. In this regard, we examined the polarization modes of GWs in a spacetime with non-metricity but no torsion in Ref.~\cite{Dong:2025ddi}. In such a spacetime, we found that the equation of relative motion satisfied by test particles carrying hypermomentum is modified compared to Eq.~\eqref{equation of geodesic deviation}. As a result, in addition to the original six polarization modes, two entirely new polarization modes may exist, which we refer to as shear modes. The relative motion of test particles under these two novel modes is illustrated in Fig.~\ref{fig2}.

In fact, the existence of shear modes is indeed possible. In Ref.~\cite{Dong:2025pre}, we studied the polarization modes of GWs in general symmetric teleparallel gravity. Our analysis revealed that when considering particles carrying hypermomentum, general symmetric teleparallel gravity always exhibits shear modes and a longitudinal mode propagating at the speed of light, regardless of the specific parameter choices. For a more detailed study of GW polarization modes in modified gravity theories within a torsionless spacetime, please refer to Refs. \cite{Dong:2025ddi,Dong:2025pre}.

%%%%%%%%%%%%%%%%%%%%%%%%%%%%%%%%%%%%%%%%%%%%%%%%%%%%%%%%%%%%%%%%%%%%%%%%%%%%%%%%%%%%%%
\section{Standard siren and cosmological parameters \label{sec:cp}}
%%%%%%%%%%%%%%%%%%%%%%%%%%%%%%%%%%%%%%%%%%%%%%%%%%%%%%%%%%%%%%%%%%%%%%%%%%%%%%%%%%%%%%
%contributor: Yungui Gong
The discovery and measurements of cosmic microwave background radiation and the discovery of accelerated expansion of the Universe through the observations of SNe Ia \cite{SupernovaSearchTeam:1998fmf,SupernovaCosmologyProject:1998vns} are among the most significant achievements in modern cosmology. 
To explain the cosmic acceleration, we usually invoke dark energy which exhibits a gravitationally repulsive force.
While the cosmological constant is the simplest candidate for dark energy, 
its theoretical estimation based on vacuum energy exceeds the observational value by 120 orders of magnitude \cite{Weinberg:1988cp}, highlighting a big challenge on the understanding of dark energy.
The inflationary scenario, along with observations of CMB, SNe Ia, baryon acoustic oscillation (BAO), gravitational lensing, and large-scale structure surveys have led to the establishment of the concord $\Lambda$CDM model.
However, the wealth of precise data not only gives highly accurate measurements of cosmological parameters, but also spots subtle tensions that suggest the presence of new physics beyond the $\Lambda$CDM model.
Notably, measurements of the CMB by the Planck satellite, based on the $\Lambda$CDM model, give the Hubble constant $H_0=67.36\pm 0.54$ km/s/Mpc \cite{Planck:2018vyg}. 
This result is in tension with the value of $H_0=73.04\pm 1.04$ km/s/Mpc determined by the SH0ES collaboration \cite{Riess:2022mme}, with a discrepancy at approximately the $5.3\sigma$ confidence level.
The CMB results rely on the $\Lambda$CDM model, highlighting the importance of studying model-independent determinations of cosmological parameters. 
Furthermore, the SNe Ia data suffers the problem of zero-point calibration, 
making it essential to develop independent distance measurement techniques to address the discrepancies in the Hubble constant.

The detection of GWs from binary mergers by LIGO, Virgo, KAGRA collaborations not only heralded the dawn of multi-messenger astronomy, 
but also opened a new observational window for probing the Universe \cite{LIGOScientific:2016aoc,LIGOScientific:2016emj,LIGOScientific:2018mvr,LIGOScientific:2020ibl,LIGOScientific:2021usb,KAGRA:2021vkt}. 
GWs can be used to measure cosmic distances, making their potential as standard sirens an independent method to address the problem of Hubble tension.
Once the host galaxy is identified, we can search for electromagnetic counterparts and use the distance-redshift relation to probe the Universe’s expansion.

Besides, GWs can be used to probe a special object which may populate our past and current Universe, gravitational atoms (GA)~\cite{Baumann:2019eav}. It is a rotating black hole (Kerr black hole) dressed with an oscillating bosonic cloud, which emits characteristic GWs with frequency at the mass of the boson~\cite{Baumann:2019eav}. In particular, if the Kerr black hole has a primordial origin as discussed in Section~\ref{sec:pbh}, the GA will become a primordial GA (PGA) and open a new window to the interplay between particle and cosmology~\cite{Kang:2024trj}. Considering that usually the PBH spin is not large, signature from PGAs provides a means to distinguish the origin of PBH. A good case in point is the PBH produced in the early matter-dominated era, which tends to yield Kerr black holes with large spin~\cite{harada2016primordial}.

\subsection{Null test and cosmological parameters}
The Data Release 1 (DR1) from the first year of observations by the Dark Energy Spectroscopic Instrument (DESI) gives measurements of the volume-averaged distance $D_V(z)/r_d$ 
at two effective redshifts $z_\text{eff}=0.30$
and $z_\text{eff}=1.48$, 
the comvoing angular diameter distances $D_M(z)/r_d$ and $D_H(z)/r_d$ at five different redshits, 
where $D_H(z)=1/H(z)$ and $r_d=r_s(z_d)$ is the sound horizon at the drag epoch $z_d$.
Assuming spatial flatness and using the Chevallier-Polarski-Linder (CPL) parametrization for the equation of state of dark energy \cite{Chevallier:2000qy,Linder:2002et}, 
the DESI BAO data yield the constraint $w_0=-0.55^{+0.39}_{-0.21}$ at the $1\sigma$ confidence level and $w_a<-1.32$ at the 95\% confidence level \cite{DESI:2024mwx}.
Combing the DESI BAO data and the {\it Planck} CMB data \cite{Planck:2018vyg}, the $1\sigma$ constraint is $w_0=-0.45^{+0.34}_{-0.21}$
and $w_a=-1.79^{+0.48}_{-1.0}$ \cite{DESI:2024mwx}. 
These results are in tension with the $\Lambda$CDM model at the $\sim 2.6\sigma$ significance level and suggest that dark energy is dynamical. \cite{DESI:2024mwx}. 
However, applying different dark energy models has shown that the evidence for a dynamical dark energy becomes weaker \cite{Gao:2024ily}. 
Furthermore, to use the DESI BAO data, we need to know the value of the cosmological parameter $r_d$ first.
This underscores the necessity for discussing model-independent methods for probing the Universe \cite{Lu:2024hvv}.

\begin{figure}[H]
\centering
\includegraphics[width=8.5cm]{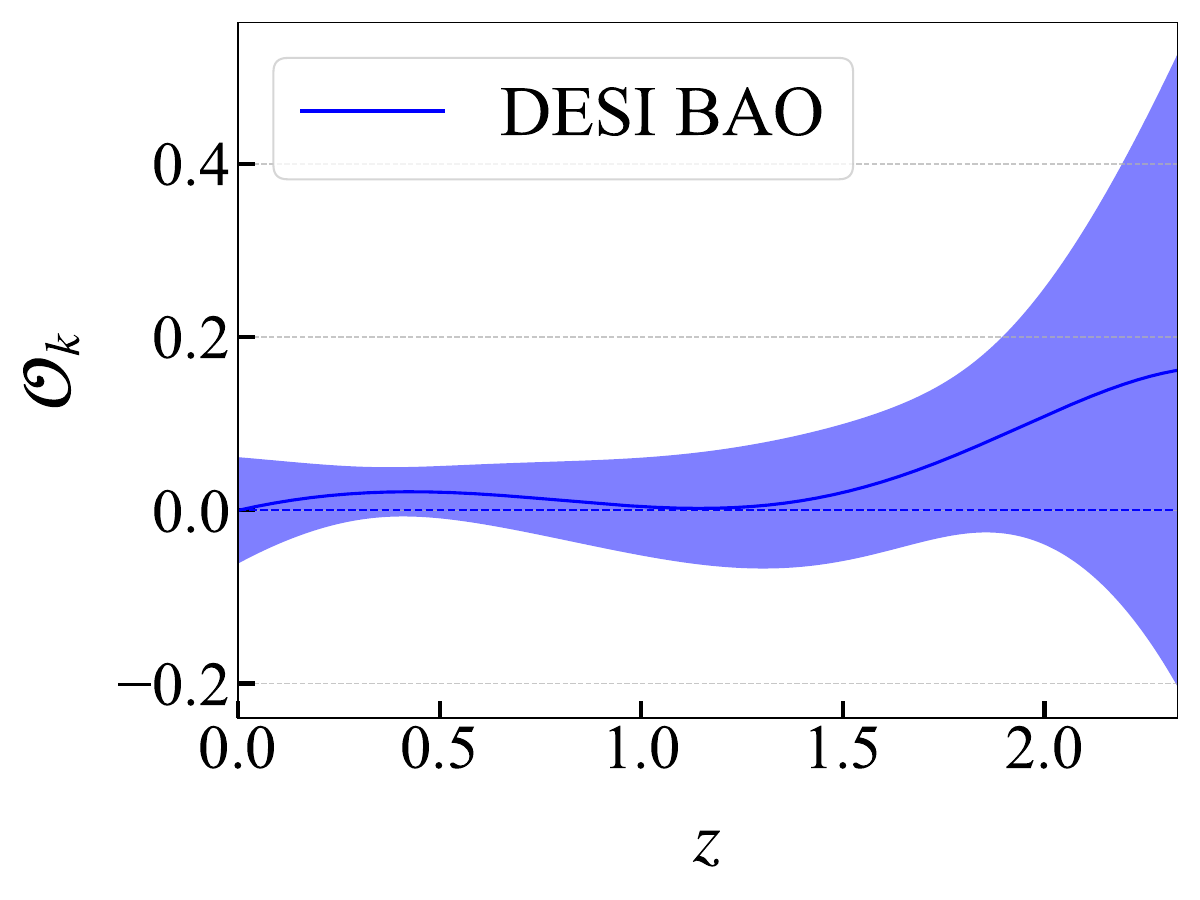}
\caption{The reconstructed $\mathcal{O}_k$ along with the $1\sigma$ error from DESI BAO data.
The solid line is for the reconstructed mean and the shaded region is for the $1\sigma$ error.
The dashed line represents $\mathcal{O}_k=0$.}
\label{fig:fap_Ok}
\end{figure}

To avoid the dependence on $r_d$ in the DESI BAO data, we use the AP parameter,
\begin{equation}
\label{apdef1}
F_{AP}(z)=\frac{D_M(z)}{D_H{z}}=\frac{D_M(z)/r_d}{D_H{z}/r_d}.
\end{equation}
Using the DESI BAO data and considering the covariance between $D_M$ and $D_H$, we derive five AP data points.
For a spatially flat Universe, the AP parameter $F_{AP}$ is
\begin{equation}
\label{eq:fap_ez}
F_{AP}=E(z)\int_0^z\frac{1}{E(x)}dx=E(z)D(z),
\end{equation}
so
\begin{equation}
E'(z)=E(z)\frac{F'_{AP}(z)-1}{F_{AP}(z)},
\end{equation}
where $E'(z)=dE(z)/dz$.
In the following, we discuss two null tests. The first one is the null test of the spatial flatness,
the $Ok$ diagnostic \cite{Clarkson:2007pz,Cai:2015pia,Gao:2025ozb},
\begin{equation}
\label{eq:fap_okz}
\mathcal{O}_k=F_{AP}\frac{D'}{D}-1=F_{AP}\frac{D_M'/r_d}{D_M/r_d}-1,
\end{equation}
where $D(z)=H_0 D_M(z)$.
For a flat Universe, $\mathcal{O}_k=0$.
Using the Gaussian Process (GP) method, we reconstruct $F_{AP}(z)$, $D_M(z)/r_d$ and $D'_M(z)/r_d$ from DESI BAO data and combine them to get $\mathcal{O}_k$ \cite{Gao:2025ozb}. 
The reconstructed $\mathcal{O}_k$ along with the $1\sigma$ error is shown in Fig. \ref{fig:fap_Ok}.
From Fig. \ref{fig:fap_Ok}, we see that a spatially flat Universe is consistent with the DESI BAO data.

\begin{figure}[H]
\centering
\includegraphics[width=8.5cm]{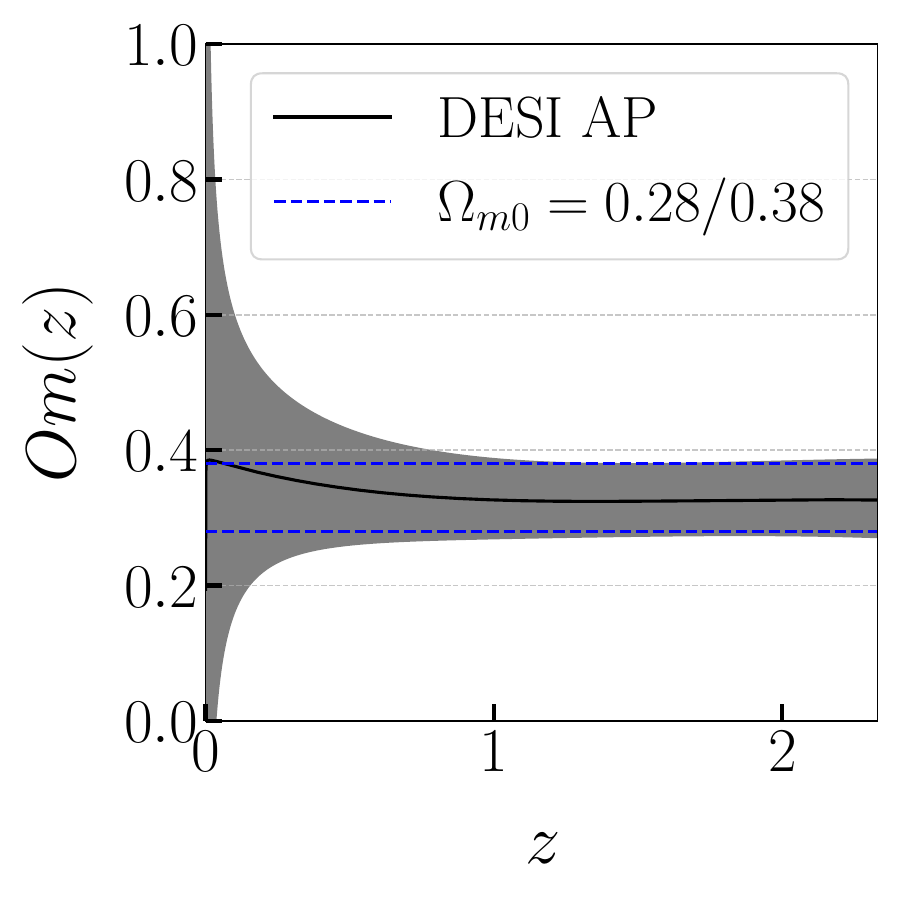}
\caption{The reconstructed $Om(z)$ along with the $1\sigma$ error from DESI BAO AP. The solid line is for the reconstructed mean and the shaded region is for the $1\sigma$ error. The blue dashed lines correspond to the results for the flat $\Lambda$CDM model with $\Omega_{m0}=0.28$ and $0.38$, respectively.}
\label{fig:fap_om}
\end{figure}

Now we discuss the $Om$ diagnostic \cite{Sahni:2008xx,Shafieloo:2012rs}
\begin{equation}
\label{om1}
Om(z)=\frac{E^2(z)-1}{(1+z)^3-1}.
\end{equation}
For the flat $\Lambda$CDM model, $Om(z)=\Omega_{m0}$ is constant, so the $Om$ diagnostic can be used as a null test of the flat $\Lambda$CDM model.
From the DESI BAO AP data, we reconstruct $E(z)$ and subsequently use this reconstruction to derive $Om(z)$, and the result is shown in Fig. \ref{fig:fap_om}.
From Fig. \ref{fig:fap_om}, it is evident that the flat $\Lambda$CDM model with $0.28\le \Omega_{m0}\le 0.38$ aligns with the DESI BAO data \cite{Gao:2025ozb}.

\subsection{Source localization}

To explore the Universe and address the Hubble tension using GWs as standard sirens \cite{Schutz:1986gp,Holz:2005df,LIGOScientific:2017adf,Chen:2017rfc,Wolf:2019hun,Riess:2019cxk},
accurate knowledge of the GW source’s position is crucial for follow-up observations of electromagnetic counterparts and statistical identification of the host galaxy when no counterpart is detected.
Since GW signals detected by ground-based GW observatory last only a few seconds to minutes,
a single detector cannot localize the source.
Because GWs propagate at the speed of light, 
a detection time delay occurs between multiple detectors.
This timing delay between widely separated ground-based observatories enables triangulation, 
helping to pinpoint the sky location of the GW source.
Therefore, three or more widely separated detectors are necessary to accurately locate GW sources using timing triangulation \cite{Fairhurst:2009tc,Fairhurst:2010is,Grover:2013sha}.

However, space-based GW detectors can observe GWs for months or even years. 
The detector's motion in space induces periodic Doppler shifts, 
causing amplitude and phase modulations in the received signals. 
These modulations carry information about both the detector’s trajectory and the source’s angular position.
As a result, a single space-based GW detector can provide accurate sky localization for the source.

There are two primary configurations and designs for space-based GW observatories. One features a geocentric orbit with three spacecraft orbiting the Earth while collectively rotating around the Sun along with Earth. TianQin adopts this design, with the normal vector of its detector plane pointing towards the source RX J0806.3+1527 \cite{TianQin:2015yph}. In contrast, LISA \cite{Danzmann:1997hm,Audley:2017drz} and Taiji \cite{Hu:2017mde,Ruan:2018tsw} use a heliocentric orbit, where their three spacecraft travel ahead of or behind the Earth by approximately $20^\circ$. Both LISA and Taiji constellations are inclined at $60^\circ$ relative to the ecliptic plane. This inclination maintains an equilateral triangular formation of the spacecraft throughout the mission. Over the course of a year, the detector plane’s normal vector rotates around the ecliptic plane’s normal, describing a conical motion with a half-opening angle of $60^\circ$.

\begin{figure*}
\centering
\includegraphics[width=0.8\textwidth]{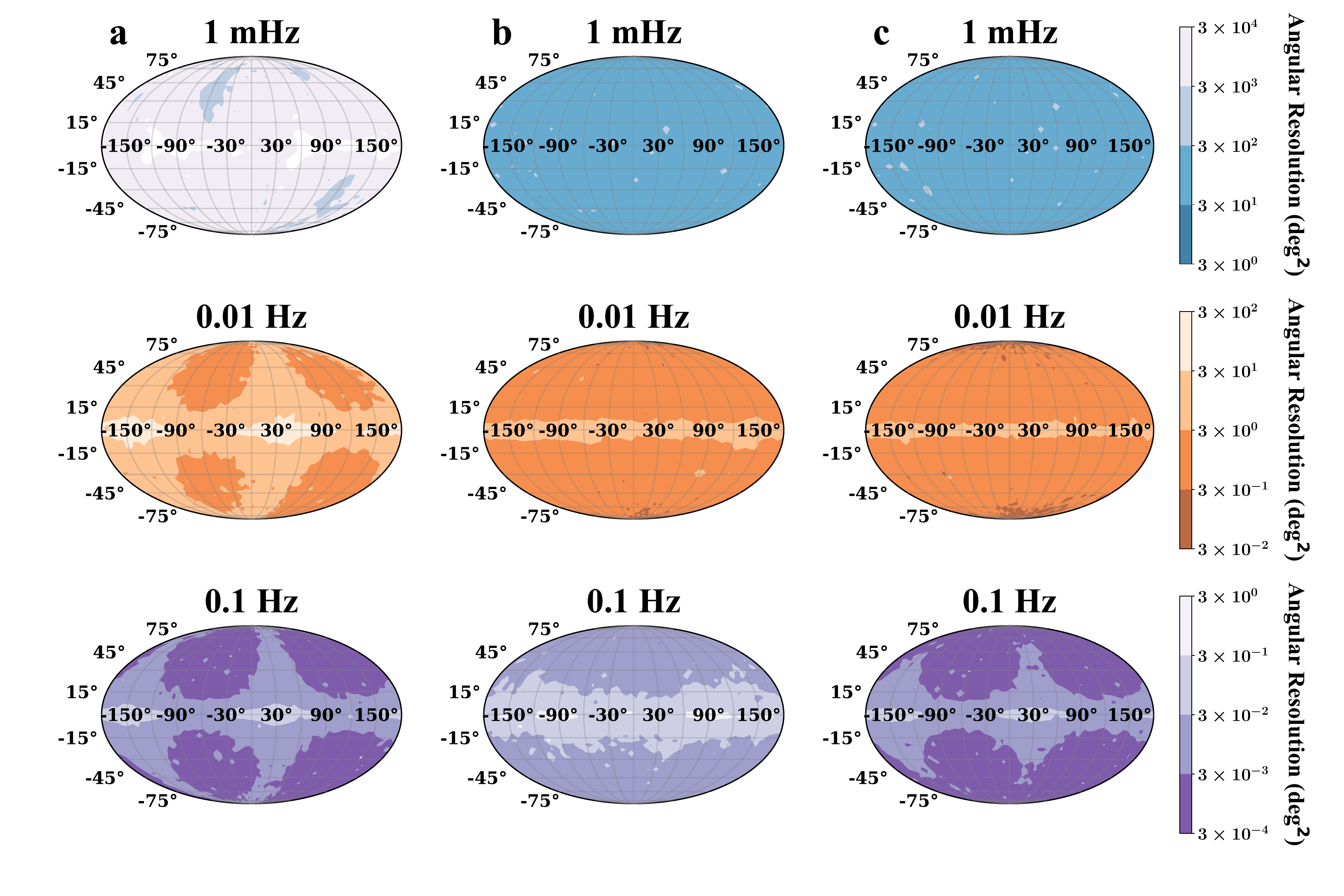}
\caption{Sky maps for monochromatic sources \cite{Gong:2021gvw}. The unit for the angular resolution is deg$^2$.
{\bf a,} Sky maps of angular resolution for TianQin at frequencies of 1 mHz (top), 0.01 Hz (middle) and 0.1 Hz (bottom).
{\bf b,} The corresponding sky maps of angular resolution for Taiji.
{\bf c,} The corresponding sky maps of angular resolution for the TianQin-Taiji network.}
\label{net1}
\end{figure*}

To investigate how the constellation of the detector, 
such as noise, arm length, time-changing orientation, and motion around the Sun,
impact source localization accuracy,
we simulate 3600 monochromatic sources uniformly distributed in the sky with $-\pi/2<\theta_s<\pi/2$ and $-\pi<\phi_s<\pi$ with frequencies $1$ mHz, $0.01$ Hz and $0.1$ Hz, respectively \cite{Zhang:2020hyx}.
The signal-to-noise ratio (SNR) of the sources is chosen to be 7 with respect to LISA.
Using the Fisher Information Matrix (FIM) approximation,
we calculate the angular resolution of the sources and the sky maps of the angular resolutions are shown in Fig. \ref{net1} \cite{Zhang:2020hyx,Gong:2021gvw}.
Our results show that the amplitude modulation caused by LISA’s annually varying detector plane orientation improves its sky localization accuracy and coverage at frequencies below several millihertz.
In contrast, this effect is negligible for TianQin, whose detector plane orientation remains fixed.
However, at frequencies above approximately 30 mHz,
TianQin outperforms LISA in sky localization.

In the low-frequency regime, the angular resolution scales as $S_n(f)/f^2$ for TianQin, whereas for LISA it scales as $S_n(f)$. This difference arises because TianQin lacks amplitude modulation, resulting in distinct localization patterns compared to LISA. In the medium-frequency range, Doppler phase modulation becomes dominant and influences only the angular resolution. Consequently, for both TianQin and LISA, the angular resolution scales as $S_n(f)/f^2$. At high frequencies, TianQin achieves smaller parameter estimation errors than LISA, primarily due to its shorter arm length \cite{Zhang:2020drf,Zhang:2020hyx}. We also discuss the effects of various first-generation time-delay interferometry (TDI) combinations on the estimation of angular resolution for monochromatic sources \cite{Jiang:2023hrw}. The inclusion of the TDI Michelson ($X, Y, Z$) combinations has little impact on the overall sky localization capability. We find that differences in localization performance among the TDI combinations are entirely due to their varying sensitivities. Of the six TDI combinations, the Michelson ($X, Y, Z$) configuration offers the most accurate source localization.

\begin{figure*}
\centering
\includegraphics[width=0.8\textwidth]{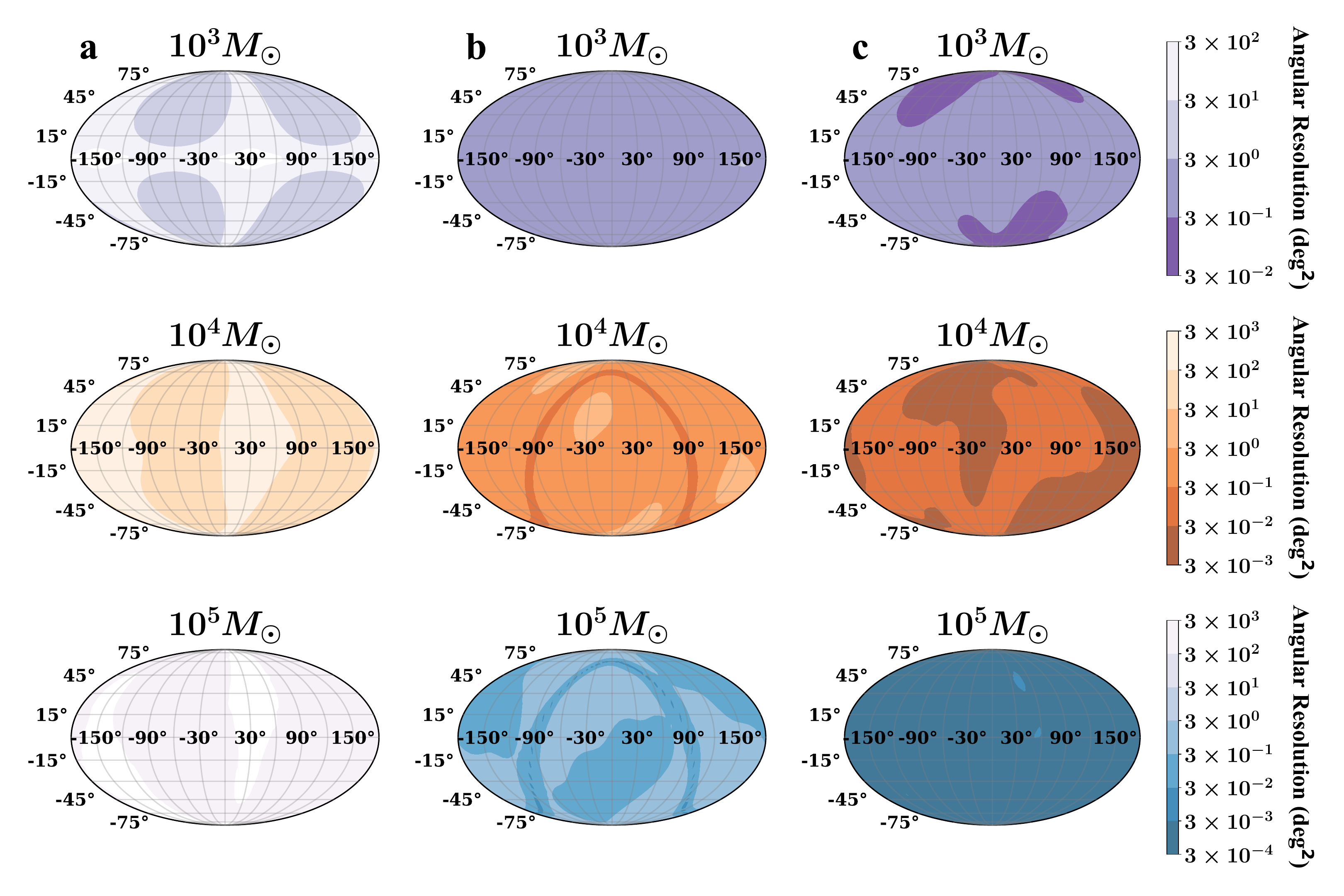}
\caption{Sky maps for coalescence sources \cite{Gong:2021gvw}. The unit for angular resolution is deg$^2$.
{\bf a,} Sky maps of angular resolution for TianQin for coalescence BBHs with equal masses of
$10^3 M_\odot$ (top), $10^4 M_\odot$ (middle) and $10^5 M_\odot$ (bottom) at $z=1$.
{\bf b,} The corresponding sky maps of angular resolution for Taiji.
{\bf c,} The corresponding sky maps of angular resolution for the TianQin-Taiji network.}
\label{net2}
\end{figure*}

For mergers of equal-mass binary black holes (BBHs) at redshift $z=1$, we consider component masses of $m_1=m_2=10^3\ M_\odot$, $10^4\ M_\odot$ and $10^5\ M_\odot$, respectively, and simulate the GW signals one year prior to coalescence. The results are presented in Fig. \ref{net2} \cite{Gong:2021gvw}. As expected, the transfer function aids Taiji in accurately localizing the sky positions of coalescing binaries with a total mass around $\sim10^5\ M_\odot$.

Space-based GW detectors are capable of detecting ringdown signals from mergers of massive BHs. For binaries with total masses $M\ge 10^5\ M_\odot$, these detectors can effectively localize sources using the ringdown signal. The transfer functions of LISA, Taiji, and TianQin enhance source localization for binaries with redshifted total masses in the ranges of $1.7\times10^5\ M_\odot\le M_z \le 1.7\times10^6\ M_\odot$, $2\times10^5\ M_\odot\le M_z \le 2\times10^6\ M_\odot$, and $1.2\times10^4\ M_\odot\le M_z \le 1.2\times10^5\ M_\odot$, respectively \cite{Zhang:2021kkh}. For BBHs at the same distance, LISA provides better localization for $M_z=6.5\times10^6\ M_\odot$,
while TianQin achieves better localization for  $M_z=3\times10^6\ M_\odot$.

As shown in Figs. \ref{net1} and \ref{net2},
the detector networks significantly improve sky localization and coverage compared to individual detectors. The TianQin-Taiji network enhances localization by nearly two orders of magnitude for BBHs with masses $m_1=m_2=10^4\ M_\odot$, and more than two orders for BBHs with masses $m_1=m_2=10^5\ M_\odot$, relative to Taiji alone, with further improvements by a factor of a few when LISA is included. The primary factor driving the improvement is the angle between the normal vectors of the detector planes \cite{Zhang:2021wwd}. Optimal angular resolution is achieved when this angle lies between $40^\circ-140^\circ$, which explains the dramatic enhancement in source localization by the LISA-Taiji, LISA-TianQin, and Taiji-TianQin networks compared to individual detectors. These networks notably improve localization for both short and long GW signals, especially when most of the SNR arises from brief signal segments. The improvement is even more pronounced for heavier sources, where the dominant SNR is concentrated in a shorter time interval. Time delays between detectors in the network further reduce localization errors by factors of a few up to an order of magnitude, except for binaries with masses $M\ge10^7\ M_\odot$ during the inspiral phase. Higher harmonics play a lesser role, while the transfer function benefits localization primarily for heavier sources \cite{Zhang:2021wwd}. These results are helpful for designing space-based detector networks to optimize localization accuracy and enhance scientific outcomes.

\subsection{Calibration of type Ia supernova with GWs}

SNe Ia suffer from the zero-point calibration problem, 
making it crucial to obtain other model-independent measurements of cosmic distances.
These measurements can then be used to calibrate SNe Ia data.
GWs provide a promising solution by offering model-independent and highly precise luminosity distance measurements.
For this reason, GWs from binary neutron star (BNS) mergers can serve as standard sirens to study the evolution of the Universe \cite{Schutz:1986gp,Holz:2005df}.
A key issue of using GWs as standard sirens is the identification of electromagnetic counterparts,
which depends heavily on the precise sky localization of GW sources. 
With accurate localization of the host galaxy of GWs from either MBBH or BNS coalescences, and if MBBH mergers and SNe Ia occur within the same host galaxy,
it becomes possible to calibrate standard candles using standard sirens \cite{Zhao:2017imr,Gupta:2019okl,Lu:2022wuk}.

To calibrate SNe Ia data using GW standard sirens, we assume that for each GW source, 
there exists an SN Ia located in the same host galaxy.
We then use the parameters ($\theta$, $\phi$, $z$) 
from the SNe Ia data to simulate corresponding GW sources. 
For the simulation, we employ the Pantheon sample of SNe Ia \cite{Pan-STARRS1:2017jku} and consider three population models for MBBHs: pop III, Q3d, and Q3nod \cite{Klein:2015hvg, Wang:2020dkc}.
Using GWs from these MBBH mergers, 
we estimate the luminosity distance error and angular resolution via the FIM method. 
The results show that the Q3nod model provides better constraints on the luminosity distance at redshift
$z\lesssim 1.5$,
while the pop III model yields better angular resolution.
Taking the median values from the Q3nod model combined with the LISA-Taiji-TianQin detector network:
$d_L\sim 1300$ Mpc, $\Delta d_L\sim 0.8$ Mpc, $\Delta \Omega_s\sim 5.1\times 10^{-5}$ deg$^2$,
we estimate the uncertainty volume of the source’s location as
$\Delta V\sim  6.7\times 10^{-8}$ Gpc$^3$ \cite{Lu:2022wuk}.
Given that the galaxy number density is approximately
$3\times 10^6$ Gpc$^{-3}$ \cite{Gupta:2019okl},
this localization error corresponds to fewer than one field galaxy within the volume.
Therefore, if an SN Ia is present in the same host galaxy as the GW source, standard candles can be effectively calibrated using standard sirens.

\begin{figure}[H]
\centering
\includegraphics[width=8.5cm]{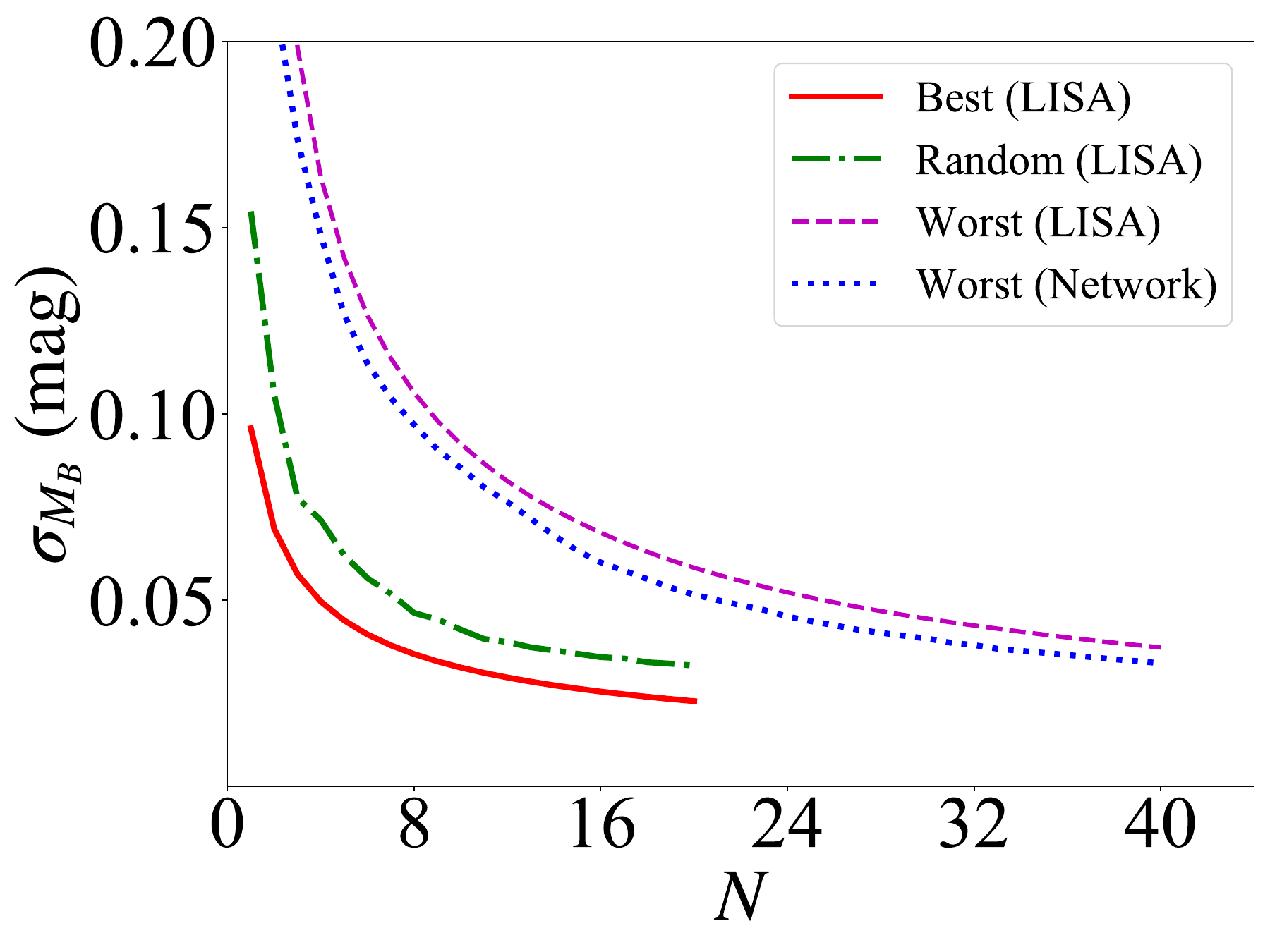}
\caption{The dependence of $\sigma_{M_{B}}$ on the number of calibrators $N$ for the pop model.
The red solid and green dash-dot lines show the estimated 1$\sigma$ error of $M_{B}$ for the best and random scenarios with LISA,respectively. The magenta dashed line represents the estimated 1$\sigma$ error for the worst scenario with LISA, while the blue dotted line corresponds to the worst scenario using the combined LISA-Taiji-TianQin network. }
\label{multipleMBfig}
\end{figure}

Assuming we can identify $N$ pairs of MBBH mergers and SNe Ia occurring in the same host galaxy, we then have
$N$ GW-calibrated SNe Ia to reduce the statistical error. 
We consider three scenarios: the best case, 
which uses SNe Ia with the smallest measurement errors in their apparent magnitudes; 
the worst case, which uses those with the largest $\sigma_{m_B}$; and a random case, 
where SNe Ia are selected randomly.
The relationship between the uncertainty
$\sigma_{M_{B}}$ and the number of calibrators
$N$ is shown in Fig.~\ref{multipleMBfig} \cite{Lu:2022wuk}. 
From this figure, it is evident that the error in
$M_{B}$ decreases as $N$ increases.
However, due to observational limitations imposed by
$\sigma_{m_B}$, 
further improvement in $\sigma_{M_B}$ becomes marginal beyond a certain number of calibrators.
For the best and random scenarios, the results obtained using LISA alone and the combined LISA-Taiji-TianQin network are similar. 
In the worst scenario, however, 
the LISA-Taiji-TianQin network achieves better constraints on
$\sigma_{M_B}$ than LISA alone. 
Although the network does not significantly reduce
$\sigma_{M_B}$ in the best and random cases, 
its substantially improved GW source localization may be crucial for reliably identifying the corresponding SN Ia counterparts.

\begin{figure}[H]
\centering
\includegraphics[width=8.5cm]{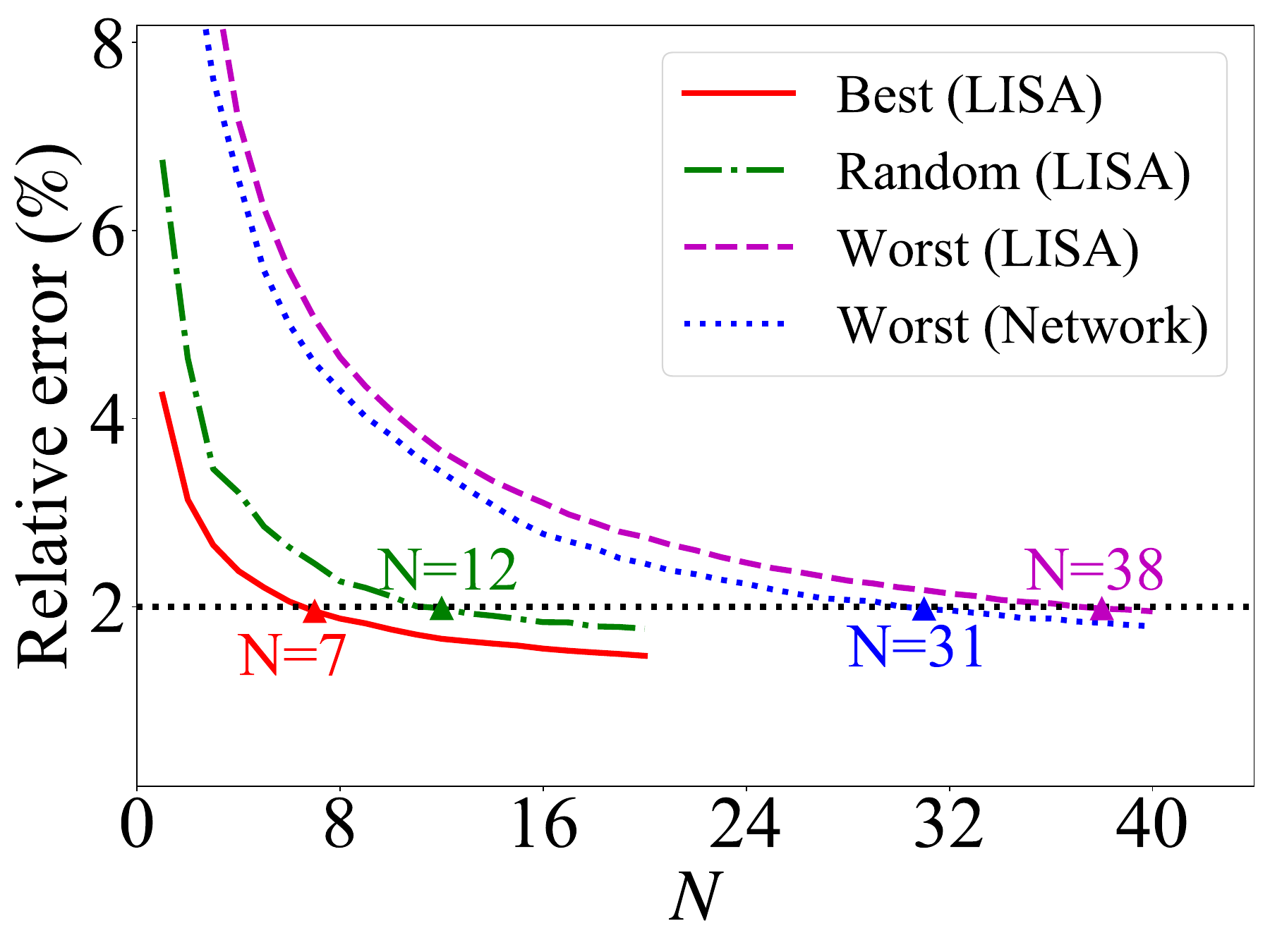}
\caption{The relative error of $H_{0}$ with the pop model.
The triangle marks the minimum number of calibrators
$N$ required for the relative error in $H_{0}$ to fall below  2\%.The red solid and green dash-dot lines show the constrained relative error of $H_{0}$ for the best and random scenarios with LISA, respectively.
The magenta dashed line represents the relative error in the worst scenario with LISA, while
the blue dotted line corresponds to the worst scenario using the LISA-Taiji-TianQin network.}
\label{multipleH0fig}
\end{figure}

Using the re-calibrated SNe Ia data, we analyze 
237 SNe Ia within the redshift range $0.023< z < 0.15$ to constrain the Hubble constant $H_0$, ensuring the result is independent of any cosmological model.
In the low-reshift regime,
we get the $d_L-z$ relation \cite{Gong:2004sd},
\begin{equation}
\label{dl*}
d_{L}(z)=\frac{c z}{H_{0}}\left[1+\frac{\left(1-q_{0}\right) z}{2}+O\left(z^{2}\right)\right],
\end{equation}
where $q_{0}$ is the deceleration parameter.
By fitting the 237 SNe Ia data with Eq.~\eqref{dl*}, 
we obtain the results shown in Fig.~\ref{multipleH0fig} \cite{Lu:2022wuk}.
These results demonstrate that, through calibrating the luminosity distances of approximately 10 SNe Ia using GWs, 
we can achieve a model-independent determination of the local Hubble constant $H_0$ with better than 2\% precision in the redshift range $0.023\le z\le 0.15$.
However, due to the inherent measurement uncertainties in the apparent magnitudes of SNe Ia, 
increasing the number of calibrated SNe Ia beyond this provides limited improvement in reducing the relative error of $H_0$.
It should be noted that since the luminosity distances of MBBHs were simulated assuming a flat $\Lambda$CDM model,
the exact central value of $H_0$ derived here may not be fully reliable. 
Nevertheless, the estimated uncertainty in $H_0$ remains model-independent. 
Once space-based GW detectors provide observations of MBBH mergers, 
this method can robustly determine the local value of
$H_0$ with precision better than 2\%.

The above simulation assumes a flat
$\Lambda$CDM model with $H_{0} = 67.27$ km/s/Mpc.
To assess the impact of different cosmological parameter choices, 
we also performed simulations using
$H_{0} = 73.00$ km/s/Mpc and $\Omega_{m0}=0.3166$ \cite{Riess:2020fzl},
finding that the results remain largely consistent.
In the best-case scenario, the relative error in $H_{0}$
falls below 2\% with just 7 calibrators using either LISA alone or the combined LISA-Taiji-TianQin network. 
For the random scenario, achieving under 2\% relative error requires 13 calibrators with LISA, 
while in the worst-case scenario, 
38 calibrators are needed.
When using the LISA-Taiji-TianQin network, 
the number of calibrators required to reach 2\% precision reduces to 12 for the random scenario and 32 for the worst scenario.

We conclude that at least seven SNe Ia, with their luminosity distances calibrated by GWs, are required to achieve a 2\% precision in determining the local Hubble constant. 
This measurement, derived from SNe Ia re-calibrated by GW standard sirens, 
is free from the zero-point calibration issues and cosmological model dependencies. 
Consequently, this model-independent determination of the local Hubble constant using GW-calibrated SNe Ia offers a promising avenue to address the Hubble tension.

\subsection{Primordial gravitational atoms}

When a spin-0 bosonic field $\Phi$ with mass $\mu_S$ is incident upon a rotating black hole (Kerr black holes) with mass $M_{\rm BH}$ and dimensionless spin $\tilde{a} \equiv \frac{M_{\rm PL}^2}{M_\text{BH}^2} J_\text{BH}$, where $J_\text{BH}$ is the angular momentum of the black hole, the dynamics of $\Phi$ in the Kerr spacetime are governed by the Klein-Gordon equation. Define the gravitational fine-structure parameter $\alpha \equiv \mu_S M_\text{BH} / M_{\rm PL}^2$. For $\alpha\ll 1$, the  gravitational bound system of black hole and $\Phi$ is like a hydrogen atom, with the bosonic cloud state labeled with $|nlm\rangle$, where $n= 1,2,3\ldots$, $l=0,1,\ldots n-1$, and $m=-l, -l+1,\ldots l$. Superradiance instability~\cite{zel1967d} may occur for the state with
\begin{align}
	\tilde{a} > \frac{4 \alpha/m}{4 \alpha^2/m^2 + 1} \ 
	\& \& \
	\alpha < \frac{m}{2}.
\end{align}
Then, such states, with $l=m=1,n=l+1=2$ the dominant one, can extract energy and angular momentum from the black hole.

\begin{figure*}
\centering
\includegraphics[width=18cm]{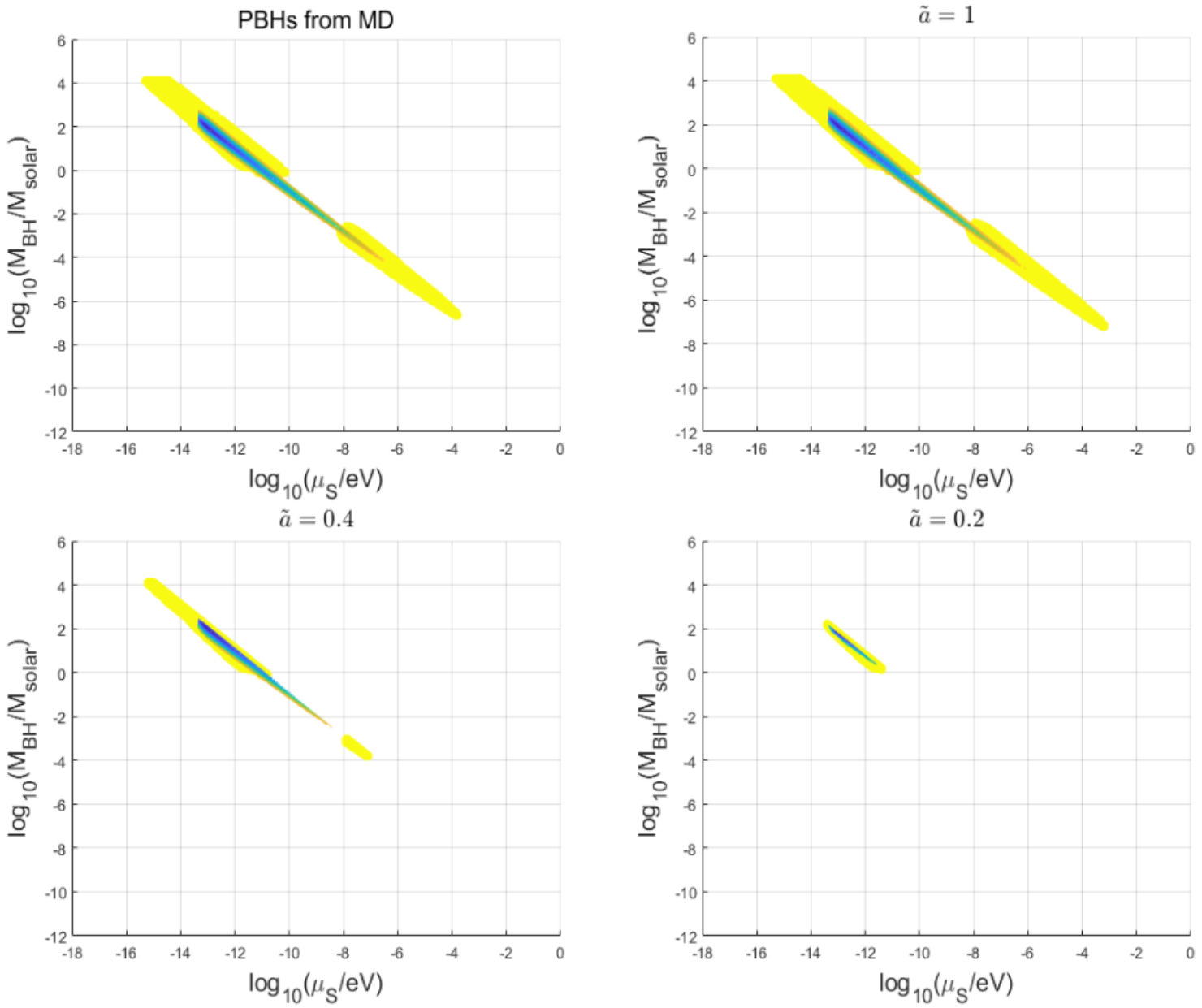} 
\caption{GWs constraints (blue and green bands) and prospects (yellow bands) of PGA with a scalar cloud, taking different spins $\tilde a$ with the first panel from a specific model where PBH is produced in a short matter-dominated era. In showing the prospects, we have collected all the available constraints on PBH, taking the maximal fraction of PBH in the total energy density of dark matter today. \textcolor{black}{Copied from Ref.~\cite{Kang:2024trj} with permission.}}
\label{paraspace_scalar}
\end{figure*}

After the cloud grows to the saturated state, it starts to decay due to the emission of GWs with an initial frequency of $\mu_S / \pi$, redshifted to $f = \frac{\mu_S}{ \pi (1+z)}$. Since each PGA generates GWs, given the randomness of their phases, they overlap to form an isotropic and homogeneous stochastic GW background in the Universe. A significant distinction between PGAs and astronomical GAs is that the GWs of PGAs contain components from the early Universe, which results in a spectrum~\cite{Kang:2024trj} 
\begin{equation}
\label{omega_estimation}
    \Omega_{\rm GW} \propto
    \begin{cases}
        f^3 & \quad f \in (f_0, f_{\rm peak}) \\
    f_{\rm peak}^4 f^{-1} & \quad f \in (f_{\rm peak}, \frac{\mu_S}{\pi}),
    \end{cases}
\end{equation}
where the peak frequency is determined by $\frac{f_{\rm peak}}{\mu_S} = \sqrt{\frac{3\tau_{\rm GW}}{\tilde{\tau}}}$ with $\tau_{\rm GW}$ the characteristic time for the consumption of the cloud and $\tilde{\tau}\sim 10^{12} \rm years$. The lower cut $f_0$, originates from GWs emitted at time $t_0$, while the higher cut $\frac{\mu_S}{\pi}$ comes from GWs emitted today. Using the current data, we can make a constraint in the $\mu_S-M_{\rm BH}$ plane, shown in Fig.~\ref{paraspace_scalar}, where the future prospect is also demonstrated. PGAs with spin-1 boson cloud can be discussed similarly and it is found that the vector boson spin significantly helps to enhance the detection prospect on this type of PGA~\cite{Kang:2024trj}. 

\section{Conclusion and outlook \label{sec:co}}

In this review paper, we have presented a comprehensive overview of recent progress in detecting and understanding cosmological GW sources, emphasizing their implications in probing cosmology and gravity. Our review article has covered a range of theoretical mechanisms for producing SGWBs, including inflationary dynamics, cosmological first-order phase transitions, evolving networks of topological defects, and the formation and merger history of PBHs. Additionally, we have also discussed the potential of GWs for exploring the nature of gravity, including testing the Lorentz and parity symmetry of gravity and detecting extra polarization modes beyond GR. Finally we also discuss the potential of GWs to measure the expansion history of the Universe, and a completely independent way to determine the Hubble constant using standard sirens. The results and topics brought together in this paper highlight the unique role of GWs in probing the early Universe, testing gravity, and providing independent measurements of cosmological parameters. 

The next generation of GW detectors, including space-based missions like LISA, Taiji, and Tianqin, as well as advanced ground-based observatories such as the ET and CE, will play a pivotal role in GW cosmology. With significant improved sensitivity and broader frequency range, these detectors will enable the detection of signals from the earliest stages of the Universe, potentially revealing the imprints of inflation, cosmological phase transitions, topological defects, and primordial black holes. They will also provide more precise measurements of the expansion history via standard sirens. By probing a much wider range of GW sources with higher redshifts, these future observations will greatly enhance our ability to test fundamental physics, distinguish between cosmological models, and deepen our understanding of the Universe's origin, evolution, and the nature of gravity itself.

%%%%%%%%%%%%%%%%%%%%%%%%%%%%%%%%%%%%%%%%%%%%%%%%%%%%%%%%%%%%%%%%%%%%%%%%%%%%%%%%%%%%%%
%\section{Conclusions \label{sec:cp}}
%%%%%%%%%%%%%%%%%%%%%%%%%%%%%%%%%%%%%%%%%%%%%%%%%%%%%%%%%%%%%%%%%%%%%%%%%%%%%%%%%%%%%%
%contributor: 

%%%%%%%%%%%%%%%%%%%%%%%%%%%%%%%%%%%%%%%%%%%%%%%%%%%%%%%%%%%%%%%%%%%%%%%%%%%%%%%%%%%%%%
\section*{Acknowledgments}
This work is supported by the National Key Research and Development Program of China Grant No. 2020YFC2201500, which consists of four projects with Grant Nos. 2020YFC2201501, 2020YFC2201502, 2020YFC2201503, and 2020YFC2201504.

%\Acknowledgements{}

\InterestConflict{The authors declare that they have no conflict of interest.}

%\Authorfootnote{}

% \newpage
%\bibliographystyle{scichina}
%\bibliographystyle{scpma}
%\bibliography{gwc}

\end{multicols}
\end{document}